\documentclass[12pt]{article}
\usepackage{amsmath}
\usepackage{graphicx}
\usepackage{enumerate}
\usepackage{natbib}
\usepackage{url} % not crucial - just used below for the URL 
\usepackage{lscape} 

\usepackage[utf8]{inputenc}
\usepackage{amsthm}
\usepackage{amssymb}
\usepackage{color}
\usepackage{comment}
\usepackage{appendix}
\usepackage{bm}
\usepackage{titlesec}
\titlelabel{\thetitle.\quad}

\usepackage{tikz}
\usetikzlibrary{shapes,snakes, graphs}

\usetikzlibrary{arrows,%
                petri,%
                topaths}%
\usepackage{pgf}
\usepackage{subfigure,tikz}
\usetikzlibrary{arrows,automata}

\PassOptionsToPackage{hyphens}{url}
\usepackage[
    colorlinks=true,
    linkcolor=blue!50!black,
    citecolor=blue!50!black,
    urlcolor=blue!50!black,
    filecolor=blue!50!black,
    breaklinks=true,
    pdfpagemode=UseNone,
    pdfstartview=FitH,
    bookmarksnumbered=true,
    bookmarksopen=true
]{hyperref}

\hypersetup{
    pdftitle={Measurement Error-Robust Causal Inference via Constructed Instrumental Variables},
    pdfauthor={Caleb H.~Miles, Linda Valeri, Brent Coull},
    pdfsubject={Statistics - stat.ME},
    pdfkeywords={Causal inference, Environmental health, Instrumental variable, Measurement error, Mediation analysis, Nutrition}
}

\usepackage{float}

\newcommand{\blind}{1}

\newtheorem{assumption}{Assumption}
\newtheorem{theorem}{Theorem}
\newtheorem{remark}{Remark}

\def\ci{\mbox{\ensuremath{\perp\!\!\!\perp}}}
\newcommand{\lv}[1]{\textcolor{magenta}{#1}}

\DeclareMathOperator*{\argmin}{arg\,min}

\allowdisplaybreaks[1]

\makeatletter
\renewcommand\NAT@open{\textcolor{black}{(}}
\renewcommand\NAT@close{\textcolor{black}{)}}
\makeatother

\begin{document}

\def\spacingset#1{\renewcommand{\baselinestretch}%
{#1}\small\normalsize} \spacingset{1}

%%%%%%%%%%%%%%%%%%%%%%%%%%%%%%%%%%%%%%%%%%%%%%%%%%%%%%%%%%%%%%%%%%%%%%%%%%%%%%

\if1\blind
{
    \title{%Working title: %Measurement Error-Robust Causal Inference via Synthetic Instrumental Variables\\
    %Alternative title: 
    Measurement Error-Robust Causal Inference via Constructed Instrumental Variables}% While Accounting for Measurement Error}% via Synthetic Instrumental Variables}
    %Measurement error robust total, direct, and indirect effect estimation via synthetic instrumental variables
    %Synthetic instruments: A measurement-error robust method for estimating average causal effects and mediated effects with error-prone confounders and exposures}
    \author{Caleb H.~Miles, Linda Valeri, Brent Coull \thanks{Caleb H. Miles (Columbia Mailman School of Public Health,
    722 W 168th Street, 6th floor, 
New York, NY 10032, USA, \url{cm3825@cumc.columbia.edu}) is Assistant Professor and Linda Valeri is Associate Professor, Department of Biostatistics, Columbia University, New York, NY 10032. Brent Coull is Professor, Departments of Biostatistics and Environmental Health, Harvard T.H. Chan School of Public Health, Boston, MA 02115.}\hspace{.2cm}}
    \date{}
    \maketitle
    %\bigskip
} \fi

\if0\blind
{
  \bigskip
  \bigskip
  \bigskip
  \begin{center}
    {\LARGE\bf Measurement Error-Robust Causal Inference via Constructed Instrumental Variables}
\end{center}
  \medskip
} \fi

\bigskip

\begin{abstract}
\noindent %While measurement error can sometimes be benign, it is often harmful when estimating causal effects. 
Measurement error can often be harmful when estimating causal effects. 
Two scenarios in which this is the case are in the estimation of (a) the average treatment effect when confounders are measured with error and (b) the natural indirect effect when the exposure and/or confounders are measured with error. Methods adjusting for measurement error typically require external data or knowledge about the measurement error distribution. Here, we propose methodology not requiring any such information. Instead, we show that when the outcome regression is linear in the error-prone variables, consistent estimation of these causal effects can be recovered using constructed instrumental variables under certain conditions. These variables, which are functions of only the observed data, behave like instrumental variables for the error-prone variables. Using data from a study of the effects of prenatal exposure to heavy metals on growth and neurodevelopment in Bangladeshi mother-infant pairs, we apply our methodology to estimate (a) the effect of lead exposure on birth length while controlling for maternal protein intake, and (b) lead exposure's role in mediating the effect of maternal protein intake on birth length. Protein intake is calculated from food journal entries, and is suspected to be highly prone to measurement error.

\end{abstract}
\noindent%
\vfill

\newpage
\spacingset{1.5} % DON'T change the spacing!
\section{Introduction}
\label{sec:intro}
Covariate measurement error is a prevalent and frequently overlooked problem in data analysis. %It is often thought to merely have an attenuating effect on estimates; however, this is not true in general. 
We consider two cases arising in causal inference in which measurement error can induce bias away from the null: estimation of the average treatment effect (ATE) in the presence of error-prone confounders, and estimation of mediated effects such as the natural indirect effect (NIE) in the presence of an error-prone exposure and/or confounders. While there are many methods to adjust for measurement error in a regression setting, most involve either some form of knowledge about the measurement error distribution, such as its variance, or external data, such as replicates of the error-prone variable, validation data, or instrumental variables (IVs). %Absent such information, neither the ATE nor the NIE are nonparametrically identified in general. 
In this article, we develop estimators for the ATE and the NIE in the presence of measurement error that do not depend on any such external information.

Our motivating example comes from an environmental epidemiology study conducted in Bangladesh, in which investigators measured maternal protein intake, lead exposure, and birth length in a cohort of mother-infant pairs. We are interested in the effects of lead exposure on birth length, as well as the role that lead exposure plays in mediating the effect of protein intake on birth length. Protein intake is thought to be measured with error, and plays the role of an error-prone exposure or confounder in these two cases. %, whereas in the second case it plays the role of an error-prone confounder.

%the familiar case in which one wishes to estimate an average treatment effect (ATE) from observational study data. The second is when the effect of measurement error can be harmful is in the context of mediation analysis. 
%We start by discussing the first case. 
One of the simplest approaches for estimating the ATE is to control for potential confounders using ordinary least squares (OLS). However, even when the no unobserved confounding assumption holds with respect to the true values of the covariates and the linear model is correctly specified, if any true confounder is measured with error, then the resulting OLS estimator of the treatment effect will be biased toward the crude estimate, %i.e., the estimate arising from omitting the error-prone confounder altogether. Since omitted-variable bias may 
which can be in either direction. %This result extends to more general models. 
In the Bangladesh study example, this means that failing to adjust for measurement error in the confounder protein intake can lead to biased estimation of the ATE of lead exposure on birth length in either direction. %, the measurement error can likewise induce bias in either direction. 
\cite{lenis2017doubly} and \cite{shu2018estimation} developed methods for estimating the ATE in the presence of confounder measurement error when external information is available, such as a validation sample or knowledge of the measurement error distribution. %In contrast, we develop methodology that does not depend on any such additional information.
%While there are many methods to adjust for measurement error in a regression setting, most involve either some form of knowledge about the measurement error distribution, such as its variance, or external data, such as replicates of the error-prone variable, validation data, or instrumental variables (IVs). Absent such information, neither the ATE nor the NIE will be nonparametrically identified in general. 
By contrast, we contribute to the literature on measurement error in the absence of any such external information (see Section 5 of \cite{schennach2016recent} for a review), and recover identifiability of the ATE when the outcome model is linear in the error-prone confounders. %and the measurement error is additive, unbiased, and nondifferential in the exposure and error-free confounders (though the measurement error restrictions can be relaxed in exchange for a more restrictive regression model). 
This is accomplished by constructing a function of the exposure and the error-free covariate measurements, and using the result as if it were an instrumental variable for the mismeasured confounders. These have been termed \emph{constructed instrumental variables} \citep{lewbel1997constructing}. In the Bangladesh study example, this is a function of lead exposure and other covariates excluding the error-prone protein intake variable. %, as they are constructed from the existing data, rather than requiring an exploration for a valid instrument. %\textcolor{blue}{Add citation for this term}

\begin{comment}
Our proposed ATE estimation methodology %, as well as some (but not all) of our implementations in the mediation setting, 
relaxes the classical measurement error model, which is used in the majority of measurement error methods. %Likewise, certain implementations of our proposed methodology for the natural indirect effect relax this model. 
%the additive measurement error model, and in some cases allows for the measurement error to be both biased and linearly associated with other covariates. 
In particular, in the linear model case, we allow %for certain estimands, 
%we also allow 
for bias in the measurement error model. Additionally, we allow for multiple covariates to be measured with error, and %Like \cite{hu2022identification}, we also allow 
for association between the error-free covariates and their corresponding measurement errors. %Therefore, we trade off stronger assumptions on the outcome model for weaker assumptions on the measurement error model, as well as the number of covariates measured with error. 
While we only consider models that are linear in the error-prone variable, it seems likely that our results can be extended to models with nonlinear link functions as in \cite{miles2018class}.
\end{comment}

The methodological development in this article is most closely related to \cite{lewbel1997constructing} and \cite{miles2018class}. \cite{lewbel1997constructing} proposed identification and estimation for regression coefficients in an error-in-variables model based on independence assumptions that yield higher-order moment restrictions. Our proposed class of estimators for the ATE when confounders are measured with error is a special case of the class of \cite{lewbel1997constructing}, where we primarily consider the first-order moment restriction, which relies on weaker assumptions. %For example, we do not rely on an assumption of homoscedasticity, which can be difficult to justify in most health applications. %, as well as many others. %applications (at least outside of economics). 
%We also consider a second-order moment restriction in Web Appendix A.
\cite{miles2018class} introduced a class of exposure model-based hypothesis test statistics that are valid in the presence of confounder measurement error and depend on neither distributional knowledge of the measurement error nor external data. %The statistics are based on testing the no unobserved confounding assumption under the sharp null hypothesis of no effect using residuals from an exposure model. %(an idea that forms the basis for $g$-estimation). That is, it tests whether the observed outcome---which is equal to the counterfactual under exposure equal to zero---is independent of the residual of the exposure after regressing it on the observed confounders. 
 %The resulting test statistic is readily extended to estimation, whereas the estimation approach in 
%\cite{miles2018class} also 
%This work primarily focused on hypothesis testing, as the corresponding estimation approach is generally unstable due to non-invertibility of the resulting estimating equations. 
The idea behind the methodology in this article is similar, but %instead of using residuals from a user-specified exposure model, 
is outcome model-based, which mitigates instability issues that arise when conducting estimation in the approach of \cite{miles2018class}.
%This issue with estimation is mitigated in this article.
%by replacing the computation of the statistic under the null of no effect with the exposure effect subtracted from the outcome, which depends on unknown causal parameters that can then be estimated.

Our proposed estimation approach for the ATE extends naturally to other estimands that depend on additional regression models involving error-prone variables. %In particular, we focus on the natural direct and indirect effects when a continuous exposure or confounder is measured with error. %, provided the outcome is linear in these error-prone variables. 
%In particular, we focus on the second case of mediation analysis with an error-prone exposure. 
In particular, we focus on estimation of the NIE \citep{robins1992identifiability, pearl2001direct} in the presence of an error-prone exposure. %The NIE is the target estimand when interest lies in the effect of the exposure on the outcome that is causally transmitted via an intermediate variable. For instance, 
In our Bangladesh study example, we are interested in whether the effect of protein intake on birth length is transmitted via lead exposure. Here, protein intake is now playing the role of the exposure rather than a confounder. \cite{le2012quantification,%tchetgen2012robust,
vanderweele2012role,valeri2014mediation,valeri2014estimation} and \cite{zhao2014covariate} %and \cite{fulcher2019estimation} 
discuss the consequences of measurement error on the mediating variable, some of whom also provide appropriate adjustments. In contrast, we focus on measurement error of a continuous exposure within a causal mediation framework, which has been relatively under-studied (\cite{cheng2023mediation,cheng2025correcting} being recent exceptions), and which \cite{valeri2017misclassified} show can severely bias estimates away from the null. 

The reason for this bias is, in fact, a consequence of the argument from the ATE setting. Estimators of the NIE typically rely on estimated regression coefficients of (a) the outcome on the baseline covariates, exposure, and intermediate variable and (b) the intermediate variable on baseline covariates and the exposure. %(and possibly other nuisance estimates). %, see e.g., \cite{tchetgen2012semiparametric}). 
In the Bangladesh study example, these come from regression models of (a) birth length on baseline covariates, protein intake, and lead exposure, and (b) lead exposure on baseline covariates and protein intake. 
The product method is one of the most popular ways of estimating the NIE, which involves simply multiplying the OLS intermediate variable coefficient estimate from the outcome model with the OLS exposure coefficient estimate in the intermediate variable model. If these linear models are correctly specified, then the second coefficient estimate will merely be attenuated when the exposure is measured with error. However, in the case of the outcome regression model, the exposure can in fact be viewed as playing the role of a confounder, and the intermediate variable as playing the role of the exposure of interest. In fact, model (a) is the same model we discussed above for studying the ATE of lead exposure on birth length when protein intake is regarded as a confounder. Thus, as discussed above, the coefficient estimate of the intermediate variable may be biased away from the null, which may, in turn, bias the final product method estimator away from the null. %For instance, in the mediation setting, 
Recently, \cite{cheng2023mediation} and \cite{cheng2025correcting} developed estimation methods for the natural direct and indirect effects in the presence of exposure measurement error using validation study data. Alternatively, \cite{valeri2017misclassified} proposed a sensitivity analysis using simulation extrapolation (SIMEX) when auxiliary information is not available. 
Our proposed constructed IV approach can be extended to this setting, where the first step is to apply the constructed IV method to estimate the effect of the intermediate variable on the outcome. We show that under an unbiased measurement error model, the measurement error variance can then be estimated and used to adjust for exposure or confounder measurement error in the mediator model.

This article makes several contributions to both the causal inference and measurement error literatures. First, we flesh out details of the constructed IV method. While \cite{lewbel1997constructing} makes no suggestion for the choice of constructed instrument functions (beyond them being nonlinear in one case), we provide general conditions on the classes of such functions and derive the most efficient functions in these classes. Second, we explicitly connect two subclasses of the consistent and asymptotically normal (CAN) estimators of \cite{lewbel1997constructing}  to the ATE in the presence of confounder measurement error without auxiliary information. %or knowledge of the measurement error distribution. 
Third, we derive a class of CAN estimators of the measurement error variance when either the conditions for the constructed IV method are satisfied, or when a valid IV is available. Finally, we develop a class of CAN estimators of the NIE when either the conditions for the constructed IV method are satisfied or a valid IV is available.

In Section \ref{sec:ate} we present our results in the context of linear models. More general results for a partially linear outcome model are given in Section \ref{sec:partially-linear}. In Section \ref{sec:me-variance} we discuss measurement error variance estimation. %and adapt our methodology to the mediation analysis setting. 
In Section \ref{sec:mediation} we adapt the methods of Sections \ref{sec:ate} and \ref{sec:me-variance} to the mediation analysis setting. %with error-prone exposure or confounder. 
In Section \ref{sec:sims} we discuss simulation studies. %demonstrating the finite-sample performance of our method. %, both when our assumptions hold and exploring sensitivity to their violations. 
In Section \ref{sec:data}, we apply the proposed approach in the context of the environmental epidemiology study %of the effect of maternal protein intake during pregnancy, potentially measured with error, on birth length mediated by lead exposure 
in a cohort of Bangladeshi mother-infant pairs. We conclude with a discussion in Section \ref{sec:discussion}.

%\cite{zhao2014covariate}: Does not take counterfactual approach to mediation. Consider both baseline covariate and mediator to be measured with classical measurement error in a survival outcome setting. Uses external replication data set. Don't think this one is sufficiently relevant given space constraints.

%\cite{lenis2017doubly,shu2018estimation}

%\cite{lenis2017doubly}: Total effect estimation with confounder measurement error. ``Mean reverting measurement error'' model with independent validation sample.

%\cite{shu2018estimation}: Total effect estimation with confounder measurement error. Assumes measurement error variance is known or estimable given external data.

%\cite{shu2019causal}: Only consider total effect estimation with outcome measurement error with external data. Don't think this one is sufficiently relevant given space constraints.

\begin{comment}asdf
\lv{what about testing? should I add a simulation setting under the null and compute type I error?}
\end{comment}

\section{Average treatment effect estimation with error-prone confounders}
\label{sec:ate}
%\textcolor{blue}{Consider using $Z$ for exposure in this section so that the notation when we move to mediation is more natural, and we can just consider $A$ to be a subset of $C_1$.}

\begin{figure}[ht]
\centering
\begin{tabular}{ccc}
\begin{tikzpicture}[->, line width=1pt, every node/.style={circle,inner sep=0pt}]
\tikzstyle{every state}=[draw=none]
\node[draw, minimum size=8mm] (C2) at (0.2679492,1) {${\bf C}_2$};
\node[draw, minimum size=8mm] (C1) at (0.2679492,-1) {${\bf C}_1$};
\node[draw, minimum size=8mm] (Z) at (2,0) {$Z$};
\node[draw, minimum size=8mm] (Y) at (4,0) {$Y$};
\node[shape=ellipse, draw, inner sep=.2mm] (S) at (2,2) {$\mathbf{S}(Z,\mathbf{C}_2)$};

  \path 	%(C2) edge [<->, color=gray, dashed] (C1)
            (C1) edge[o-o] (C2)
			(C2) edge (Z)
			(C2) edge [bend left] (Y)
			(Z) edge (Y)
			(C1) edge (Z)
			(C1) edge [bend right] (Y)
            (C2) edge [color=red] (S)
            (Z) edge [color=red] (S)
					  ;
\end{tikzpicture}
& 
&
\begin{tikzpicture}[->, line width=1pt, every node/.style={circle,inner sep=0pt}]
\tikzstyle{every state}=[draw=none]
\node[draw, minimum size=8mm] (C) at (0.2679492,1) {${\bf C}$};
\node[draw, minimum size=8mm] (A) at (0.2679492,-1) {$A$};
\node[draw, minimum size=8mm] (Z) at (2,0) {$Z$};
\node[draw, minimum size=8mm] (Y) at (4,0) {$Y$};
\node[shape=ellipse, draw, inner sep=.2mm] (S) at (2,2) {$\mathbf{S}(Z,\mathbf{C})$};

  \path 	(C) edge (A)
			(C) edge (Z)
			(C) edge [bend left] (Y)
			(Z) edge (Y)
			(A) edge (Z)
			(A) edge [bend right] (Y)
            (C) edge [color=red] (S)
            (Z) edge [color=red] (S)
					  ;
\end{tikzpicture}
\\
(a)	&	&	(b) \\
\\
\end{tabular}
\caption{(a) The partial ancestral graph representing the causal structure among the variables ${\bf C}_1$, ${\bf C}_2$, $Z$, and $Y$, and the constructed IV $\mathbf{S}(Z,\mathbf{C}_2)$. The edge with circles at either end represents a possible direct causal relationship between ${\bf C}_1$ and ${\bf C}_2$ and/or the presence of hidden common causes of ${\bf C}_1$ and ${\bf C}_2$. The presence of this edge has no bearing on the sufficient adjustment set for nonparametric identification of the effect of $Z$ on $Y$. The red arrows represent a deterministic relationship. (b) The directed acyclic graph representing the causal structure among the variables ${\bf C}$, $A$, $Z$, and $Y$, and the constructed IV $\mathbf{S}(Z,\mathbf{C})$. The red arrows represent a deterministic relationship.}
\label{fig:pag}
\end{figure}
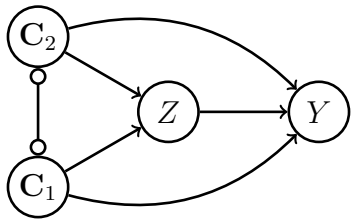

Suppose there exist $n$ independent, identically-distributed realizations of $({\bf C},Z,Y)$ from an observational study, where $Y$ is a continuous outcome, %(may extend to other types later), 
$Z$ is the exposure, which may be discrete or continuous, and ${\bf C}$ are confounders, which can be partitioned into ${\bf C}_1$, which are continuous and subject to measurement error, and ${\bf C}_2$, which may be continuous or discrete and are always correctly measured. In the Bangladesh study example, $Y$ is birth length, $Z$ is a measure of maternal lead exposure, ${\bf C}_1$ is the true maternal protein intake, and ${\bf C}_2$ consists of other baseline covariates, which are listed in full in Section \ref{sec:data}. %For visualization purposes, 
Figure \ref{fig:pag}.a displays a partial ancestral graph (PAG) %\citep{zhang2008causal} 
demonstrating the assumed causal structure of these variables. %(though we will focus on a subset of counterfactual independencies implied by a causal model corresponding to this graph rather than a full corresponding causal model). 
Throughout, we will use an asterisk superscript to denote a variable that has been measured with error, while the absence of this superscript indicates its corresponding true value. 
%, where $C$ is correctly measured, but $X$ is an error-prone measurement of the true continuous covariate value $X^*$. %and $E(\varepsilon^*\mid C,A,Y)=0$ (figure out how much these can be relaxed later). 

Let $Y(z)$ denote the potential outcome, or counterfactual value the outcome would have taken had (possibly contrary to fact) there been an intervention assigning $Z$ to $z$. %\citep{splawa1990application,rubin1974estimating}. 
Causal effects can be characterized in a number of ways. The dose-response curve (typically used for ordinal and continuous exposures) is the mapping $z\mapsto E\{Y(z)\}$ over a given domain of levels of the exposure. The average treatment effect (typically used for binary exposures) contrasting two particular levels of $Z$, say $z'$ and $z''$, is 
%We define the average treatment effect (ATE) for a $\Delta$-unit increase in $Z$ to be 
$ATE(z', z'')\equiv E\{Y(z')-Y(z'')\}$. Nonparametric identification of each of these is implied by nonparametric identification of the quantities $E\{Y(z)\}$ for the levels of $z$ in question. For a given $z$, when all variables are correctly measured, $E\{Y(z)\}$ will be nonparametrically identified with respect to the observed data distribution under standard causal assumptions, %viz., consistency, no unobserved confounding, and positivity. These 
which are stated explicitly in Section \ref{sec:causal-assumptions}. 
\begin{comment}
In particular, we will assume the following hold throughout for all levels $z$ of interest:
\begin{assumption}[Consistency] %For all $z\in\mathrm{supp}(Z)$,
\label{assn:consistency}
$Z=z$ implies $Y=Y(z)$ almost everywhere.
\end{assumption}
\begin{assumption}[No unobserved confounding]
\label{assn:nuca}
$Y(z)\ci Z\mid {\bf C}$.% for all $z\in\mathrm{supp}(Z)$.
\end{assumption}
\begin{assumption}[Positivity]
\label{assn:positivity}
%For continuous $Z$, 
$f_{Z\mid {\bf C}}(z\mid {\bf C})>\delta>0$ %and $f_{Z\mid C}(z+\Delta\mid C)>\delta$ 
almost everywhere in ${\bf C}$, where $f_{Z\mid {\bf C}}$ is the conditional probability mass (density) function for discrete (continuous) $Z$. %or the conditional probability density function for continuous $Z$. %For discrete $Z$, $\mathrm{Pr}(Z=z\mid C)>\delta$ and $\mathrm{Pr}(Z=z+\Delta\mid C)>\delta$ almost everywhere in $C$ for some $\delta >0$.
\end{assumption}
\end{comment}
Under these causal assumptions %\ref{assn:consistency}--\ref{assn:positivity} 
and no measurement error, $E\{Y(z)\}$
%the ATE 
is nonparametrically identified by $E\{E(Y\mid Z=z,{\bf C})\}$.
%$ATE(z,\Delta)=E\{E(Y\mid Z=z+\Delta,C)-E(Y\mid Z=z,C)\}$. 
Suppose, however, that instead of directly observing the realizations $({\bf C},Z,Y)$, we observe $n$ i.i.d.~copies of ${\bf O}\equiv ({\bf C}_1^*,{\bf C}_2,Z,Y)$. In the Bangladesh study example, ${\bf C}_1^*$ is a measure of maternal protein intake based on a food frequency questionnaire. 
When ${\bf C}_1$ is measured with error, $E\{Y(z)\}$ is no longer nonparametrically identified, and requires additional information or assumptions for identification.
%$\{C,X^*\}$ are sufficient to control for confounding, such that $Y(a)\ci A\mid C,X^*$, as well as consistency, i.e., $A=a$ implies $Y=Y(a)$ almost everywhere. Further, we assume positivity (or covariate overlap), which states that for every $a$ in the support of $A$, $\mathrm{Pr}(A=a\mid C,X^*)>0$ almost everywhere in $(C,X^*)$.

%We assume unbiased measurement error, such that $X=X^*+\varepsilon^*$, and that $\varepsilon^*$ is mean-independent of $\{C,A\}$, i.e., $E(\varepsilon^*\mid C,A)=0$. 

\subsection{Linear model setting with a valid IV}
%To gain intuition for our proposed methodology, 
Here we consider identification of $E\{Y(z)\}$ and $ATE(z',z'')$ when the outcome regression model is linear with only main effect terms. We describe an extension to a more general model in Section \ref{sec:partially-linear}. %In this setting, 
We begin by assuming the following measurement error model. 
\begin{assumption}
\label{assn:me1}
%\begin{align}
%\label{eqn:me1}
$E({\bf C}_1^*\mid Z,{\bf C}_2)=E({\bf C}_1\mid Z,{\bf C}_2)$.  
%\end{align}
\end{assumption}
In the Bangladesh study example, this assumption will hold if the mean of the measured protein intake is the same as that of the true underlying protein intake within strata of lead exposure and the other baseline covariates. Assumption \ref{assn:me1} will hold when there is additive, unbiased measurement error $\varepsilon$ that is mean independent of $\{Z,{\bf C}_1,{\bf C}_2\}$, under either 
the classical measurement error model ${\bf C}_1^*={\bf C}_1+\varepsilon$ or the Berkson/regression calibration model ${\bf C}_1={\bf C}_1^*+\varepsilon$ as 
discussed in \cite{carroll2006measurement}. %, where $E(\varepsilon\mid Z,{\bf C}_1,{\bf C}_2)={\bf 0}$ in either case.

%$E(X\mid A,C)=\gamma_0+\Gamma_1 E(X^*\mid A,C)+\Gamma_2 C$, where $\gamma_0$ is a vector of dimension $\dim(X)$, $\Gamma_1$ is an $\dim(X)\times\dim(X)$ nonsingular matrix (we consider $X$ and $X^*$ to have the same dimension), and $\Gamma_2$ is a $\dim(X)\times\dim(C)$ matrix. For instance, this will hold when $X=\gamma_0+\Gamma_1X^*+\varepsilon^*$ and $E(\varepsilon^*\mid A,C)=\Gamma_2C$. That is, $X$ can be linear in $X^*$, and the measurement error must be mean independent of $A$, but can be linear in $C$. This is a considerable relaxation of the typical classical measurement error model, in which $X=X^*+\varepsilon^*$, and $\varepsilon^*$ is independent of $X^*$ and the other observed variables. Later when we consider more general outcome models, there will be a trade-off between the outcome and measurement error models, such that we will require a more restrictive measurement error model.

Suppose the linear outcome regression model $E(Y\mid Z,{\bf C};\theta) = \theta_0 + \theta_{{\bf C}_1}^T{\bf C}_1 + \theta_{{\bf C}_2}^T{\bf C}_2 + \theta_Z Z$ is correctly specified. %This model will not be appropriate for non-ordinal polytomous $Z$, which will be accommodated by the generalized model in the following subsection. 
If this model's coefficients are known, %and consistency, no unobserved confounding, positivity, 
and standard causal assumptions and the measurement error model in Assumption \ref{assn:me1} all hold, %the ATE of a one-unit increase in $Z$, $E\{Y(z+1) - Y(z)\}$ is constant in $z$, and 
then $E\{Y(z)\}$ is identified by $\theta_0 + \theta_{{\bf C}_1}^TE({\bf C}_1^*) + \theta_{{\bf C}_2}^TE({\bf C}_2) + \theta_Z z$. Were we to observe ${\bf C}_1$ directly, the OLS estimator solving the normal equations
\begin{comment}
\[
\frac{1}{n}\sum_{i=1}^n\left[
\begin{array}{c}
1 \\
C_{1,i} \\
{\bf C}_{2,i} \\
Z_i 
\end{array}
\right](Y_i - \theta_0 -  \theta_{{\bf C}_1}^TC_{1,i} - \theta_{{\bf C}_2}^T{\bf C}_{2,i} - \theta_Z Z_i) = 0
\]
\end{comment}
$
n^{-1}\sum_{i=1}^n\left[1,{\bf C}_{1,i}, {\bf C}_{2,i}, Z_i \right]^T(Y_i - \theta_0 -  \theta_{{\bf C}_1}^T{\bf C}_{1,i} - \theta_{{\bf C}_2}^T{\bf C}_{2,i} - \theta_Z Z_i) = {\bf 0}
$
would yield a CAN estimator for $\theta\equiv (\theta_0, \theta_{{\bf C}_1}^T, \theta_{{\bf C}_2}, \theta_Z)^T$, and so for $E\{Y(z)\}$ as well. However, since we observe ${\bf C}_1^*$ rather than ${\bf C}_1$, the normal equations %based on the observed data 
will be biased. %due to the presence of the product of ${\bf C}_1^*$ with itself, and hence the variance of the measurement error. %Under certain assumptions about the measurement error, 

Suppose an IV for ${\bf C}_1^*$ of length $\dim({\bf C}_1)$, say ${\bf V}$, were available. For ${\bf V}$ to be a valid IV for ${\bf C}_1^*$, it must satisfy the exclusion restriction $E(Y\mid Z,{\bf C},{\bf V})=E(Y\mid Z,{\bf C})$ 
%$E({\bf V}\mid A,C)=E({\bf V}\mid A,C_2)$ 
and the condition $E({\bf C}_1^*\mid {\bf V})=E({\bf C}_1\mid {\bf V})$. The latter is implied by mean independence of the measurement error in an additive measurement error model: $E(\varepsilon\mid {\bf V})=E(\varepsilon)={\bf 0}$. When such a variable is available, replacing the ${\bf C}_1^*$ in the vector of the normal equations with ${\bf V}$ will recover unbiased estimating equations 
\begin{comment}
To see why, observe that, due to the properties of the IV and the measurement error model, the elements of the normal equations that were previously inducing bias will become
\begin{align*}
    &  \; E\{{\bf V}(Y-\theta_0^{\dag}-\theta_{{\bf C}_1}^{\dag T}{\bf C}_1^*-\theta_{{\bf C}_2}^{\dag T}{\bf C}_2-\theta_Z^{\dag}Z)\}\\
    =& \;  E[{\bf V}\{E(Y\mid Z,{\bf C},{\bf V})-\theta_0^{\dag}-\theta_{{\bf C}_1}^{\dag T}{\bf C}_1^*-\theta_{{\bf C}_2}^{\dag T}{\bf C}_2-\theta_Z^{\dag}Z\}]\\
    =&  \; E[{\bf V}\{E(Y\mid Z,{\bf C})-\theta_0^{\dag}-\theta_{{\bf C}_1}^{\dag T}{\bf C}_1^*-\theta_{{\bf C}_2}^{\dag T}{\bf C}_2-\theta_Z^{\dag}Z\}]\\
    =& \; E\{{\bf V}(\theta_{{\bf C}_1}^{\dag T}{\bf C}_1-\theta_{{\bf C}_1}^{\dag T}{\bf C}_1^*)\}\\
    =& \; E\left[{\bf V}\theta_{{\bf C}_1}^{\dag T}\left\{E({\bf C}_1\mid {\bf V})-E({\bf C}_1^*\mid {\bf V})\right\}\right]= 0,
\end{align*}
where dagger superscripts denote the true value of the parameters throughout. 
\end{comment}
%Thus, the resulting estimating equations will be unbiased 
at the true parameter value $\theta^{\dag}\equiv (\theta_0^{\dag}, \theta_{{\bf C}_1}^{\dag T}, \theta_{{\bf C}_2}^{\dag T}, \theta_Z^{\dag})^T$, where we will use dagger superscripts to denote the true value of a parameter throughout.

If we are only interested in estimating $\mathrm{ATE}(z',z'')$, which is identified by $\theta_Z^{\dag}(z'-z'')$, rather than estimating $E\{Y(z)\}$ for one level $z$ or the entire dose-response curve, then it is not necessary to estimate the rest of $\theta^{\dag}$ consistently. Instead, we can use the following model. %, and we can in fact relax the measurement error model to the following and retain consistent estimation. %of $\mathrm{ATE}(z',z'')$. 
%Specifically, we can relax the measurement error model to be the following.
\begin{assumption}
\label{assn:me2}
%\begin{align}
%\label{eqn:me2}
$E({\bf C}_1^*\mid Z,{\bf C}_2)=\gamma_0+{\bf\Gamma}_1 E({\bf C}_1\mid Z,{\bf C}_2)+{\bf\Gamma}_2 {\bf C}_2$,    
%\end{align}
where $\gamma_0$ is a $\dim({\bf C}_1)$-vector, ${\bf\Gamma}_1$ is a $\dim({\bf C}_1)\times\dim({\bf C}_1)$ nonsingular matrix, and ${\bf\Gamma}_2$ is $\dim({\bf C}_1)\times\dim({\bf C}_2)$.
\end{assumption}
In the Bangladesh study example, this will hold if the mean of the measured protein intake is a linear function of that of the true underlying protein intake within strata of lead exposure and other baseline covariates, and the intercept of this function can vary linearly in the other covariates, but not in lead exposure. Neither Assumption \ref{assn:me1} nor \ref{assn:me2} is testable without additional data. 
We assume ${\bf C}_1^*$ and ${\bf C}_1$ have common dimension. %, though this could potentially be relaxed. 
Assumption \ref{assn:me2} will hold when ${\bf C}_1^*=\gamma_0+{\bf\Gamma}_1{\bf C}_1+\varepsilon$ and $E(\varepsilon\mid Z,{\bf C}_2)={\bf\Gamma}_2{\bf C}_2$, i.e., ${\bf C}_1^*$ can be linear in ${\bf C}_1$, and the measurement error must be mean independent of $Z$, but can be linear in ${\bf C}_2$. This is a special case of the error calibration model (1.2) in \cite{carroll2006measurement} where the coefficient for our $Z$ is restricted to be zero. This is a considerable relaxation of the additive, unbiased measurement error model in Assumption \ref{assn:me1} with which we began. %The generality of this measurement error model reflects how IVs can be used to deal with more general forms of endogeneity, such as that induced by omitted exogenous variables. Measurement error can be thought of as a special case of such an omitted variable, as the measurement error $\varepsilon$ can be viewed as the omitted variable that, if observed, recovers exogeneity of $Z$. %This suggests it may also be possible to extend the constructed IV methodology described in the following subsection to settings with 
%This suggests that our constructed IV methodology described in the following subsection may also be possible to extend to settings with 
%unobserved confounding rather than confounder measurement error under linear outcome regression models. 
When considering more general outcome models, there is a trade-off between the outcome and measurement error models, such that we require the more restrictive unbiased measurement error model. % that we considered initially.

In addition to being unbiased, %of the estimating equations alone is insufficient for consistent estimation of $\theta$ or $\theta_Z$. 
the IV estimating equations must additionally have a unique solution $\theta^*\equiv (\theta_0^*, \theta_{{\bf C}_1}^{*T}, \theta_{{\bf C}_2}^{*T}, \theta_Z^{\dag})^T$ for some $\theta_0^*$, $\theta_{{\bf C}_1}^{*}$, and $\theta_{{\bf C}_2}^{*}$ in order for $\theta_Z^{\dag}$ to be identified, and $\theta^*$ must be equal to $\theta^{\dag}$ for the full parameter $\theta^{\dag}$ to be identified. As stated in the Wikipedia article for ``instrumental variable'', one commonly-invoked sufficient condition is: ``The instrument must be correlated with the endogenous explanatory variables [in our case, ${\bf C}_1^*$], conditionally on the other covariates'' %\citep{ enwiki:1182042304} 
(see also: chapter 4.1 of \cite{angrist2009mostly}, for example). However, this is stronger than necessary; it is sufficient for the instrument to be correlated with the residual of the projection of ${\bf C}_1^*$ onto the linear span of $1$, $Z$, and ${\bf C}_2$. %(i.e., the terms in the outcome regression model). 
This distinction will be crucial in what follows. Further, this correlation must be sufficiently strong for reasonable asymptotic approximations---a condition known as the \emph{strong first-stage condition}. It is common to test this condition using an $F$-test of the partial correlation between ${\bf C}_1^*$ and ${\bf V}$ in the linear model regressing ${\bf C}_1^*$ on $Z$, ${\bf C}_2$, and ${\bf V}$ \citep{bound1995problems}.

%To see why, let $\theta_0^*=\theta_0^{\dag}-({\bf\Gamma}_1^{-1}\gamma_0)^T\theta_X^{\dag}$, $\theta_C^*=\theta_C^{\dag}-({\bf\Gamma}_1^{-1}{\bf\Gamma}_2)^T\theta_X^{\dag}$, and $\theta_X^*=({\bf\Gamma}_1^{-1})^T\theta_X^{\dag}$. Then it is simple to show that $\theta^*\equiv (\theta_0^*, \theta_C^{*T}, \theta_X^{*T}, \theta_A^{\dag})$ solves 

%For instance, consider additive, unbiased measurement error that is mean independent of $A$ and $C$, such that $E(X\mid A,C)=E(X^*\mid A,C)$. The measurement error in the residual part of the normal equations will then vanish.

\subsection{Linear model setting with a constructed IV}

In contrast to the standard IV approach, instead of searching for a valid instrument external to the original set of variables of interest, we propose using a function ${\bf s}:\mathrm{supp}(Z,{\bf C}_2)\rightarrow \mathbb{R}^{\dim({\bf C}_1)}$ %of $Z$ and ${\bf C}_2$ 
for this purpose. %, where $\mathrm{supp}(\cdot)$ denotes the support of its argument. 
Clearly, ${\bf s}(Z,{\bf C}_2)$ satisfies the exclusion restriction criterion, $E\{Y\mid {\bf s}(Z,{\bf C}_2), Z,{\bf C}\}=E(Y\mid Z,{\bf C})$. %$E\{{\bf s}(Z,{\bf C}_2)\mid Z,C_1,{\bf C}_2\}=E\{{\bf s}(Z,{\bf C}_2)\mid Z,{\bf C}_2\}$
Under %the measurement error model in 
Assumption \ref{assn:me1}, it also satisfies $E\{{\bf C}_1^*\mid {\bf s}(Z,{\bf C}_2)\}=E\{{\bf C}_1\mid {\bf s}(Z,{\bf C}_2)\}$, hence ${\bf s}(Z,{\bf C}_2)$ can be used as an IV to recover unbiasedness of the estimating equations for $\theta^{\dag}$. We also show that under the relaxed measurement error model in Assumption \ref{assn:me2}, %(\ref{eqn:me2}), %$E\{{\bf s}(A,{\bf C}_2)\mid A,{\bf C}_2,\varepsilon\}=E\{{\bf s}(A,{\bf C}_2)\mid A,{\bf C}_2\}$, 
the normal equations with ${\bf s}(Z,{\bf C}_2)$ replacing the ${\bf C}_1^*$ in the instrument vector remain unbiased for $\theta_Z^{\dag}$ (but not necessarily the remaining coefficients).
\begin{comment}

To demonstrate why this is, we first observe that $E(Y-\theta_Z^0Z\mid {\bf C},X^*,Z) = E(Y\mid C,X^*,A=0) = E\{Y(0)\mid C,X^*,A=0\} = E\{Y(0)\mid C,X^*\}$, and hence $Y-\theta_A^0A$ is conditionally mean independent of $A$ given $\{C,X^*\}$. \textcolor{blue}{question: can we put in the context of the IV assumptions and discuss how functions of A can make it a stronger instrument? I guess related to the question you ask below!}, at the true value of $\theta_A$, the model can be reformulated as regressing $Y-\theta_AA$ on $\{C,X^*\}$. For this model, $A$ will be an IV (in the mean independence sense) for $X^*$. We may then replace $X$ in the vector of the observed-data normal equations with $A$. However, this repeats an equation in the normal equations, so to retain linear independence, we can instead replace $X$ with $Af(C)$, where $f$ is some arbitrary, potentially vector-valued function. The identity function is a simple example of a choice of $f$. While the IV property of $A$ only holds for $Y-\theta_A^0A$ at the true $\theta_A^0$, we have sufficiently many estimating equations to consistently estimate this parameter simulataneously with the rest of the regression coefficients. 

\end{comment}
%When $f$ is real-valued, the parameters are just-identified; when $f$ is vector valued, they are over-identified. 

Next, we must consider whether these estimating equations have a unique solution. %The necessary and sufficient rank condition for this to hold is that
\begin{comment}
\[E\left\{\left[\begin{array}{c}
1 \\
s(A,{\bf C}_2)    \\
{\bf C}_2 \\
A
\end{array}\right]
\left[1, C_1^*, {\bf C}_2, A\right]
\right\}\]
\end{comment}
The condition that $E\{[1, {\bf s}(Z,{\bf C}_2)^T, {\bf C}_2^T, Z]^T[1, {\bf C}_1^{*T}, {\bf C}_2^T, Z]\}$ is nonsingular is necessary and sufficient for uniqueness. This implies that no element of ${\bf s}(Z,{\bf C}_2)$ is in the linear span of 1, $Z$, and ${\bf C}_2$, and that ${\bf C}_1^*$ is correlated with the residual of the projection of ${\bf s}(Z,{\bf C}_2)$ onto the linear span of 1, $Z$, and ${\bf C}_2$. Thus, $E({\bf C}_1^*\mid Z, {\bf C}_2)$ must be nonlinear in $Z$ and ${\bf C}_2$ for $\theta^{\dag}$ or $\theta_Z^{\dag}$ to be identified. Whether this condition is met depends on the setting. %If ${\bf C}_1$ is a true confounder, then ${\bf C}_1^*$ will be associated with $Z$, but this does not necessarily imply that the relationship is nonlinear.

Generalized method of moments (GMM) estimation theory dictates that the solution to
\begin{comment}
\[
\frac{1}{n}\sum_{i=1}^n\left[
\begin{array}{c}
1 \\
s(M_i,{\bf C}_{2,i})  \\
{\bf C}_{2,i} \\
M_i \\
\end{array}
\right](Y_i - \theta_0 - \theta_{{\bf C}_1}^T{\bf C}^*_{1,i} - \theta_{{\bf C}_2}^T{\bf C}_{2,i} - \theta_M M_i) = {\bf 0}
\]
\end{comment}
\[
\frac{1}{n}\sum_{i=1}^n\left[1, {\bf s}(Z_i,{\bf C}_{2,i})^T, {\bf C}_{2,i}^T, Z_i\right]^T(Y_i - \theta_0 - \theta_{{\bf C}_1}^T{\bf C}^*_{1,i} - \theta_{{\bf C}_2}^T{\bf C}_{2,i} - \theta_Z Z_i) = {\bf 0}
\]
will be CAN for the solution to its population mean analog. This estimator is simply the IV estimator with IV vector $\{1,{\bf s}(Z,{\bf C}_2)^T,{\bf C}_2^T,Z\}$. %It is identical to the classical two-stage least squares estimator. %, which is the estimator resulting from first performing a linear regression of ${\bf C}_1^*$ on ${\bf s}(Z,{\bf C}_2)$, $Z$, and ${\bf C}_2$, then performing the linear regression of $Y$ on $Z$, ${\bf C}_2$, and the vector of fitted values from the first-stage regression.
This class of estimators, indexed by ${\bf s}$, can be generalized to a larger class of estimators, indexed by %$\hat{W}$ and 
${\bf S}$, where %$\hat{W}$ is a convergent $\dim(\theta)\times\dim(\theta)$ matrix and 
${\bf S}(Z,{\bf C}_2)$ is a 
%\dim({\bf C}_1^*)+\dim({\bf C}_2)+2$
$\dim(\theta)$-dimensional vector-valued function. Specifically, letting
\[{\bf U}_1({\bf O}_i;\theta)\equiv {\bf S}(Z_i,{\bf C}_{2,i})(Y_i - \theta_0 - \theta_{{\bf C}_1}^T{\bf C}^*_{1,i} - \theta_{{\bf C}_2}^T{\bf C}_{2,i} - \theta_Z Z_i),\]
we define the GMM estimator to be the solution $\hat{\theta}$ to %the estimating equations 
$\sum_{i=1}^n {\bf U}_1( {\bf O}_i;\theta) = {\bf 0}$. 

\begin{comment}
We can then %also augment this set of estimating equations to directly 
estimate the dose response curve. Let $\mu_j\equiv E\{Y(z_j)\}$ for ordinal $Z$ with $J$ levels $\{z_1,\ldots,z_J\}$. The parameters $\mu_j$ can be estimated by $n^{-1}\sum_{i=1}^n(\hat{\theta}_0 + \hat{\theta}_{C_1}^T{\bf C}^*_{1,i} + \hat{\theta}_{C_2}^T{\bf C}_{2,i} + \hat{\theta}_Z z_j)$ for each $j\in\{1,\ldots,J\}$. %, where $\hat{\theta}\equiv (\hat{\theta}_0, \hat{\theta}_{C_1}^T, \hat{\theta}_{C_2}^T, \hat{\theta}_Z)^T$ is the GMM estimator solving the above estimating equations. 
The corresponding joint asymptotic distribution can be obtained using the delta method. Alternatively, the dose-response curve and its asymptotic distribution can be obtained by augmenting the estimating equations as follows:
\[U_2(O_i,\bm{z};\theta,\bm{\mu})\equiv \left[\begin{array}{c}
    U_1(O_i;\theta)\\ 
    \theta_0 + \theta_{{\bf C}_1}^T{\bf C}^*_{1,i} + \theta_{{\bf C}_2}^T{\bf C}_{2,i} + \theta_Z z_1-\mu_1\\
    \vdots \\
    \theta_0 + \theta_{{\bf C}_1}^T{\bf C}^*_{1,i} + \theta_{{\bf C}_2}^T{\bf C}_{2,i} + \theta_Z z_J-\mu_J
    \end{array}\right].
\]
\end{comment}

For continuous or ordinal $Z$, the dose-response curve %and its estimator 
will be a linear function of $z$ under the linear outcome regression model assumed in this subsection. Its slope can be estimated by $\hat{\theta}_Z$ and, under the measurement error model in Assumption \ref{assn:me1}, its intercept by $n^{-1}\sum_{i=1}^n(\hat{\theta}_0 + \hat{\theta}_{C_1}^T{\bf C}^*_{1,i} + \hat{\theta}_{C_2}^T{\bf C}_{2,i})$. The latter can be estimated using %$U_2(O_i,0;\theta,\mu_0)$, 
${\bf U}_2({\bf O}_i;\theta,\mu)\equiv [%\begin{array}{c}
    {\bf U}_1({\bf O}_i;\theta)^{\top}, 
    \theta_0 + \theta_{{\bf C}_1}^T{\bf C}^*_{1,i} + \theta_{{\bf C}_2}^T{\bf C}_{2,i} - \mu
    %\end{array}
    ]^{\top},
$
where $\mu^{\dag}\equiv E\{E(Y\mid Z=0,{\bf C})\}$ (the true value of $\mu$) is the intercept of the dose-response curve. 
\begin{comment}
The asymptotic distribution of the intercept can be obtained by the delta method or by augmenting the estimating equations as follows: %by $\theta_0 + \theta_{{\bf C}_1}^T{\bf C}^*_{1,i} + \theta_{{\bf C}_2}^T{\bf C}_{2,i} - \mu_0$, 
\[{\bf U}_3(O_i;\theta,\mu_0)\equiv \left[\begin{array}{c}
    {\bf U}_1(O_i;\theta)\\ 
    \theta_0 + \theta_{{\bf C}_1}^T{\bf C}^*_{1,i} + \theta_{{\bf C}_2}^T{\bf C}_{2,i} - \mu_0 \\
    \end{array}\right],
\]
where $\mu_0$ is the intercept of the dose-response curve.
\end{comment}
A simultaneous confidence region for the slope and intercept can then be transformed into a confidence band for the dose-response curve.
\begin{comment}
\[{\bf U}_3(O_i;\theta)\equiv \left[\begin{array}{c}
    {\bf U}_1(O_i;\theta)\\ 
    \theta_0 + \theta_{{\bf C}_1}^T{\bf C}^*_{1,i} + \theta_{{\bf C}_2}^T{\bf C}_{2,i} - \mu_0 \\
    \end{array}\right].
\]
\end{comment}
%corresponding to whether $\mathrm{ATE}(z',z'')$ or $\mu(z)\equiv E\{Y(z)\}$, respectively, is the causal parameter of interest. 

For a given %$\hat{W}$, and 
${\bf S}$, we define the corresponding GMM estimators to be the solution %$\hat{\theta}$ to the estimating equations $n^{-1}\sum_{i=1}^n{\bf U}_1(O_i;\theta)=0$, the solution 
$[\hat{\theta}^T,\hat{\mu}]^T$ to the estimating equations $\sum_{i=1}^n{\bf U}_2({\bf O}_i;\theta,\mu)={\bf 0}$. %and the solution $[\hat{\theta}^T,\hat{\mu}_0]^T$ to the estimating equations $n^{-1}\sum_{i=1}^n{\bf U}_3( {\bf O}_i;\theta,\mu_0)=0$. 
The estimator $\hat{\theta}$ is identical for estimating equations based on both ${\bf U}_1$ and ${\bf U}_2$; ${\bf U}_2$ simply allows for joint estimation and inference of $\mu$ along with $\theta$.
\begin{comment}
\begin{align}
    \hat{\theta}=\arg\min_{\theta}\left\{\frac{1}{n}\sum_{i=1}^nU({\bf O}_i;\theta)\right\}\hat{W}\left\{\frac{1}{n}\sum_{i=1}^nU({\bf O}_i;\theta)\right\}.
\end{align}
\end{comment}
Let $\phi_1\equiv \theta$, $\phi_2\equiv (\theta^T,\mu)^T$, and ${G}_1\equiv E\left\{{\bf S}(Z,{\bf C}_2)[1,C_1^{*T},{\bf C}_2^T,Z]\right\}$, and for vector ${\bf x}$, let ${\bf x}^{\otimes 2}\equiv {\bf xx}^T$. The following assumption is the rank condition ensuring uniqueness.
\begin{assumption}
\label{assn:rank-linear}
    ${G}_1$ is nonsingular.
\end{assumption}
In the Bangladesh data example, this assumption implies that the conditional mean of the measured protein intake must be nonlinear in lead exposure and/or any of the baseline covariates, and the elements of our choice of the constructed IV $\bf S$ must be linearly independent functions of lead exposure and baseline covariates. 
The following result gives the CAN property of estimators in this class for $\phi_1^*=(\theta_0^*,\theta_{{\bf C}_1}^{*T},\theta_{{\bf C}_2}^{*T},\theta_Z^{\dag})^T$ under Assumption \ref{assn:me2} %(\ref{eqn:me2}) 
($\theta_0^*$, $\theta_{{\bf C}_1}^{*}$, and $\theta_{{\bf C}_2}^{*}$ are specified in the proof) and $\phi_2^*=(\theta_0^{\dag},\theta_{{\bf C}_1}^{\dag T},\theta_{{\bf C}_2}^{\dag T},\theta_Z^{\dag}, \mu^{\dag})^T$ under Assumption \ref{assn:me1}. %, and $\phi_3\equiv (\theta^T,\mu_0)^T$.

%\color{blue}NOTE: REMOVE THESE TWO THEOREMS AND JUST STATE RESULTS AS CONSEQUENCES OF THEOREMS FROM SECTION 2.3(?)\color{black}

\begin{theorem}\label{thm:can}
Suppose %(i) $\hat{W}$ converges in probability to a positive semi-definite matrix $W$, 
(a) the measurement error model in Assumption \ref{assn:me2} %(\ref{eqn:me2}) 
holds (case k=1) or the measurement error model in Assumption \ref{assn:me1} holds (case k=2), (b) $\phi_k^*$ is in the interior of a compact subset of the parameter space for $k=1,2$, %$\mathrm{B}$, 
(c) ${\bf C}_1$ and all elements of ${\bf O}$ have finite fourth moment and $E\{\lVert {\bf S}(Z,{\bf C}_2)\rVert^4\}<\infty$, and (d) Assumption \ref{assn:rank-linear} holds.
%${G}_1\equiv E\left\{{\bf S}(Z,{\bf C}_2)[1,C_1^{*T},{\bf C}_2^T,Z]\right\}$ is nonsingular. 
Then for the estimators $\hat{\phi}_k$ %$\hat{\theta}$, $(\hat{\theta}^T,\hat{\bm{\mu}}^T)^T$, and $(\hat{\theta}^T,\hat{\mu}_0)^T$ 
solving $\sum_{i=1}^n{\bf U}_k({\bf O}_i;\phi_k)={\bf 0}$ for $k=1,2$, %$n^{-1}\sum_{i=1}^n{\bf U}_2({\bf O}_i;\theta,\bm{\mu})=0$, and $n^{-1}\sum_{i=1}^n{\bf U}_3({\bf O}_i;\theta,\mu_0)=0$, respectively,
$\sqrt{n}(\hat{\phi}_k-\phi_k^*) \rightsquigarrow \mathcal{N}\{0,{G}_k^{-1}{\mathit\Omega}_k({G}_k^{-1})^T\}$,
where $G_2$ is the block diagonal matrix with blocks $G_1$ and the 1$\times$1 matrix $[1]$, 
%\[{G}_2\equiv \left[\begin{array}{cc}
%     {G}_1 & {\bf 0}  \\
%     {\bf 0}^T & 1 
%\end{array}\right],
%\; {\bf G}_3\equiv \left[\begin{array}{cc}
%     {\bf G}_1 & {\bf 0}_d  \\
%     {\bf 0}_d^T & 1 
%\end{array}\right],
%\]% ${\bf G}_3\equiv E\left\{{\bf S}(Z,{\bf C}_2)[1,C_1^{*T},{\bf C}_2^T,Z,1]\right\}$, 
%$b\equiv \dim(\theta)$, 
and ${\mathit\Omega}_k\equiv E\{{\bf U}_k({\bf O};\phi_k^*)^{\otimes 2}\}$ for $k=1,2$.
\end{theorem}

All proofs and additional theoretical results are available in Section \ref{sec:proofs}. Define $\sigma^2(Z,{\bf C}_2)\equiv E\{(Y-\theta_0^*-\theta_{{\bf C}_1}^{* T}{\bf C}_1^*-\theta_{{\bf C}_2}^{* T}{\bf C}_2-\theta_Z^{\dag}Z)^2\mid Z,{\bf C}_2\}$. %and $L$ to be the lower triangular matrix in the Cholesky decomposition $E\{\sigma^{-2}(A,{\bf C}_2)[1, E({\bf C}_1^*\mid A,{\bf C}_2)^T, {\bf C}_2^T, A]^T[1, E({\bf C}_1^*\mid A,{\bf C}_2)^T, {\bf C}_2^T, A]\}^{-1}=LL^T$. 
The following theorem gives the optimal choice of ${\bf S}$ with respect to asymptotic variance. 
\begin{theorem}\label{thm:efficiency-linear}
Under Theorem \ref{thm:can} conditions (a)--(d) for $k=1$,  ${\bf S}^*(Z,{\bf C}_2)\equiv \sigma^{-2}(Z,{\bf C}_2)[1,\allowbreak E({\bf C}_1^*\mid Z,{\bf C}_2)^T, {\bf C}_2^T, Z]^T$ %and $W^*\equiv E\{\sigma^{-2}(A,{\bf C}_2)[1, E({\bf C}_1^*\mid A,C_1)^T, {\bf C}_2^T, A]^T[1, E({\bf C}_1^*\mid A,C_1)^T, {\bf C}_2^T, A]\}^{-1}$ 
yields the estimator %$\hat{\phi}_1$ 
with the minimum asymptotic variance $[E\{\sigma^{-2}(\allowbreak Z,\allowbreak {\bf C}_2)[1,E({\bf C}_1^{*}\mid Z,{\bf C}_2)^T,{\bf C}_2^T,Z]^{T\otimes 2}\}]^{-1}$ in the class of GMM estimators indexed by ${\bf S}$.% and $W$. 
\end{theorem}
%One can show that under the exclusion restriction and rank conditions, the most efficient estimator in this class of estimators indexed by $s$ is $\sigma^{-2}(A,{\bf C}_2)[1,E({\bf C}_1^*\mid A,{\bf C}_2)^T,{\bf C}_2^T,A]^T$, where $\sigma^2(A,{\bf C}_2)\equiv E\{(Y-\theta_0^{\dag}-\theta_{{\bf C}_1}^{\dag T}{\bf C}_1^*-\theta_{{\bf C}_2}^{\dag T}{\bf C}_2-\theta_A^{\dag}A)^2\mid A,{\bf C}_2\}$. 
\begin{comment}
\begin{remark}
The efficient estimator based on ${\bf S}^*$ can be identically computed by
\begin{align*}
    \hat{\theta}=\arg\min_{\theta}\left\{\frac{1}{n}\sum_{i=1}^nU(O_i;\theta)\right\}W\left\{\frac{1}{n}\sum_{i=1}^nU(O_i;\theta)\right\},
\end{align*}
where 
\begin{align*}
U(O_i;\theta) &= \sigma^2(A,{\bf C}_2)[1,E({\bf C}_1^*\mid A,{\bf C}_2)^T,{\bf C}_2^T,A]^T(Y_i - \theta_0 - \theta_{{\bf C}_1}^T{\bf C}^*_{1,i} - \theta_{{\bf C}_2}^T{\bf C}_{2,i} - \theta_A A_i)\\
W &= E\{\sigma^{-2}(A,{\bf C}_2)[1, E({\bf C}_1^*\mid A,{\bf C}_2)^T, {\bf C}_2^T, A]^T[1, E({\bf C}_1^*\mid A,{\bf C}_2)^T, {\bf C}_2^T, A]\}^{-1}.
\end{align*}
\end{remark}
\end{comment} 
Throughout, we restrict our attention to efficiency with respect to the class of GMM estimators. Additional efficiency gains may be possible if the assumed regression and measurement error models jointly imply additional restrictions on the observed data distribution beyond those captured in the GMM moment restrictions. %on which the GMM estimators are based. 
When the residuals in the true outcome regression model are homoscedastic and the measurement error model in Assumption \ref{assn:me1} holds, $\sigma^2(Z,{\bf C}_2)$ is constant in $Z$ and ${\bf C}_2$, and this reduces to the the previous IV vector with $E({\bf C}_1^*\mid Z,{\bf C}_2)$ as the IV for ${\bf C}_1^*$. Thus, when heteroscedasticity is mild, $[1,E({\bf C}_1^*\mid Z,{\bf C}_2)^T,{\bf C}_2^T,Z]^T$ will be a reasonable choice for ${\bf S}(Z,{\bf C}_2)$. %Alternatively, one could estimate $\sigma^2(Z,{\bf C}_2)$ as is done for weighted least squares, and substitute this estimate into the final estimating equations. Either way, 
However, the estimator based on such an ${\bf S}$ is clearly infeasible, as $E({\bf C}_1^*\mid Z,{\bf C}_2)$ is unknown and must also be estimated. %It need not be estimated consistently to preserve asymptotic normality, though an inconsistent estimate will result in sub-optimal asymptotic variance of the constructed IV estimator.

When estimating $E({\bf C}_1^*\mid Z,{\bf C}_2)$, it is imperative that it be fit nonlinearly in $Z$ and ${\bf C}_2$ so that it is not in the linear span of $Z$ and ${\bf C}_2$. There are two general approaches one might take to estimating $E({\bf C}_1^*\mid Z,{\bf C}_2)$. First, one might estimate this regression function parametrically, in which case, one would need to include nonlinear and/or interaction terms if an identity link is used. %Inference could then be obtained by stacking the estimating equations used for estimating $E({\bf C}_1^*\mid Z,{\bf C}_2)$ together with the constructed IV estimating equations, and solving them jointly. Namely, for estimating equations $\psi(Z,{\bf C}_1^*,{\bf C}_2;\alpha)$ for a nonlinear parametric model $E({\bf C}_1^*\mid Z,{\bf C}_2;\alpha)$, $\theta_Z$ is consistently and asymptotically normally estimated by the solution $\hat{\theta}_Z$ to
\begin{comment}
\begin{align*}
    \frac{1}{n}\sum_{i=1}^n\left[\begin{array}{c}
        \left[\begin{array}{c}
             1 \\
             E({\bf C}_1^*\mid Z_i,{\bf C}_{2,i};\alpha) \\
             {\bf C}_{2,i}    \\
             Z_i    \\
        \end{array}\right](Y_i - \theta_0 - \theta_{{\bf C}_1}^T{\bf C}_{1,i}^* - \theta_{{\bf C}_2}^T{\bf C}_{2,i} - \theta_Z Z_i)  \\
         \psi(Z_i,{\bf C}_{1,i}^*,{\bf C}_{2,i};\alpha)
    \end{array}\right]=0.
\end{align*}
\end{comment}
Alternatively, one could estimate $E({\bf C}_1^*\mid Z,{\bf C}_2)$ nonparametrically, %for instance by using %a kernel smoother, or by 
%some flexible machine learning estimator. This approach 
which has the advantage of naturally capturing potential nonlinearity of $E({\bf C}_1^*\mid Z,{\bf C}_2)$ without affecting the convergence rate. The only restriction is that the nuisance parameter estimator must either be contained in a Donsker class, or be fit using sample splitting. Sample splitting involves partitioning subjects into $K\geq 2$ roughly evenly-sized sets, $B_1, \ldots, B_K$. Let $p:[n]\rightarrow [K]$ denote the function mapping each subject to the set to which it belongs, such that $i\in B_{p(i)}$ for all $i$. For each subject $i=1,\ldots,n$, fit a model for $E({\bf C}_1^*\mid Z,{\bf C}_2)$ using only the subjects in $[n]\setminus B_{p(i)}$, then evaluate the resulting fit at $(Z_i, {\bf C}_{2,i})$. The result is then substituted for $E({\bf C}_1^*\mid Z_i,{\bf C}_{2,i})$ for each $i$ in the constructed IV estimator. 
%Inference can be performed using the bootstrap if the $E({\bf C}_1^*\mid Z,{\bf C}_2)$ estimator is a sufficiently smooth function of the empirical distribution (viz., Hadamard differentiable). This will (likely?) be the case for kernel smoothers, but is not well understood for many machine learning estimators. 
Under either approach, the uncertainty induced by the nuisance parameter estimation will not affect the asymptotic distribution of the causal parameter of interest (see Theorem 6.2 in \cite{newey1994large}). The asymptotic variance will be affected by the probability limit of the nuisance parameter estimator such that an inconsistent estimator of $E({\bf C}_1^*\mid Z,{\bf C}_2)$ will produce an inefficient estimator of the causal parameter of interest; however, such an estimator will remain CAN. The parametric approach to modeling $E({\bf C}_1^*\mid Z,{\bf C}_2)$ has the disadvantage of having to guess where the nonlinearities might be. %and including them as terms in the model. %The trade-off is that it will then be more straightforward to check whether these nonlinear terms are indeed unnecessary in the outcome regression model, which is required for identifiability to hold.

As mentioned previously, traditional best practices for IV estimation include performing a first-stage $F$-test of the IV relevance condition. We propose a modified $F$-test when the constructed IV is an estimate of the efficient choice of constructed IV in Section \ref{sec:F-test}. In Section \ref{sec:partially-linear}, we extend the results of this section to the outcome regression model
\begin{align}
\label{eqn:om4}
E(Y\mid {\bf C},Z;\theta)=g_1(Z,{\bf C}_2;\theta_1)+{\bf g}_2(Z,{\bf C}_2;\theta_2)^T{\bf C}_1,
\end{align}
where $g_1$ is real-valued, ${\bf g}_2$ is vector-valued with dimension $\dim({\bf C}_1)$, and both functions are known. Results in the following sections will be based on this more general model.

\section{Measurement error variance estimation}
\label{sec:me-variance}
Given a consistent estimator of the outcome regression coefficients, it then becomes possible to estimate the variance of the measurement error of a scalar error-prone covariate under an additive measurement error model. %While it may be possible that this can be extended to estimate the covariance matrix of the measurement error of a vector of error-prone covariates, it is not obvious how to do so. 
Estimating the measurement error variance %can serve multiple purposes. It 
can be of interest in its own right, or it can be used to adjust for the measurement error in any analysis involving the error-prone variable, either using the same data set or in future studies. %if data is sampled from the same %(or a similar enough) 
%population and measured in the same way. %In the following section, we will discuss 
For instance, we can adjust for exposure measurement error when estimating a mediated effect. In such a setting, an estimate of the measurement error variance is useful for adjusting for the measurement error in the mediator model, even as the measurement error is adjusted for in the outcome model using the approach described in the previous section. Such a strategy may also be useful when estimating quantities requiring more than one model involving the same error-prone variable. %, such as in principal stratification, estimating longitudinal causal effects with time-varying confounding, or for multiply-robust estimators in general.
%When $\theta$ is consistently estimated, it is possible to also consistently estimate the measurement error variance. (or covariance matrix for multiple error-prone variables). %Consider the classical measurement error model that satisfies measurement error model (1), with $A$ replacing $C_1$ and $C$ replacing ${\bf C}_2$: $A^*=A+\varepsilon$ and $E(\varepsilon\mid Z,C)=0$. 
%In particular, 
Suppose the following measurement error model holds.
\begin{assumption}
\label{assn:me3}
%\begin{align}
%\label{eqn:me5}
    $C_1^*=C_1+\varepsilon$, $E(\varepsilon\mid C_1, {\bf C}_2, Z, Y)=0$, and $\sigma^{2\dag}_{\varepsilon}\equiv E(\varepsilon^2) = E(\varepsilon^2\mid {\bf C}_2, Z)$.
    %C_1^*=C_1+\varepsilon; \;\; E(\varepsilon\mid C_1, {\bf C}_2, Z, Y)=0; \;\; E(\varepsilon\varepsilon^T\mid {\bf C}_2, Z)=E(\varepsilon\varepsilon^T).% E(C\varepsilon)=0; \;\; E(Y\mid Z,C_1,{\bf C}_2,\varepsilon)=E(Y\mid Z,C_1,{\bf C}_2).
%\end{align}
\end{assumption}
In our Bangladesh data example, this will hold if the protein intake measurement error is additive and mean independent of the true underlying protein intake, other baseline covariates, lead exposure, and birth length, and its variance does not depend on lead exposure or the other covariates. As with Assumptions \ref{assn:me1} and \ref{assn:me2}, this assumption is not testable without additional data. 
%and let $\sigma^{2\dag}_{\varepsilon}\equiv E(\varepsilon^2)$. 
Under Assumption \ref{assn:me3}, we can consistently estimate the entire coefficient vector $\theta$ in model (\ref{eqn:om4}) using an estimator from the previous section, which will in turn allow us to estimate $\sigma^{2\dag}_{\varepsilon}$. Let $T(z,{\bf c}_2)$ be a real-valued function. %if we assume $\varepsilon$ to be mean independent of $A$ and conditionally mean independent of $Y$ given $Z$, $A$, and $C$. 
For the true parameter value $\theta^{\dag}$, it can be shown that if $E\{T(Z,\allowbreak {\bf C}_2)\allowbreak g_2(Z,\allowbreak {\bf C}_2;\allowbreak \theta_2^{\dag})\}\allowbreak \neq\allowbreak  0$,
\begin{comment}
\[ E\left[T(Z,{\bf C}_2){\bf C}_1^{*}\left\{Y-g_1(Z,{\bf C}_2;\theta_1^{\dag})-g_2(Z,{\bf C}_2;\theta_2^{\dag})C_1^*\right\}\right]
= -E\left\{T(Z,{\bf C}_2)g_2(Z,{\bf C}_2;\theta_2^{\dag})\right\}\sigma^{2\dag}_{\varepsilon},\]
so that 
\end{comment}
$\sigma^{2\dag}_{\varepsilon}$ is identified by 
\[
-E\left\{T(Z,{\bf C}_2)g_2(Z,{\bf C}_2;\theta_2^{\dag})\right\}^{-1}
E\left[T(Z,{\bf C}_2){\bf C}_1^{*}\left\{Y-g_1(Z,{\bf C}_2;\theta_1^{\dag})-g_2(Z,{\bf C}_2;\theta_2^{\dag})C_1^*\right\}\right].
\]
\begin{comment}
\begin{align*}
&E\left[T(Z,{\bf C}_2)C_1^{*T}\left\{Y-g_1(Z,{\bf C}_2;\theta_1^{\dag})-{\bf g}_2(Z,{\bf C}_2;\theta_2^{\dag})^TC_1^*\right\}\right]\\
%=& E\left[T(Z,{\bf C}_2){C_1}^T\left\{Y-g_1(Z,{\bf C}_2;\theta_1^{\dag})-{\bf g}_2(Z,{\bf C}_2;\theta_2^{\dag})^TC_1\right\}\right]\\
%&-E\left\{T(Z,{\bf C}_2){C_1}^T{\bf g}_2(Z,{\bf C}_2;\theta_2^{\dag})^T\varepsilon\right\}\\
%&+ E\left[T(Z,{\bf C}_2)\varepsilon^T\left\{Y-g_1(Z,{\bf C}_2;\theta_1^{\dag})-{\bf g}_2(Z,{\bf C}_2;\theta_2^{\dag})^TC_1
%\right\}\right]\\
%&- E\left\{T(Z,{\bf C}_2)\varepsilon^T{\bf g}_2(Z,{\bf C}_2;\theta_2^{\dag})^T\varepsilon\right\}\\
=& -E\left\{T(Z,{\bf C}_2){\bf g}_2(Z,{\bf C}_2;\theta_2^{\dag})^T\right\}E(\varepsilon\varepsilon^T),
\end{align*}
so that $E(\varepsilon\varepsilon^T)$ is identified by 
\begin{align*}
&-E\left\{T(Z,{\bf C}_2){\bf g}_2(Z,{\bf C}_2;\theta_2^{\dag})^T\right\}^{-1}\\
&\times E\left[T(Z,{\bf C}_2)C_1^{*T}\left\{Y-g_1(Z,{\bf C}_2;\theta_1^{\dag})-{\bf g}_2(Z,{\bf C}_2;\theta_2^{\dag})^TC_1^*\right\}\right]
\end{align*}
\end{comment}
Thus, $\sigma^{2\dag}_{\varepsilon}$ can be estimated by substituting empirical means for the population means in the above expression. Alternatively, it could be estimated by constructing an estimating equation in terms of the unknown $\sigma^{2}_{\varepsilon}$, which could then be stacked with estimating equations for $\theta^{\dag}$ if it is also unknown. The next theorem shows the estimator will be asymptotically normal.

Define $\dot{G}\equiv E\{T(Z,{\bf C}_2)g_2(Z,{\bf C}_2;\theta_2^{\dag})\}$, $\dot{\Omega}\equiv E\{\dot{U}({\bf O};\sigma^{2\dag}_{\varepsilon})^{2}\}$, and
$\dot{U}({\bf O};\sigma_{\varepsilon}^2)\allowbreak \equiv\allowbreak T(Z,\allowbreak {\bf C}_2)\allowbreak [C_{1}^{*}\allowbreak \{Y-g_1(Z,{\bf C}_2;\theta_1^{\dag})-g_2(Z,{\bf C}_2;\theta_2^{\dag}){\bf C}_1^*\}+g_2(Z,{\bf C}_2;\theta_2^{\dag})\sigma^2_{\varepsilon}]$.
%\begin{align*}
%\dot{U}({\bf O};\sigma_{\varepsilon}^2) \equiv T(Z,{\bf C}_2)\left[C_{1}^{*}\left\{Y-g_1(Z,{\bf C}_2;\theta_1^{\dag})-g_2(Z,{\bf C}_2;\theta_2^{\dag}){\bf C}_1^*\right\}+g_2(Z,{\bf C}_2;\theta_2^{\dag})\sigma^2_{\varepsilon}\right].
%\end{align*}
%, and let $\mathbb{S}$ be the space of all $\dim(C_1)\times\dim(C_1)$ positive definite matrices. 
For a given $T$, we define the corresponding estimator to be the solution $\hat{\sigma}_{\varepsilon}^2$ to the estimating equation $\sum_{i=1}^n\dot{U}({\bf O}_i;\sigma^{2}_{\varepsilon})=0$.

%We can then define the following class of estimators of $\sigma^2_{\varepsilon}$ indexed by $T$:
%\[\hat{\sigma}^2_{\varepsilon}\equiv\argmin_{\sigma^2_{\varepsilon}\in\mathbb{R}^+}\left\{\frac1n\sum_{i=1}^n\dot{U}\left({\bf O}_i;\sigma^2_{\varepsilon}\right)\right\}^2.\]
%The following theorem shows that for fixed, known $\theta^{\dag}$, $\hat{\sigma}^2_{\varepsilon}$ is consistent and asymptotically normal.\\

%NOTE: How to deal with $G$ if this is just-identified, i.e., $G$ is not square? Alternatively, how to deal with zero entries?\\

\begin{theorem}\label{thm:sigma_can}
Suppose the following all hold: (a) outcome model (\ref{eqn:om4}) and the measurement error model in Assumption \ref{assn:me3} are both correctly-specified, (b) $0<\delta_{\ell}<\sigma^{2\dag}_{\varepsilon}< \delta_u<\infty$ for some known $\delta_{\ell}$ and $\delta_u$, (c) $\dot{G}\neq 0$, (d) $E\lvert T(Z,{\bf C}_2)g_2(Z,{\bf C}_2;\theta_2^{\dag})\rvert< \infty$, (e) $E\{\dot{U}({\bf O};\sigma^{2\dag}_{\varepsilon})^2 \}<\infty$. Then for the estimator $\hat{\sigma}^{2}_{\varepsilon}$ solving $\sum_{i=1}^n\dot{U}({\bf O}_i;\sigma^{2}_{\varepsilon})=0$,
$n^{1/2}(\hat{\sigma}^{2}_{\varepsilon}-\sigma^{2\dag}_{\varepsilon})\rightsquigarrow N(0,\dot{\Omega}/\dot{G}^{2})$.
%(\dot{G}^TW\dot{G})^{-1}\dot{G}^TW\dot{\Omega} W\dot{G}(\dot{G}^TW\dot{G})^{-1}
%Then $\widehat{\mathrm{Var}}(\varepsilon)\xrightarrow{p}\mathrm{Var}(\varepsilon)$. Suppose further that $\hat{\theta}$ is asymptotically normal with asymptotic variance $\Sigma$. Then
%\[\sqrt{n}\left\{\widehat{\mathrm{Var}}(\varepsilon)-\mathrm{Var}(\varepsilon)\right\}\rightsquigarrow N(0,?)\]
\end{theorem}
%\begin{remark}
The condition that $\sigma^{2\dag}_{\varepsilon}$ is bounded away from zero %in a compact subset of $\mathbb{R}^+$ 
is more stringent than the analogous conditions %on the true parameter 
in the previous theorems, as this 
means that asymptotic normality will not hold if $C_1$ truly has no measurement error. Furthermore, convergence is not uniform over all of $\mathbb{R}^+$. %, meaning that for any given sample size, there will be a neighborhood around zero in which the asymptotic distribution will be a poor approximation of the true sampling distribution of $\hat{\sigma}^{2}_{\varepsilon}$. %, since $\Sigma$ would not be positive definite. 
%This is generally the case with variance estimation on the boundary of the parameter space. %However, $\hat{\Sigma}$ will remain consistent for $\Sigma^{\dag}$ near the boundary of the space of positive definite matrices. 
%\end{remark}
Recall that Theorem \ref{thm:sigma_can} holds for known $\theta^{\dag}$. When $\theta$ is estimated, %(as will generally be the case), 
the uncertainty from this estimate can be accounted for by applying the delta method or by stacking the estimating equations in $\dot{U}$ with the estimating equations used to estimate $\theta$, if available, or alternatively by bootstrapping the entire procedure. %Unfortunately, the optimal choice of $T$ is a complicated function and depends on the unknown $\Sigma^{\dag}$, even for scalar $C_1$, so we refrain from presenting it here. 
In practice, an initial estimate of $g_2(Z,{\bf C}_2;\theta_2^{\dag})$ can serve as a reasonable choice for $T$, as it makes $\dot{G}>0$ likely to hold. %For $\hat{W}$, one can use $n^{-1}\sum_{i=1}^n\dot{U}({\bf O}_i;\dot{L})^{\otimes 2}$ using a preliminary estimator $\dot{L}$ of $L$. If this preliminary estimator is consistent, then $\hat{W}$ will be consistent for $\dot{\Omega}$, yielding an asymptotic variance of $(\dot{G}^T\dot{\Omega}\dot{G})^{-1}$.

\section{Mediation analysis with exposure measured with error}
\label{sec:mediation}
%Define new set of variables -- different notation? (Maybe $T$ for exposure in prev section, $A$ for exposure here?)\lv{I would keep the same notation as before. We can focus on the case that now A is latent and $A^*$ error-prone.}\\
We now outline our approach when a mediated effect is of interest, which applies the approaches described in the previous two sections. %in a modular fashion. 
Section \ref{sec:supp-mediation} gives a comprehensive discussion. We now use $A$ to denote a continuous exposure of interest, $Z$ to denote a mediator of the effect of $A$ on $Y$, and ${\bf C}$ to denote baseline covariates. We focus on $A$ being the variable subject to measurement error, though a similar approach can be applied to correct for an error-prone element of ${\bf C}$. Estimating the mediated effect typically involves estimating the effect of $Z$ on $Y$, hence $A$ can be viewed as playing the role of ${\bf C}_1$ in the previous sections. %, i.e., it plays the role of a confounder subject to measurement error for this part of the estimation problem. 
The DAG in Figure \ref{fig:pag}.b illustrates the causal relationship between ${\bf C}$, $A$, $Z$, and $Y$. Comparing this with %the PAG in 
Figure \ref{fig:pag}.a, %we can see 
$A$ replaces the role of ${\bf C}_1$ and ${\bf C}$ replaces the role of ${\bf C}_2$, and the causal relationship between the two is restricted to hold in one direction %, i.e., 
with ${\bf C}$ affecting $A$.

For comparing exposure levels $a'$ and $a''$, the natural direct effect (NDE) and natural indirect effect (NIE) are defined as $\mathrm{NDE}(a',a'')\equiv E[Y\{a',Z(a'')\}]-E[Y\{a'',Z(a'')\}]$ and $\mathrm{NIE}(a',a'')\equiv E[Y\{a',Z(a')\}]-E[Y\{a',Z(a'')\}]$, where the nested counterfactual $Y\{a',Z(a'')\}$ is the outcome we would have seen had (potentially contrary to fact) $A$ been assigned to $a'$, and had $Z$ been assigned to the value it would have taken (possibly contrary to fact) had $A$ been assigned to $a''$. %These effects have the useful property of decomposing the average treatment effect: $\mathrm{ATE}(a',a'')=\mathrm{NIE}(a',a'')+\mathrm{NDE}(a',a'')$. We will focus on estimation of the NIE; estimation of the NDE follows analogously. %Identification of these effects is discussed in Section S5 of the supporting web materials.
\begin{comment}
%Nonparametric identification formula?\\

The following assumptions are standard for nonparametric identification of mediated effects:
\begin{assumption}[Consistency] For all $z\in\mathrm{supp}(Z)$,
$Z=z$ implies $Y=Y(z)$ almost everywhere. For $a\in\{a',a''\}$, $A=a$ implies $Y=Y(a)$ and $Z=Z(a)$ almost everywhere. For $a\in\{a',a''\}$ and $z\in\mathrm{supp}(Z)$, $A=a$ and $Z=z$ implies $Y=Y(a,z)$ almost everywhere. 
\end{assumption}
\begin{assumption}[Counterfactual conditional independence]
For all $z\in\mathrm{supp}(Z)$ and $a\in\{a',a''\}$, $Z(a)\ci A\mid {\bf C}$; $Y(a,z)\ci A\mid {\bf C}$; $Y(a,z)\ci Z\mid {\bf C},A=a$; and $Y(a',z)\ci Z(a'')\mid {\bf C},A=a'$.
\end{assumption}
\begin{assumption}[Positivity]
For continuous $Z$, $f_{Z\mid A,{\bf C}}(z\mid A, {\bf C})>\delta$ almost everywhere in ${\bf C}$ and $A$, and for discrete $Z$, $\mathrm{Pr}(Z=z\mid A, {\bf C})>\delta$ almost everywhere in ${\bf C}$ and $A$ for some $\delta >0$. For $a\in\{a',a''\}$, $f_{A\mid {\bf C}}(a\mid {\bf C})>\delta$ almost everywhere in ${\bf C}$ for some $\delta>0$.
\end{assumption}
\end{comment}
Under additional causal assumptions given in Section \ref{sec:supp-mediation}, 
\begin{comment}
$\mathrm{NIE}(a',a'')$ is nonparametrically identified by
\begin{align*}{%\scriptstyle
%\mathrm{NIE}(a',a'')= 
E\left[E\left\{E(Y\mid Z,A=a',{\bf C})\mid A=a',{\bf C}\right\}-E\left\{E(Y\mid Z,A=a',{\bf C})\mid A=a'',{\bf C}\right\}\right].}
\end{align*}
\end{comment}
these effects can then be estimated by fitting models for $Y$ and $Z$. %Consider the following linear models:
Under the linear models, 
%\begin{align*}
    $E(Y\mid Z,A,{\bf C};\theta) = \theta_0+\theta_C^T{\bf C}+\theta_AA+\theta_ZZ$ and 
    $E(Z\mid A,{\bf C};\beta) = \beta_0+\beta_C^T{\bf C}+\beta_AA$,
%\end{align*}
$\mathrm{NIE}(a',a'')$ is identified by $\beta_A\theta_Z(a'-a'')$. %Substituting regression coefficient estimates for $\beta_A$ and $\theta_Z$ to estimate the NIE is known as the product method \citep{baron1986moderator}. 
We consider more general models allowing for interactions, nonlinearity in $\bf{C}$, and basis expansion in $Z$, in Section \ref{sec:supp-mediation}. %In the case of basis expansion of $Z$, the model for $Z$ above is replaced by models for each term in the basis expansion. %This method predates the causal mediation literature, and is by far the most popular method for estimating mediated/indirect effects. %\cite{vanderweele2009conceptual} demonstrated the consistency of the product method for the NIE under the above linear models. 
%However, the nonparametric identification formula accommodates more general models, allowing for nonlinearity and interactions, including exposure-mediator interactions, which can otherwise be mistaken for mediation even when none is present. 
\begin{comment}
More generally, consider an outcome model taking the following form:
\begin{align*}
    E(Y\mid Z,A,{\bf C};\theta) &= \sum_{k=1}^{k^*} g_k({\bf C};\theta_k^{\dag})b_k(Z) + A\sum_{k=k^*+1}^{K}g_k({\bf C};\theta_k^{\dag}) b_k(Z),
\end{align*}
where $b_k$ are user-specified functions of $Z$. Since this is a submodel of the parametric partially linear model (\ref{eqn:om4}), the constructed IV method described in Section \ref{sec:partially-linear} can be applied to estimate $\theta$. 

Measurement error model (\ref{eqn:me5}) applied to this setting becomes
\begin{align}
\label{eqn:me6}
    A^*=A+\varepsilon; \;\; E(\varepsilon\mid A, {\bf C}, Z, Y)=0; \;\; E(\varepsilon^2\mid {\bf C}, Z)=E(\varepsilon^2).
    %A^*=A+\varepsilon; \;\; E(\varepsilon\mid A, {\bf C}_2, Z, Y)=0; \;\; E(\varepsilon\varepsilon^T\mid {\bf C}_2, Z)=E(\varepsilon\varepsilon^T).% E(C\varepsilon)=0; \;\; E(Y\mid Z,A,{\bf C}_2,\varepsilon)=E(Y\mid Z,A,{\bf C}_2).
\end{align}
\end{comment}

Under the above outcome model and the measurement error model in Assumption \ref{assn:me3} applied to $A$ in this setting, the outcome model parameters as well as the exposure measurement error variance can be estimated using the methods described in the previous two sections. These can then be combined to estimate mediated effects. The primary differences between this setting and the ATE setting are that (a) here, we need to account for the presence of measurement error in both outcome and mediator models rather than just one, and (b) the measurement error models in Assumptions \ref{assn:me1} and \ref{assn:me2} are insufficient; we instead use the classical measurement error model in Assumption \ref{assn:me3}. The measurement error variance estimate can then be used to correct for the measurement error in the mediator model(s). In Section \ref{sec:supp-mediation}, we provide a class of CAN estimators of model coefficients given the measurement error variance, as well as the efficient estimator in this class. In the linear model case, this class of estimators contains the classical method-of-moments estimator \citep{fuller2009measurement} (which is equivalent to regression calibration in this setting \citep{carroll2006measurement}) as a special case. Alternatively, it may be more reasonable to exchange the roles of the mediator and outcome models, and use the constructed IV approach to estimate the mediator model and estimate the measurement error variance, then use the latter estimate to estimate the outcome model. One might also be willing to use the constructed IV approach for both models rather than relying on estimating the measurement error variance. Which approach to take depends on which set of assumptions seem most plausible in the setting of interest. If a valid IV is available, this can of course replace the constructed IV in any of these approaches. 

\section{Simulation study}
\label{sec:sims}
We performed a series of simulations to compare performance of the constructed IV approach with traditional sensitivity analyses for measurement error correction and under potential violation of our assumptions. %in which we evaluated bias, variance, and confidence interval coverage probability. 
The data generating process mimics 
the environmental epidemiology study that motivates our work. Full details are available in Section Section \ref{sec:sims-supp} of the supporting web materials. We simulated moderate to severe measurement error, setting the conditional reliability ratio (the ratio of the conditional variances of the true and mismeasured exposure) to 90\%, 80\%, and 70\%. We considered a setting in which the outcome is additive in the exposure and mediator %(``no exposure-mediator-interaction'') 
as well as a scenario in which the effect of the exposure is modified by the mediator. %(``exposure-mediator-interaction''). 
We also considered a scenario in which the mismeasured variable is a linear function of the covariates, violating the key rank condition of the constructed IV approach. We used generalized additive models (GAMs) with penalized splines and data-driven selection of degrees of freedom in R package mgcv to recover the constructed IV. %The sample size of each simulated dataset was set to n = 1000. 
In each setting, we performed 100 replications with a sample size of 1000. We implemented the constructed IV approach in three ways: (a) constructed IV regression for both outcome and mediator models (IVZ-IVY), (b) constructed IV regression for the outcome model and our %measurement error 
adjusted estimator described in Section \ref{sec:supp-mediation}, which corresponds to the method-of-moments (MoM) adjustment %(which is equivalent to regression calibration in this linear model setting) 
for the mediator model, using the estimated measurement error variance from the IV outcome regression (MoMZ-IVY), (c) constructed IV regression for the mediator model and MoM adjusted regression for the outcome model, using the estimated measurement error variance from the IV mediator regression (IVZ-MoMY). We compared the constructed IV approach with a sensitivity analysis using MoM for measurement error correction assuming a range of reliability ratios of 0.7--0.9. 
Results from the no-interaction setting are presented in Tables \ref{tab:sims-holds} and \ref{tab:sims-fails}. Section \ref{sec:additional-tabs} contains results from the interaction setting, a weak IV setting with larger measurement error, and with varying functional forms of the GAM used to fit the constructed IV. It also contains results for ATE estimation under the measurement error model in Assumption \ref{assn:me2}.
\begin{table}
\caption{\label{tab:sims-holds}Simulation results in the no interaction case with rank condition holding using constructed IV functional form 1 (NDE = 1.5, NIE = -0.15, TE = 1.35).}
\centering
\begin{tabular}{ll r@{.}l r@{.}l r@{.}l r@{.}l r@{.}l r@{.}l r@{.}l r@{.}l r@{.}l }
\\
\hline\hline
\multicolumn{2}{l}{Reliability ratio:}          & \multicolumn{6}{c}{70\%}    & \multicolumn{6}{c}{80\%}    & \multicolumn{6}{c}{90\%}    \\
 &  & \multicolumn{2}{c}{Bias}  & \multicolumn{2}{c}{S.D.} & \multicolumn{2}{c}{C.P.} & \multicolumn{2}{c}{Bias}  & \multicolumn{2}{c}{S.D.} & \multicolumn{2}{c}{C.P.} & \multicolumn{2}{c}{Bias}  & \multicolumn{2}{c}{S.D.} & \multicolumn{2}{c}{C.P.} \\
 \hline
Naive     & NDE      & -0&47 & 0&23     & 0&56     & -0&32 & 0&25     & 0&84     & -0&17 & 0&27     & 0&89     \\
          & NIE      & 0&08 & 0&06     & 0&77     & 0&05 & 0&08     & 0&87     & 0&03 & 0&09     & 0&91      \\
          & TE       & -0&39  & 0&24     & 0&64     & -0&26 & 0&25     & 0&85     & -0&13 & 0&27     & 0&93     \\
IVZ-IVY & NDE      & -0&02 & 0&35      & 0&95     & -0&01 & 0&35      & 0&95     & -0&01 & 0&35      & 0&95     \\
          & NIE      & 0&01  & 0&10      & 0&89     & 0&01 & 0&10      & 0&89     & 0&00 & 0&10      & 0&89     \\
          & TE       & -0&01  & 0&32     & 0&97     & -0&01  & 0&32     & 0&97     & 0&00  & 0&32     & 0&96     \\
IVZ-MoMY  & NDE      & 0&01 & 0&34      & 0&97     & 0&00 & 0&32      & 0&96     & 0&00 & 0&31      & 0&95     \\
          & NIE      & 0&01  & 0&09      & 0&91     & 0&00 & 0&09      & 0&91     & 0&00 & 0&10      & 0&91      \\
          & TE       & 0&01  & 0&34     & 0&97     & 0&01  & 0&32     & 0&96     & 0&00  & 0&30     & 0&96     \\
MoMZ-IVY  & NDE      & -0&02 & 0&35     & 0&95     & -0&01 & 0&35     & 0&95     & -0&01 & 0&35      & 0&95     \\
          & NIE      & 0&00 & 0&16     & 0&94     & -0&01 & 0&12     & 0&93     & 0&00 & 0&12     & 0&93      \\
          & TE       & -0&02 & 0&32     & 0&96     & -0&02 & 0&31     & 0&95     & -0&01     & 0&31      & 0&94     \\
MoM-70   & NDE      & 0&03  & 0&35     & 0&96     & 0&40  & 0&41     & 0&87     & 0&84   & 0&49     & 0&67     \\
          & NIE      & 0&00     & 0&10     & 0&94     & -0&07 & 0&13     & 0&92     & -0&18  & 0&16     & 0&79     \\
          & TE       & 0&03  & 0&34     & 0&95     & 0&32  & 0&39     & 0&88     & 0&66   & 0&44     & 0&78     \\
MoM-80   & NDE      & -0&23 & 0&29      & 0&93     & 0&01  & 0&32     & 0&96     & 0&28  & 0&37     & 0&86     \\
          & NIE      & 0&04 & 0&08     & 0&91     & 0&00     & 0&10     & 0&93     & -0&05 & 0&12     & 0&93     \\
          & TE       & -0&19 & 0&29      & 0&93     & 0&02  & 0&32     & 0&95     & 0&23  & 0&35     & 0&90     \\
MoM-90   & NDE      & -0&37 & 0&25     & 0&75     & -0&19 & 0&28     & 0&92     & 0&00     & 0&31     & 0&96     \\
          & NIE      & 0&06  & 0&07     & 0&84     & 0&03  & 0&08     & 0&91     & 0&00     & 0&10     & 0&92     \\
          & TE       & -0&31 & 0&26     & 0&80     & -0&16 & 0&28     & 0&93     & 0&00     & 0&30     & 0&95    \\
\hline
\end{tabular}
\end{table}

\begin{table}
\caption{\label{tab:sims-fails} Simulation results in the no interaction case with rank condition \emph{not} holding (NDE = 1.5, NIE = -0.15, TE = 1.35).}
\centering
\begin{tabular}{ll r@{.}l r@{.}l r@{.}l r@{.}l r@{.}l r@{.}l r@{.}l r@{.}l r@{.}l }
\\
\hline\hline
\multicolumn{2}{l}{Reliability ratio:}          & \multicolumn{6}{c}{70\%}    & \multicolumn{6}{c}{80\%}    & \multicolumn{6}{c}{90\%}    \\
 &  & \multicolumn{2}{c}{Bias}  & \multicolumn{2}{c}{S.D.} & \multicolumn{2}{c}{C.P.} & \multicolumn{2}{c}{Bias}  & \multicolumn{2}{c}{S.D.} & \multicolumn{2}{c}{C.P.} & \multicolumn{2}{c}{Bias}  & \multicolumn{2}{c}{S.D.} & \multicolumn{2}{c}{C.P.} \\
 \hline
Naive     & NDE & -0&47 &	0&22 &	0&39 &	-0&31 &	0&25 &	0&75 &	-0&15 &	0&27 &	0&87   \\
%-0&47  & 0&22  & 0&00       & -0&31  & 0&24  & 0&00       & -0&15  & 0&26  & 0&86    \\
                           & NIE & 0&09 &	0&06 &	0&73 &	0&06 &	0&07 &	0&85 &	0&04 &	0&08 &	0&92 \\
                           %-0&08  & 0&06  & 0&71    & -0&06  & 0&07  & 0&84    & -0&03  & 0&08  & 0&92    \\
                           & TE  & -0&38 &	0&23 &	0&55 &	-0&25 &	0&25 &	0&76 &	-0&12 &	0&26 &	0&88  \\
                           %-0&37  & 0&23  & 0&55    & -0&25  & 0&24  & 0&76    & -0&11  & 0&26  & 0&88    \\
IVZ-IVY & NDE & -0&24 &	4&72 &	1&00 &	-0&19 &	5&21 &	1&00 &	-1&19 &	6&84 &	1&00 \\
%-0&02  & 1&08  & 1&00       & 0&16   & 1&14  & 1&00       & 0&32   & 1&20   & 1&00       \\
                           & NIE & 0&14 &	0&32 &	1&00 &	0&06 &	0&42 &	1&00 &	0&24 &	1&45 &	1&00 \\
                           %0&06   & 0&11  & 0&98    & 0&02   & 0&13  & 1&00       & -0&01  & 0&15  & 1&00       \\
                           & TE  & -0&11 &	4&55 &	1&00 &	-0&13 &	4&89 &	1&00 &	-0&96 &	5&90 &	1&00  \\
                           %0&03   & 1&03  & 1&00       & 0&19   & 1&08  & 1&00       & 0&31   & 1&12  & 1&00       \\
IVZ-MoMY  & NDE & -0&41 &	0&56 &	0&95 &	-0&23 &	0&62 &	0&99 &	-0&05 &	0&67 &	1&00   \\
%-0&02  & 1&08  & 1&00       & 0&16   & 1&14  & 1&00       & 0&32   & 1&20   & 1&00       \\
                           & NIE & 0&01 &	0&12 &	1&00 &	-0&04 &	0&14 &	1&00 &	-0&09 &	0&19 &	1&00   \\
                           %-0&61  & 1&13  & 0&97    & -0&57  & 0&93  & 0&98    & 1&36   & 7&69  & 0&98    \\
                           & TE  & -0&41 &	0&49 &	1&00 &	-0&27 &	0&52 &	1&00 &	-0&14 &	0&55 &	1&00  \\
                           %-0&63  & 1&24  & 0&96    & -0&40   & 1&04  & 0&96    & -1&04  & 7&79  & 0&99    \\
MoMZ-IVY  & NDE & -0&24 &	4&72 &	1&00 &	-0&19 &	5&21 &	1&00 &	-1&19 &	6&84 &	1&00   \\
%-0&4   & 0&61  & 0&95    & -0&17  & 0&85  & 0&98    & -0&04  & 0&64  & 0&99    \\
                           & NIE & -0&96 &	6&32 &	0&99 &	-0&36 &	5&46 &	0&98 &	-0&23 &	10&99 &	0&98 \\
                           %-0&01  & 0&29  & 1&00       & -0&10   & 0&61  & 1&00       & -0&10   & 0&21  & 1&00       \\
                           & TE  & -1&21 &	5&24 &	0&93 &	-0&55 &	2&52 &	0&96 &	-1&42 &	8&60 &	1&00  \\
                           %-0&41  & 0&46  & 1&00       & -0&27  & 0&50   & 1&00       & -0&14  & 0&51  & 1&00       \\
MoM-70   & NDE & 0&05 &	0&34 &	0&94 &	0&43 &	0&40 &	0&79 &	0&89 &	0&48 &	0&54   \\
%0&04   & 0&36  & 0&94    & 0&29   & 0&37  & 0&85    & 0&54   & 0&40   & 0&73    \\
                           & NIE & 0&00 &	0&10 &	0&94 &	-0&09 &	0&12 &	0&89 &	-0&22 &	0&15 &	0&69   \\
                           %0&00      & 0&09  & 0&92    & -0&05  & 0&11  & 0&92    & -0&11  & 0&12  & 0&90     \\
                           & TE  & 0&05  &	0&33  &	0&91  &	0&34  &	0&38  &	0&79  &	0&67  &	0&43  &	0&60 \\
                           %0&04   & 0&33  & 0&93    & 0&23   & 0&35  & 0&85    & 0&42   & 0&37  & 0&74    \\
MoM-80   & NDE & -0&22 &	0&28 &	0&85 &	0&03 &	0&32 &	0&94 &	0&31 &	0&36 &	0&82  \\
%-0&17  & 0&28  & 0&88    & 0&03   & 0&31  & 0&95    & 0&24   & 0&34  & 0&85    \\
                           & NIE & 0&05 &	0&08 &	0&92 &	0&00 &	0&09 &	0&91 &	-0&06 &	0&11 &	0&93    \\
                           %-0&04  & 0&08  & 0&93    & 0&00      & 0&09  & 0&92    & -0&04  & 0&11  & 0&90     \\
                           & TE  & -0&17  &	0&28  &	0&87  &	0&03  &	0&31  &	0&91  &	0&25  &	0&34  &	0&82    \\
                           %-0&13  & 0&28  & 0&91    & 0&03   & 0&3   & 0&92    & 0&20    & 0&32  & 0&86    \\
MoM-90   & NDE & -0&37 &	0&25 &	0&68 &	-0&18 &	0&27 &	0&88 &	0&02 &	0&30 &	0&92 \\
%-0&34  & 0&25  & 0&71    & -0&16  & 0&27  & 0&88    & 0&02   & 0&30   & 0&94    \\
                           & NIE & 0&07 &	0&07 &	0&80 &	0&04 &	0&08 &	0&92 &	0&00 &	0&09 &	0&91 \\
                           %0&06   & 0&07  & 0&82    & 0&04   & 0&08  & 0&93    & 0&00      & 0&09  & 0&92    \\
                           & TE  & -0&30  &	0&25  &	0&72  &	-0&14  &	0&27  &	0&87  &	0&02  &	0&29  &	0&90   \\
                           %-0&27  & 0&25  & 0&71    & -0&12  & 0&27  & 0&91    & 0&02   & 0&29  & 0&92    \\
\hline
\end{tabular}
\end{table}

When the assumptions of the constructed IV approach hold, our proposed estimators dramatically reduce bias and have coverage closer to 95\% than the traditional sensitivity analysis approach, except for when the reliability ratio is known. Performances of the three proposed versions of the constructed IV approach are similar across all scenarios. However, when the assumption of a nonlinear relationship between covariates and exposure is violated, the constructed IV approach breaks down, displaying a marked increase in bias and variance. %, as expected. 
In this case, confidence intervals are seen to over-cover, yielding conservative inference. %(See simulationresults.pdf). 
%\\

\section{Data analysis}
\label{sec:data}
Maternal nutrition during pregnancy is associated with fetal growth in children. In contrast, metal exposures, lead (Pb) in particular, are harmful to child physical and cognitive development. %(cite). asdf
In Bangladesh, malnutrition and exposure to heavy metals, such as lead, are of particular concern. %(cite). asdf
The effect of maternal nutrition on child development is likely a combination of a direct positive impact of beneficial nutrients and an indirect adverse impact of chemical exposures obtained through diet. %, and an antagonistic interaction between nutritional factor and the metal, whereby the beneficial effect of nutritional intake can be reduced by lead exposure. 
Figure \ref{fig:pag}.b represents the hypothesized mediating mechanism in this study, where $A$ denotes the maternal nutritional factor as measured by the protein intake score obtained from food frequency questionnaire, $Y$ is a measure of fetal growth as measured by birth length in cm, and $Z$ is maternal exposure to lead as measured by log transformed and centered lead in cord blood.  The disentangling of these effects is complicated by error incurred by dietary surveys. Indeed, dietary surveys are well-known to be prime sources of error-prone data and have motivated the development of many measurement error methods (e.g., see \cite{rosner1989correction}, and running examples throughout \cite{carroll2006measurement}). Furthermore, the measurement error may be biased as in Assumption \ref{assn:me2} if %measurement error model (\ref{eqn:me2}). %, thereby violating the classical measurement error model that is typically assumed when adjusting for measurement error. 
%For instance, 
mothers over- or underestimate the amount of food containing protein they eat. %when responding to the questionnaire. 

%We apply our proposed methodology to this data to estimate both the ATE of maternal lead exposure on birth length as well as the NIE of protein intake on birth length through lead exposure. Since estimation of the NIE involves estimating the relationship between lead exposure and birth length, we can view estimation of the ATE as something of an intermediate step to estimation of the NIE, where in the former case, the error-prone protein intake variable plays the role of a confounder, and in the latter, it plays the role of the exposure.

In a sample of 764 Bangladeshi mother-infant pairs enrolled from two clinics in the Sirajdikhan and Pabna districts, we aimed to (a) estimate the ATE of lead on birth length, a proxy for fetal grown, and (b) explain potentially lead-induced adverse effects of protein intake, a proxy for maternal diet, on birth length by estimating the NIE of protein intake on birth length through lead. In the former case, the error-prone protein variable plays the role of a confounder, %(for the exposure lead), 
whereas in the latter, it plays the role of the exposure. Both require estimation of a model for birth length regressed on lead, protein, and potential confounders. The mediation case further requires estimation of a model regressing lead (the mediator) on protein and potential confounders. 
%quantify and explain the beneficial effects of protein intake, a proxy for maternal diet, on birth length, a proxy for fetal grown. We further sought to quantify the harmful effects of lead, the potentially lead-induced adverse effects of protein intake (i.e., the indirect effect), and the potential interaction between lead and protein intake. 
%The outcome of interest, birth length, was obtained at delivery. 
%Maternal protein intake, was obtained during pregnancy via a questionnaire on meat, fish, vegetable, and egg intake. Lead was measured in cord blood, which is thought to reflect maternal exposure during pregnancy. 
The following variables were adjusted for as confounders, both for the ATE of lead and %, as well as for the exposure-outcome, exposure-mediator, and mediator-outcome relationships in 
the mediation analysis: clinic, mother's age, education, smoking exposure, and quality of the home (a marker of SES). We used the constructed IV approach to correct for measurement error in protein intake without auxiliary information, as there is no data on the magnitude of the exposure error available in this population. %Protein intake is both the exposure of interest as well as a confounder of the lead--birth length relationship.

We modeled the mismeasured protein intake as a nonlinear function of lead and confounders using generalized additive models to produce the constructed IV. %, as done in the simulation study. 
We detected significant nonlinear terms: a nonlinear interaction between lead with the site indicator, a nonlinear association with lead, and interaction among socioeconomic confounding factors (Web Table S2). 
We proceeded by correcting the lead and birth length models by running instrumental variable regressions using the residuals from the exposure regressions as instruments (Web Table S4, IVZ-IVY approach). For the average treatment effect of lead on birth length, we estimated a non-statistically significant reduction of
-0.255 (95\% CI: -0.589, 0.076) cm per unit change in log lead level. Unlike the mediation analysis, this estimate is robust to biased measurement error of the form specified in Assumption \ref{assn:me2}. %measurement error model (\ref{eqn:me2}).

For the mediation analysis, in addition to direct IV regression adjustment, we estimated the measurement error variance and used this to obtain measurement error-corrected mediator and outcome regressions via a MoM approach, as in the simulation study. We refer to these approaches as IVZ-MoMY or MoMZ-IVY when the measurement error variance was recovered from the mediator IV or outcome IV regression, respectively. Our instrumental variable approach estimated a reliability ratio for the nutritional factor of $92\%$ from the outcome model and of $80\%$ from the mediator model (mean across 500 bootstrap replicates). We compared these with a sensitivity analysis using a MoM correction applied to both mediator and outcome regressions assuming a range of reliability from 70\% to 90\%. The constructed IV regression analyses as well as the MoM corrected regression analyses indicated that the relationships between exposure and outcome, mediator and outcome, and exposure and mediator might be underestimated by the analyses that ignore exposure measurement error (Web Tables S4 and S5). 
Total, direct, and indirect effects were then estimated as functions of regression parameters from the IVZ-IVY, IVZ-MoMY, MoMZ-IVY, and MoM analyses. Standard errors were obtained via bootstrapping.

Table \ref{tab:analysis} shows the results of the mediation analysis for the naive, constructed IV, the hybrid constructed IV and MoM approaches and the sensitivity analysis using MoM. 
\begin{table}[]
\caption{\label{tab:analysis} Results of measurement error-naive and measurement error-adjusted mediation analyses of Bangladeshi study in Sirajdikhan and Pabna clinics (excluding exposure-mediator interaction).}
\centering
\begin{tabular}{ll r@{.}l@{ } r@{.}l@{ } r@{.}l r@{.}l@{ } r@{.}l@{ } r@{.}l r@{.}l@{ } r@{.}l@{ } r@{.}l}
\hline \hline
Site   &   Estimator & \multicolumn{6}{c}{NDE} & \multicolumn{6}{c}{NIE} & \multicolumn{6}{c}{TE}  \\
\hline 
Sirajdikhan   &   Naive     & 0&30    & (0&19,    & 0&40)   & 0&014   & (0&004,   & 0&03)  & 0&31   & (0&21,    & 0&41)    \\
   &IVZ-IVY   & 0&46   & (0&004,   & 0&94)   & 0&012   & (-0&03,   & 0&09)  & 0&47   & (0&03,    & 0&96)    \\
   &IVZ-MoMY  & 0&28   & (-0&35,   & 1&05)   & 0&012   & (-0&04,   & 0&11)  & 0&30    & (-0&35,   & 1&02)    \\
   &MoMZ-IVY  & 0&46   & (0&004,   & 0&94)   & 0&012   & (-0&001,  & 0&03)  & 0&30    & (-0&35,   & 1&02)    \\
   &MoM 70\%  & 1&07   & (0&60,    & 1&52)   & -0&10    & (-0&56,   & 0&11)  & 0&94   & (0&44,    & 1&42)    \\
   &MoM 80\%  & 0&62   & (0&38,    & 0&85)   & 0&05    & (-0&004,  & 0&12)  & 0&65   & (0&43,    & 0&91)    \\
   &MoM 90\%  & 0&41   & (0&27,    & 0&54)   & 0&05    & (0&01,    & 0&10)  & 0&45   & (0&31,    & 0&58)   \\
   \hline
Pabna   &   Naive     &  0&15   & (0&10,    & 0&19)   & 0&002  & (-0&001,  & 0&006)  & 0&15   & (0&10,    & 0&19)  \\
   &IVZ-IVY   &  0&21   & (-0&06,   & 0&42)   & 0&002  & (-0&03,   & 0&03)   & 0&21   & (0&06,    & 0&42)  \\
   &IVZ-MoMY  &  0&14   & (-0&19,   & 0&60)   & 0&001  & (-0&04,   & 0&04)   & 0&14   & (-0&20,   & 0&58)  \\
   &MoMZ-IVY  &  0&21   & (-0&06,   & 0&42)   & 0&001  & (-0&001,  & 0&01)   & 0&14   & (-0&20,   & 0&61)  \\
   &MoM 70\%  &  0&53   & (0&31,   & 0&78)   & -0&20   & (-0&63,   & 0&02)   & 0&31   & (-0&00,   & 0&62)  \\
   &MoM 80\%  &  0&30    & (0&19,    & 0&39)   & -0&02  & (-0&08,   & 0&03)   & 0&28   & (0&15,    & 0&36)  \\
   &MoM 90\%  &  0&20    & (0&13,    & 0&25)   & 0&002  & (-0&016,  & 0&02)   & 0&20    & (0&13,    & 0&26) \\
\hline
\end{tabular}
\end{table}
The adjusted mediation analyses lead to larger estimates of total and direct effects in both clinics along with wider confidence intervals, yet we found weak evidence of a direct effect. The results from the hybrid IVZ-MoMY approach produced much wider confidence intervals. This is expected because the mediator regression IV displayed a lower $F$ statistic than the one obtained from the outcome regression IV than the outcome regression IV (2.2 vs.~3.4). Although both $F$ statistics were  statistically significant, their values being
%The $F$-tests for both lead and birth length regression constructed IVs took values of 2.6 and 3.7, respectively---both 
less than the conventional threshold of 10 %, and while this is an arbitrary threshold, this indicates, 
suggests that the constructed instruments may still be weak. The practical implication is that convergence of the corresponding IV-based estimators may be slow such that the normal approximations are poor when sample size is not especially large, hence inference may be less reliable. Our weak IV simulation study results in Web Table S7 show this can result in inflated bias and variance, though coverage probabilities tended to be conservative. Thus, while our point estimates using the constructed IV methods may be less reliable, we may still have confidence in the findings of a significant NDE in Sirajdikhan based on the IVY methods and a significant TE based on the IVZ-IVY method in both clinics.

Results from MoM adjusted sensitivity analyses based on specified measurement errors more markedly indicated a potential under-estimation of direct and total effects, displaying much tighter confidence intervals but also confirmed weak evidence of mediation. %Assuming a range of reliability from 70\% to 100\%, the lower bound of the sensitivity analysis still excludes the null hypothesis for both direct and total effects (except for the total effect in Pabna), while in the IV analyses, the confidence intervals for the total and direct effects includes the null in most cases.  
%To check for sensitivity to potential model mis-specification, we re-ran the 
IV analyses including an exposure-mediator interaction did not alter our conclusions (Web Table S6). %\newline

%Although the analyses show potential underestimation of a negative lead-protein intake interaction, the size of the interaction coefficients were low ($\hat{\theta}^{\mathrm{naive}}_{Z*A}=-0.00\hspace{0.1in} (-0.04,0.01)$, $\hat{\theta}^{IVZ-IVY}_{Z*A}=-0.06\hspace{0.1in} (-0.14,0.02)$), and including this additional non-linearity did not change our conclusions (Table S6). 

%The results of the constructed IV analyses are in line with the sensitivity analyses. However, the $F$-statistics indicate that the constructed IV approach might be relying on weak instruments resulting in less efficiency than the MoM sensitivity analyses and potential bias in the estimation of the measurement error variance. A small $F$-statistic might be the result of model mis-specification or insufficient sample size. The conclusions from this epidemiological analysis should be taken with caution due to potential residual unmeasured confounding. 

%\begin{comment}asdf
\section{Discussion}
\label{sec:discussion}
%Recap\\
%Limitations\\
%\lv{practical recommendations}\\
%\lv{extensions/future work}
%\end{comment}

We have developed measurement error-robust estimation of causal effects in the absence of auxiliary information %data or knowledge of the measurement error distribution 
by using constructed instrumental variables, which are functions of data only coming from the main study. In particular, we developed estimators for the ATE on a continuous outcome when confounders are measured with error, as well as the NDE/NIE when a continuous exposure is measured with error. We additionally proposed estimators of the measurement error variance. %, which may be used to correct for measurement error in other models or studies, or may be of independent interest itself. 
This approach builds on the higher-order moment restriction methodology of \cite{lewbel1997constructing} by explicitly connecting this approach to causal estimands as well as deriving asymptotically efficient estimators within the classes of estimators generated by first- and second-order moment restrictions. 

\begin{comment}
In our application of the proposed methodology to the Bangladeshi study, protein intake was found to be potentially measured with error, leading to underestimation of the total and direct effect of protein intake on birth length.  After accounting for measurement error we found suggestive evidence of lead contamination of dietary intake in the Pabna clinic; however, we did not find strong evidence of indirect effects through lead contamination nor of antagonistic interaction of maternal protein intake with lead contamination.
\end{comment}

The proposed methodology requires correct specification of the conditional mean outcome and that this model is linear in the error-prone variable. Thus, one important direction for future work is to extend this methodology to outcomes with nonlinear link functions, as in \cite{miles2018class}. Furthermore, this methodology relies on a rank condition that may not be satisfied in many applications. In the linear model case, this requires that the conditional mean of the error prone variable is nonlinear in the other covariates. When there is insufficient nonlinearity, the constructed IV will behave like a weak instrument, which is known to yield unstable estimation. For this reason, it is prudent to perform a first-stage $F$-test, which we developed a modified version of tailored to our setting. %Although this test did reject in our data analysis, the $F$ statistic did not meet the tradition threshold of 10 (though this is a somewhat arbitrary rule of thumb).

Given these limitations, we do not necessarily advocate for the proposed approach to replace estimators that take advantage of additional information %(e.g., traditional IVs or replication or validation study data) 
when available. For example, when a validation study is available, one can apply the methods proposed in \cite{cheng2023mediation,cheng2025correcting}. However, when such information is unavailable %(as is often the case) 
and our assumptions seem plausible, we recommend using this methodology in concert with other complementary approaches such as sensitivity analysis to gain a more comprehensive %and multifaceted 
perspective of the data, as we have demonstrated in our application to the Bangladeshi study data.

%  The \backmatter command formats the subsequent headings so that they
%  are in the journal style.  Please keep this command in your document
%  in this position, right after the final section of the main part of 
%  the paper and right before the Acknowledgements, Supplementary Materials,
%  and References sections. 

%\backmatter

%  This section is optional.  Here is where you will want to cite
%  grants, people who helped with the paper, etc.  But keep it short!

\section*{Acknowledgements}

The authors are grateful to Maitreyi Mazumdar for providing the Bangladesh study data.\vspace*{-8pt}

%  If your paper refers to supplementary web material, then you MUST
%  include this section!!  See Instructions for Authors at the journal
%  website http://www.biometrics.tibs.org

\section*{Supplementary Materials}
Web Appendices, Tables, and Figures referenced in Sections \ref{sec:intro}, \ref{sec:ate}, \ref{sec:mediation}, \ref{sec:sims}, and \ref{sec:data} as well as simulation R code are available with this paper at the Biometrics website on Oxford Academic.

\section*{Data Availability}
The data underlying this article were provided by Maitreyi Mazumdar by permission. Data will be shared on request to the corresponding author, with permission of Maitreyi Mazumdar and David Christiani.

%Web Appendix A, referenced in Sections \ref{sec:intro}, \ref{sec:ate}, \ref{sec:mediation}, \ref{sec:sims}, and \ref{sec:data} is available with
%this paper at the Biometrics website on Wiley Online Library.
%\vspace*{-8pt}

\bibliographystyle{apalike}

\bibliography{references}

\newpage

\appendix

\renewcommand{\thepage}{S\arabic{page}}  
\renewcommand{\thesection}{S\arabic{section}}   
\renewcommand{\thetable}{S\arabic{table}}   
\renewcommand{\thefigure}{S\arabic{figure}}
\renewcommand{\thetheorem}{S\arabic{theorem}}
\renewcommand{\theequation}{S\arabic{equation}}
\renewcommand{\theassumption}{S\arabic{assumption}}
\renewcommand{\theremark}{S\arabic{remark}}
\setcounter{page}{1}
\setcounter{section}{0}
\setcounter{table}{0}
\setcounter{figure}{0}
\setcounter{theorem}{0}
\setcounter{equation}{0}
\setcounter{assumption}{0}
\setcounter{remark}{0}

\section{Constructed instrumental variables depending on ${C}_1^*$}
\label{sec:c1-dependent}
We can recover identifiability of causal parameters under a different set of assumptions using a constructed instrumental variable of the form ${\bf S}_1(Z,{\bf C}_2){C}^*_1+{\bf S}_2(Z,{\bf C}_2)$, where ${\bf S}_1$ and ${\bf S}_2$ have the same dimension as $\theta$, and $E\left\{{\bf S}_1(Z,{\bf C}_2)\right\} = {\bf 0}$. Here we consider the case where ${C}_1$ is a scalar as is done in \cite{lewbel1997constructing} and \cite{lewbel2012using}, though it may be possible to extend to a vector of covariates measured with error. If
\begin{align}
\label{eqn:me3}
    {C}_1^* = {C}_1+\varepsilon;\;
    E(\varepsilon\mid Z,{\bf C}) = {0};\;
    E(Y\mid Z,{\bf C},\varepsilon) = E(Y\mid Z,{\bf C});\;
    E(\varepsilon^2\mid Z,{\bf C}_2) = E(\varepsilon^2)
\end{align}
all hold, then $E[\{{\bf S}_1(Z,{\bf C}_2){C}^*_1+{\bf S}_2(Z,{\bf C}_2)\}(Y-\theta_0^{\dag}-\theta_{{C}_1}^{\dag}{C}_1^*-\theta_{{\bf C}_2}^{\dag T}{\bf C}_2-\theta_Z^{\dag}Z)]={\bf 0}$.
Further, if $E[\{{\bf S}_1(Z,{\bf C}_2){C}_1^*+{\bf S}_2(Z,{\bf C}_2)\}[1,{C}_1^*,{\bf C}_2^T,Z]]$ is nonsingular, then the above estimating equations will have a unique solution, and hence $\theta$ will be identified. This rank condition implies that ${C}_1^*$ must be heteroscedastic in $(Z,{\bf C}_2)$. Let $\check{\bf U}_1({\bf O};\theta)\equiv \left\{{\bf S}_1(Z,{\bf C}_2){C}^*_1+{\bf S}_2(Z,{\bf C}_2)\right\}(Y-\theta_0-\theta_{{C}_1}{C}_1^*-\theta_{{\bf C}_2}^T{\bf C}_2-\theta_ZZ)$ and \[\check{\bf U}_2({\bf O};\theta,\mu)\equiv \left[\begin{array}{c}
    \check{\bf U}_1({\bf O};\theta)\\ 
    \theta_0 + \theta_{{C}_1}{C}^*_1 + \theta_{{\bf C}_2}^T{\bf C}_2 - \mu
    \end{array}\right].
\]
The following theorem gives the consistency and asymptotic normality of the GMM estimator solving $\sum_{i=1}^n\check{\bf U}_k({\bf O}_i;\theta)={\bf 0}$ for $k\in\{1,2\}$.
\begin{theorem}\label{thm:can_c1}
Suppose (a) the measurement error conditions in Assumption 4 hold, (b) $\phi_k^*$ is in the interior of a compact subset of the parameter space for $k=1,2$, (c) ${C}_1$ and all elements of ${\bf O}$ have finite fourth moment and $E\{\lVert {\bf S}_1(Z,{\bf C}_2){C}_1^*+{\bf S}_2(Z,{\bf C}_2)\rVert^4\}<\infty$, and (d) $\check{{G}}_1\equiv E[\{{\bf S}_1(Z,{\bf C}_2){C}^*_1+{\bf S}_2(Z,{\bf C}_2)\}[1,{C}_1^{*},{\bf C}_2^T,Z]]$ is nonsingular. Then for the estimators $\check{\phi}_k$ solving $\sum_{i=1}^n\check{\bf U}_k({\bf O}_i;\phi_k)={\bf 0}$ for $k=1,2$,
$\sqrt{n}(\check{\phi}_k-\phi_k^*)\rightsquigarrow \mathcal{N}\{0,\check{{G}}_k^{-1}\check{{\mathit\Omega}}_k(\check{{G}}_k^{-1})^T\}$,
where \[\check{{G}}_2\equiv \left[\begin{array}{cc}
     \check{{G}}_1 & {\bf 0}  \\
     {\bf 0} & 1 
\end{array}\right],\]
and $\check{{\mathit\Omega}}_k\equiv E\{\check{\bf U}_k({\bf O};\phi_k^*)^{\otimes 2}\}$ for $k=1,2$.
\end{theorem}
%In fact, this theorem demonstrates that the standard OLS estimator is consistent and asymptotically normal for $\theta$ when its conditions are met, since the form of the instrument vector contains the left-multiplying vector in the normal equations. In other words, this is a case in which measurement error is benign with respect to bias. Further, the robust standard error estimator used in the error-free case will remain consistent for $\mathrm{Var}(\hat{\theta})$. 
%However, the weighted least squares estimator is no longer the most efficient estimator in the class of estimators indexed by ${\bf S}_1$ and ${\bf S}_2$. 
The following theorem gives the most efficient estimator in the class of estimators indexed by ${\bf S}_1$ and ${\bf S}_2$ under homoscedasticity of $Y$ in $\{Z,{C}_1^*,{\bf C}_2\}$.
\begin{theorem}\label{thm:efficiency-nonlinear}
Suppose that $E\{(Y-\theta_0^{\dag}-\theta_{{C}_1}^{\dag}{C}_1^*-\theta_2^{\dag T}{\bf C}_2-\theta_Z^{\dag}Z)^2\mid Z,{C}_1^*,{\bf C}_2\}$ is constant in $Z$, ${C}_1^*$, and ${\bf C}_2$. The under Theorem \ref{thm:can_c1} conditions (a)--(d) for $k=1$, the instrument vector
\[\left[1,\left[1-\frac{\mathrm{Var}({C}_1^*\mid Z,{\bf C}_2)^{-1}}{E\{\mathrm{Var}({C}_1^*\mid Z,{\bf C}_2)^{-1}\}}\right]{C}_1^*,{\bf C}_2^T,Z\right]^T.\]
%\[%{\bf S}_1(Z,{\bf C}_2){\bf C}_1^*+{\bf S}_2(Z,{\bf C}_2)=
%\sigma^{-2}(Z,{\bf C}_2)\left[1, \left[1-\frac{E\{\sigma^{-2}(Z,{\bf C}_2)\}\mathrm{Var}({\bf C}_1^*\mid Z,{\bf C}_2)^{-1}}{E\{\sigma^{-2}(Z,{\bf C}_2)\mathrm{Var}({\bf C}_1^*\mid Z,{\bf C}_2)^{-1}\}}\right]{\bf C}_1^*, {\bf C}_2^T,Z\right]^T\]
%and $W^*\equiv E\{\sigma^{-2}(A,{\bf C}_2)[1, E({\bf C}_1^*\mid A,C_1)^T, {\bf C}_2^T, A]^T[1, E({\bf C}_1^*\mid A,C_1)^T, {\bf C}_2^T, A]\}^{-1}$ 
produces the estimator $\check{\phi}_1$ attaining the minimum asymptotic variance, 
\[\sigma^{2}E\left[\left\{{\bf S}_1^*(Z,{\bf C}_2){C}_1^*+{\bf S}_2^*(Z,{\bf C}_2)\right\}^{\otimes 2}\right]\] 
in the class of GMM estimators indexed by $({\bf S}_1^T,{\bf S}_2^T)^T$.%, where $\sigma^{*2}(Z,{\bf C}_1^*,{\bf C}_2) \equiv E\{(Y-\theta_0^{\dag}-\theta_{{\bf C}_1}^{\dag}{\bf C}_1^*-\theta_2^{\dag T}{\bf C}_2-\theta_Z^{\dag}Z)^2\mid Z,{\bf C}_1^*,{\bf C}_2\}$.% and $W$. 
\end{theorem}
The general efficient estimator under heteroscedasticity of $Y$ is also shown in the proof, as well as the efficient estimator under homoscedasticity of $Y$ only with respect to ${C}_1^*$. However, these are more complex and less practical for implementation than the above estimator.
\begin{comment}
The optimal estimator in terms of asymptotic variance within the class of GMM estimators indexed by $({\bf S}_1,{\bf S}_2)$ solving the above estimating equations uses the instrument vector
\[%{\bf S}_1(Z,{\bf C}_2){\bf C}_1^*+{\bf S}_2(Z,{\bf C}_2)=
\sigma^{-2}(Z,{\bf C}_2)\left[1, \left[1-\frac{E\{\sigma^{-2}(Z,{\bf C}_2)\}\mathrm{Var}({\bf C}_1^*\mid Z,{\bf C}_2)^{-1}}{E\{\sigma^{-2}(Z,{\bf C}_2)\mathrm{Var}({\bf C}_1^*\mid Z,{\bf C}_2)^{-1}\}}\right]{\bf C}_1^*, {\bf C}_2^T,Z\right]^T.\]
\end{comment}
\begin{comment}
If $Y$ is homoscedastic in $(Z,C)$, under (3), $\sigma^2(Z,{\bf C}_2)$ will be constant, and the optimal constructed IV reduces to
\[\left[1,\left[1-\frac{\mathrm{Var}({\bf C}_1^*\mid Z,{\bf C}_2)^{-1}}{E\{\mathrm{Var}({\bf C}_1^*\mid Z,{\bf C}_2)^{-1}\}}\right]{\bf C}_1^*,{\bf C}_2^T,Z\right]^T.\]
\end{comment}
Similarly to the previous case, $\mathrm{Var}({C}_1^*\mid Z,{\bf C}_2)$ must be estimated, though in this case, it may be done using a linear model. However, if there is little to no variation in $Z$ and ${C}_1^*$, the element containing ${C}_1^*$ will be nearly zero, and will be a weak instrument. As such, it remains prudent to conduct a weak instrument test prior to deploying this constructed IV.

\section{Causal assumptions for the mean counterfactual outcome}
\label{sec:causal-assumptions}
We assume the following hold throughout for all levels $z$ of interest:
\begin{assumption}[Consistency] %For all $z\in\mathrm{supp}(Z)$,
\label{assn:consistency}
$Z=z$ implies $Y=Y(z)$ almost everywhere.
\end{assumption}
\begin{assumption}[No unobserved confounding]
\label{assn:nuca}
$Y(z)\ci Z\mid {\bf C}$.% for all $z\in\mathrm{supp}(Z)$.
\end{assumption}
\begin{assumption}[Positivity]
\label{assn:positivity}
%For continuous $Z$, 
$f_{Z\mid {\bf C}}(z\mid {\bf C})>\delta>0$ %and $f_{Z\mid C}(z+\Delta\mid C)>\delta$ 
almost everywhere in ${\bf C}$, where $f_{Z\mid {\bf C}}$ is the conditional probability mass (density) function for discrete (continuous) $Z$. %or the conditional probability density function for continuous $Z$. %For discrete $Z$, $\mathrm{Pr}(Z=z\mid C)>\delta$ and $\mathrm{Pr}(Z=z+\Delta\mid C)>\delta$ almost everywhere in $C$ for some $\delta >0$.
\end{assumption}

\section{A modified first-stage $F$-test of the relevance condition for the constructed IV}
\label{sec:F-test}
We propose a modified $F$-test when the constructed IV is an estimate of the efficient choice of constructed IV in the linear model case. Instead of the standard $F$-test, which is an $F$-test of the regression models of the endogenous variable with and without the IV, one performs an $F$-test comparing the linear model of the endogenous variable ${\bf C}_1^*$ on $A$ and ${\bf C}_2$ and a larger model with nonlinear terms in which the former model is nested. If one is using a parametric model to estimate the constructed IV, one would use this same model as the larger model in the $F$-test. When estimating the constructed IV nonparametrically, one can select any linear model with nonlinear terms, such as basis expansion terms. In this latter case, even though the larger parametric model is not being used to produce the constructed IV, this still produces a valid test of the strength of the IV provided the nonparametric estimator is flexible enough to contain the larger, alternative parametric model.

\section{Partially linear parametric model setting}
\label{sec:partially-linear}
%Having illustrated our methodology in the simple linear model setting, 
Here, we consider the more general partially linear parametric model under which constructed IVs can be applied. %For simplicity, we only focus on extending constructed instrumental variables not depending on ${\bf C}_1^*$, but the results of Section \ref{sec:c1-dependent} could be extended to the model in this subsection as well. 
Suppose the measurement error model in Assumption 1 holds, and that the outcome regression model (\ref{eqn:om4}), i.e.,
\begin{align*}
%\label{eqn:om4}
E(Y\mid {\bf C},Z;\theta)=g_1(Z,{\bf C}_2;\theta_1)+{\bf g}_2(Z,{\bf C}_2;\theta_2)^T{\bf C}_1,
\end{align*}
is correctly specified, where $g_1$ is real-valued and ${\bf g}_2$ is vector-valued with dimension $\dim({\bf C}_1)$. This model allows for any functional form of the regression function with the sole exception that it must be linear in ${\bf C}_1$. Specifically, it allows for arbitrary nonlinearity in $Z$ and ${\bf C}_2$, and arbitrary interaction between $Z$ and ${\bf C}_2$, as well as $Z$ and ${\bf C}_2$ with the linear trend in ${\bf C}_1$. It is also appropriate for non-ordinal polytomous $Z$ as it allows for the use of dummy variables for levels of $Z$. However, in contrast with how the term ``partially linear model'' is typically used (i.e., as a semiparametric model), our model does require that the functional forms of $g_1$ and ${\bf g}_2$ are known. Under this model, which is potentially nonlinear in $Z$, if $Z$ is continuous, one may prefer to estimate the dose-response curve $\mu(z)\equiv E\{Y(z)\}$. Given the true ${\bf C}_1$, $\mu(z)$ would be nonparametrically identified by $E\{E(Y\mid {\bf C},Z=z)\}$ under Assumptions 1--3, with positivity required to hold at all $z$ in the domain of interest for the dose-response curve.

Consider a constructed IV vector ${\bf S}(Z,{\bf C}_2)$, with dimension $\dim(\theta)$, and define
\[\tilde{\bf U}_1({\bf O};\theta)\equiv {\bf S}(Z,{\bf C}_2)\left\{Y-g_1(Z,{\bf C}_2;\theta_1)-{\bf g}_2(Z,{\bf C}_2;\theta_2)^T{\bf C}_1^*\right\}.\]
The next theorem shows that the GMM estimator solving $\sum_{i=1}^n\tilde{\bf U}_1({\bf O}_i;\theta)={\bf 0}$ is consistent and asymptotically normal provided a rank condition and other regularity conditions hold. This can be used to generate a consistent and asymptotically normal estimator for $\mu(z)$ at any $z$, and in turn a dose-response curve. The parameters $\mu_j\equiv E\{Y(z_j)\}$ can be estimated by $n^{-1}\sum_{i=1}^n\{g_1(z_j,{\bf C}_{2,i};\hat{\theta}_1)+{\bf g}_2(z_j,{\bf C}_{2,i};\hat{\theta}_2)^T{\bf C}_{1,i}^*\}$ for each value $z_j$ of interest. %, where $\hat{\theta}\equiv (\hat{\theta}_0, \hat{\theta}_{C_1}^T, \hat{\theta}_{{\bf C}_2}^T, \hat{\theta}_Z)^T$ is the GMM estimator solving the above estimating equations. 
The corresponding joint asymptotic distribution can be obtained using the delta method. Alternatively, the dose-response curve approximated at finitely many $z_j$ and its asymptotic distribution can be obtained by augmenting the estimating equations as follows:
\[\tilde{\bf U}_2({\bf O}_i,\bm{z};\theta,\mu)\equiv \left[\begin{array}{c}
    \tilde{\bf U}_1({\bf O}_i;\theta)\\ 
    m(z_1,{\bf C}^*_{1,i},{\bf C}_{2,i};\theta)-\mu_1\\
    \vdots \\
    m(z_J,,{\bf C}^*_{1,i},{\bf C}_{2,i};\theta)-\mu_J
    \end{array}\right],
\]
where $m(z,{\bf c}_1,{\bf c}_2;\theta)\equiv g_1(z,{\bf c}_{2};\theta_1)+{\bf g}_2(z,{\bf c}_{2};\theta_2)^T{\bf c}_1$. Of course, when $g_1$ or ${\bf g}_2$ is a nonlinear function of $z$, the dose-response curve for a continuous exposure is no longer characterized by the intercept and slope as in the linear model case, and so the above approach can be taken for a grid of points in the domain of interest for $z$ in order to provide a summary.
\begin{theorem}
\label{thm:partially-linear-can}
Suppose the following hold: (a) the measurement error model in Assumption 1 and outcome regression model (\ref{eqn:om4}) are both correctly specified, (b) $\theta^{\dag}$ is in the interior of a compact subset of the parameter space, (c) $g_1(z,{\bf c}_2;\theta_1)$ and ${\bf g}_2(z,{\bf c}_2;\theta_2)$ are continuously differentiable in a neighborhood $\mathcal{N}$ of $\theta^{\dag}$ with probability approaching one, (d) $E\{\lVert \tilde{\bf U}_k({\bf O};\theta^{\dag})\rVert^2 \}<\infty$ for $k=1,2$, (e) $E\{\sup_{\theta\in\mathcal{N}}\lVert\nabla_{\theta}\tilde{\bf U}_k({\bf O};\theta^{\dag})\rVert\}<\infty$ for $k=1,2$, and (f) $\tilde{{G}}_1\equiv E\{\nabla_{\theta}\tilde{\bf U}_1({\bf O};\theta^{\dag})\}$ is nonsingular. Let $\tilde{\phi}_1\equiv(\tilde{\theta}_1^T,\tilde{\theta}_2^T)^T$ and $\tilde{\phi}_2\equiv(\tilde{\theta}_1^T,\tilde{\theta}_2^T,\tilde{\mu}^T)^T$ be the estimators solving the constructed IV equations: $\sum_{i=1}^n\tilde{\bf U}_1({\bf O}_i;\theta)={\bf 0}$ and $\sum_{i=1}^n\allowbreak \tilde{\bf U}_2({\bf O}_i;\allowbreak \theta,\allowbreak \mu)\allowbreak =\allowbreak {\bf 0}$, respectively. Then 
% \begin{align*}
%g_1(m+\Delta,{\bf C}_{2,i};\tilde{\theta}) - g_1(m,{\bf C}_{2,i};\tilde{\theta})\right.\\
%&\left.+ \left\{{\bf g}_2(m+\Delta,{\bf C}_{2,i};\tilde{\theta})-{\bf g}_2(m,{\bf C}_{2,i};\tilde{\theta}\right\}{\bf C}_{1,i}^*)\right]
%m(m+\Delta,{\bf C}_{1,i}^*,{\bf C}_{2,i};\tilde{\theta})-m(a,{\bf C}_{1,i}^*,{\bf C}_{2,i};\tilde{\theta})\right\}\]
%\left[g_1(a+\Delta,{\bf C}_2;\tilde{\theta}_1)-g_1(a,{\bf C}_2;\tilde{\theta}_1)+\left\{{\bf g}_2(a+\Delta,{\bf C}_{2,i};\tilde{\theta}_2)^T-{\bf g}_2(a,{\bf C}_{2,i};\tilde{\theta}_2)^T\right\}C_{1,i}\right]\]
%\end{align*}
$n^{1/2}(\tilde{\phi}_k-\phi^{\dag}_k)\rightsquigarrow \mathcal{N}\{0,\tilde{{G}}_k^{-1}\tilde{{\mathit \Omega}}_k(\tilde{{G}}_k^{-1})^T\}$ for $k=1,2$, %and
%$n^{1/2}\{\tilde{\mu}(z)-\mu(z)\}\rightsquigarrow N\left(0,\sigma_{\mu}^2\right)$ for each $z$, 
where %$\tilde{\mu}(z)\equiv  n^{-1}\sum_{i=1}^nE(Y_i\mid z,{\bf C}_{1,i}^*,{\bf C}_{2,i};\tilde{\theta})$ and 
$\tilde{{\mathit\Omega}}_k\equiv E\{\tilde{\bf U}_k({\bf O};\theta^{\dag})^{\otimes 2}\}$,
\[\tilde{{G}}_2\equiv \left[\begin{array}{cc}
     \tilde{{G}}_1 & {0}  \\
     {0} & {I}_J 
\end{array}\right],\]
and ${I}_J$ is the $J\times J$ identity matrix.

%$\sigma^2_{\mu}\equiv \mathrm{Var}\{D^T{\bf G}U({\bf O};\theta^{\dag})+m({\bf O};\theta^{\dag})\}$
\end{theorem}
\begin{remark}
For nonlinear functions $g_1$ and ${\bf g}_2$ of $\theta$, the rank condition (f) can be challenging to demonstrate, and effectively amounts to the less-primitive condition that the estimating equations have a unique solution. Whether this holds will depend on the functional form of the model, and as such practitioners should exercise caution when using nonlinear models in this setting. One precaution that can be taken %when it is unclear if this condition holds 
is to try different initial values when solving the nonlinear system of equations to ensure distinct solutions are not found, though of course this cannot be taken as proof that the solution is unique. When $g_1$ and ${\bf g}_2$ are both linear functions of $\theta$, $\tilde{{G}}_1$ will be invariant to $\theta^{\dag}$, and hence will be easier to interpret.
\end{remark}

The following theorem gives the optimal choice of ${\bf S}$ with respect to asymptotic variance. 
\begin{theorem}\label{thm:partially-linear-efficiency}
Under Theorem \ref{thm:partially-linear-can} conditions (a)--(f),  \[{\bf S}^*(Z,{\bf C}_2)\equiv \sigma^{-2}(Z,{\bf C}_2)[\nabla_{\theta_1}g_1(Z,{\bf C}_2)^T,E({\bf C}_1^*\mid Z,{\bf C}_2)^T\nabla_{\theta_2}^T{\bf g}_2(Z,{\bf C}_2)]^T\] %and $W^*\equiv E\{\sigma^{-2}(A,{\bf C}_2)[1, E({\bf C}_1^*\mid A,C_1)^T, {\bf C}_2^T, A]^T[1, E({\bf C}_1^*\mid A,C_1)^T, {\bf C}_2^T, A]\}^{-1}$ 
produces the estimator attaining the minimum asymptotic variance,
\begin{align*}
    &\left[E\left\{\sigma^{-2}(Z,{\bf C}_2)[\nabla_{\theta_1}g_1(Z,{\bf C}_2)^T,E({\bf C}_1^*\mid Z,{\bf C}_2)^T\nabla_{\theta_2}^T{\bf g}_2(Z,{\bf C}_2)]^{\otimes 2}\right\}\right]^{-1},
    %&\times\left.\left.[\nabla_{\theta_1}g_1(Z,{\bf C}_2)^T,E({\bf C}_1^*\mid Z,{\bf C}_2)^T\nabla_{\theta_2}^T{\bf g}_2(Z,{\bf C}_2)]\right\}\right]^{-1},
\end{align*}
in the class of GMM estimators indexed by ${\bf S}$.% and $W$. 
\end{theorem}

\section{Mediation analysis with exposure measured with error}
\label{sec:supp-mediation}
\begin{figure}[ht]
\centering
\begin{tikzpicture}[->, line width=1pt, every node/.style={circle,inner sep=0pt}]
\tikzstyle{every state}=[draw=none]
\node[draw, minimum size=8mm] (C) at (0.2679492,1) {${\bf C}$};
\node[draw, minimum size=8mm] (A) at (0.2679492,-1) {$A$};
\node[draw, minimum size=8mm] (Z) at (2,0) {$Z$};
\node[draw, minimum size=8mm] (Y) at (4,0) {$Y$};

  \path 	(C) edge (A)
			(C) edge (Z)
			(C) edge [bend left] (Y)
			(Z) edge (Y)
			(A) edge (Z)
			(A) edge [bend right] (Y)
					  ;
\end{tikzpicture}
\caption{The directed acyclic graph representing the causal structure among the variables ${\bf C}$, $A$, $Z$, and $Y$.}
\label{fig:dag}
\end{figure}
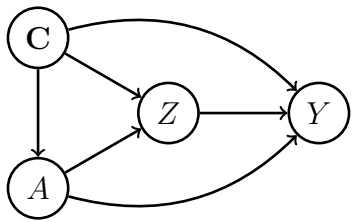

%Define new set of variables -- different notation? (Maybe $T$ for exposure in prev section, $A$ for exposure here?)\lv{I would keep the same notation as before. We can focus on the case that now A is latent and $A^*$ error-prone.}\\
%We now consider a setting in which a mediated effect is of interest. 
Here we will use $A$ to denote a continuous exposure of interest, $Z$ to denote the potential mediator of the effect of $A$ on $Y$, and ${\bf C}$ to denote baseline covariates. We will now focus on $A$ being the variable subject to measurement error, though a similar approach can be applied to correct for an error-prone scalar element of ${\bf C}$ instead. Estimating the mediated effect typically involves estimating the effect of $Z$ on $Y$. As such, $A$ can be viewed as playing the role of ${\bf C}_1$ in the previous sections, i.e., it plays the role of a confounder subject to measurement error for this part of the estimation problem. The DAG in Figure \ref{fig:dag} illustrates the causal relationship between ${\bf C}$, $A$, $Z$, and $Y$. Comparing this with the PAG in Figure 1, we can see $A$ replacing the role of ${\bf C}_1$ and ${\bf C}$ replacing the role of ${\bf C}_2$, and the causal relationship between the two being restricted to hold in one direction, i.e., ${\bf C}$ (potentially) affecting $A$.

%Define nested counterfactuals, NIE, effect decomposition\\

In addition to the counterfactual $Y(z)$ we considered previously, we will also need to assume the existence of the counterfactual/potential outcome based on an intervention setting $A$ to $a$, $Y(a)$, the counterfactual/potential outcome based on a joint intervention setting $A$ to $a$ and $Z$ to $z$, $Y(a,z)$, and the counterfactual/potential mediator $Z(a)$. These are all interpreted analogously to $Y(z)$. Additionally, causal mediation analysis assumes the existence of \emph{nested} counterfactuals. In particular, we will assume the existence of the nested counterfactual $Y\{a',Z(a'')\}$, i.e., the outcome we would have seen had (potentially contrary to fact) $A$ been assigned to $a'$, and had $Z$ been assigned to the value it would have taken (again possibly contrary to fact) had $A$ been assigned to $a''$. For comparing exposure levels $a'$ and $a''$, the natural (pure) direct effect (NDE) and natural indirect effect (NIE) are defined as $\mathrm{NDE}(a',a'')\equiv E[Y\{a',Z(a'')\}]-E[Y\{a'',Z(a'')\}]$ and $\mathrm{NIE}(a',a'')\equiv E[Y\{a',Z(a')\}]-E[Y\{a',Z(a'')\}]$, and have the useful property of decomposing the average treatment effect: $\mathrm{ATE}(a',a'')=\mathrm{NIE}(a',a'')+\mathrm{NDE}(a',a'')$. We will focus on estimation of the NIE; estimation of the NDE follows analogously.

%Nonparametric identification formula?\\

The following assumptions are standard for nonparametric identification of mediated effects:
\begin{assumption}[Consistency] For all $z\in\mathrm{supp}(Z)$,
$Z=z$ implies $Y=Y(z)$ almost everywhere. For $a\in\{a',a''\}$, $A=a$ implies $Y=Y(a)$ and $Z=Z(a)$ almost everywhere. For $a\in\{a',a''\}$ and $z\in\mathrm{supp}(Z)$, $A=a$ and $Z=z$ implies $Y=Y(a,z)$ almost everywhere. 
\end{assumption}
\begin{assumption}[Counterfactual conditional independence]
For all $z\in\mathrm{supp}(Z)$ and $a\in\{a',a''\}$, $Z(a)\ci A\mid {\bf C}$; $Y(a,z)\ci A\mid {\bf C}$; $Y(a,z)\ci Z\mid {\bf C},A=a$; and $Y(a',z)\ci Z(a'')\mid {\bf C},A=a'$.
\end{assumption}
\begin{assumption}[Positivity]
For continuous $Z$, $f_{Z\mid A,{\bf C}}(z\mid A, {\bf C})>\delta$ almost everywhere in ${\bf C}$ and $A$, and for discrete $Z$, $\mathrm{Pr}(Z=z\mid A, {\bf C})>\delta$ almost everywhere in ${\bf C}$ and $A$ for some $\delta >0$. For $a\in\{a',a''\}$, $f_{A\mid {\bf C}}(a\mid {\bf C})>\delta$ almost everywhere in ${\bf C}$ for some $\delta>0$.
\end{assumption}
Under Assumptions 4--6, the NIE is nonparametrically identified by
\begin{align*}
\mathrm{NIE}(a',a'')= E\left[E\left\{E(Y\mid Z,A=a',{\bf C})\mid A=a',{\bf C}\right\}-E\left\{E(Y\mid Z,A=a',{\bf C})\mid A=a'',{\bf C}\right\}\right].
\end{align*}
This can then be estimated by fitting models for $Y$ and $Z$. Consider the following linear models:
\begin{align*}
    E(Y\mid Z,A,{\bf C};\theta) &= \theta_0+\theta_C^T{\bf C}+\theta_AA+\theta_ZZ \\
    E(Z\mid A,{\bf C};\beta) &= \beta_0+\beta_C^T{\bf C}+\beta_AA.
\end{align*}
Under these models, the identification formula for $\mathrm{NIE}(a',a'')$ reduces to $\beta_A\theta_Z(a'-a'')$. Plugging in regression coefficient estimates for $\beta_A$ and $\theta_Z$ to estimate the NIE is known as the product method \citep{baron1986moderator}. This method predates the causal mediation literature, and is by far the most popular method for estimating mediated/indirect effects. \cite{vanderweele2009conceptual} demonstrated the consistency of the product method for the NIE under the above linear models. However, the nonparametric identification formula accommodates more general models, allowing for nonlinearity and interactions, including exposure-mediator interactions, which can otherwise be mistaken for mediation even when none is present. More generally, consider an outcome model taking the following form:
\begin{align*}
    E(Y\mid Z,A,{\bf C};\theta) &= \sum_{k=1}^{k^*} g_k({\bf C};\theta_k^{\dag})b_k(Z) + A\sum_{k=k^*+1}^{K}g_k({\bf C};\theta_k^{\dag}) b_k(Z),
\end{align*}
where $b_k$ are user-specified functions of $Z$. Since this is a submodel of the parametric partially linear model (\ref{eqn:om4}), the constructed IV method described in Section \ref{sec:partially-linear} can be applied to estimate $\theta$. 

The measurement error model in Assumption 4 applied to this setting becomes
\begin{align}
\label{eqn:me6}
    A^*=A+\varepsilon; \;\; E(\varepsilon\mid A, {\bf C}, Z, Y)=0; \;\; E(\varepsilon^2\mid {\bf C}, Z)=E(\varepsilon^2).
    %A^*=A+\varepsilon; \;\; E(\varepsilon\mid A, {\bf C}_2, Z, Y)=0; \;\; E(\varepsilon\varepsilon^T\mid {\bf C}_2, Z)=E(\varepsilon\varepsilon^T).% E(C\varepsilon)=0; \;\; E(Y\mid Z,A,{\bf C}_2,\varepsilon)=E(Y\mid Z,A,{\bf C}_2).
\end{align}
Under these models, the exposure measurement error variance can be estimated using the method described in Section 3. Alternatively, if a valid IV is available, this can be used for both of these tasks instead of the constructed IV. Under this model, the identification formula for $\mathrm{NIE}(a',a'')$ becomes
\begin{align*}
E\left(\sum_{k=1}^Ka'^{I(k>k^*)}g_k({\bf C};\theta_k^{\dag})\left[E\left\{b_k(Z)\mid A=a',{\bf C}\right\}-E\left\{b_k(Z)\mid A=a'',{\bf C}\right\}\right]\right).
\end{align*}
Thus, we also need to estimate $E\left\{b_k(Z)\mid A=a,{\bf C}\right\}$ for each $k$. 

Now consider the mediator model
\begin{align}
\label{eqn:om7}
    E\{b(Z)\mid A, {\bf C}\} = h_1({\bf C};\beta_1^{\dag}) + h_2({\bf C};\beta_2^{\dag})A
\end{align}
for a given $b_k$ (here we suppress the dependence on $k$ for notational simplicity, though separate models with separate parameters must be fit for each $k$). If the true exposure measurement error variance $\sigma^{2\dag}_{\varepsilon}$ is known, we can correct for it directly, as shown in our next result. Consider known functions $\bar{{\bf S}}_1({\bf C})$ and $\bar{{\bf S}}_2({\bf C})$, both with the same dimension as $\beta$, and define $\bar{\bf U}({\bf O};\beta)$ to be
\begin{align*}%\scriptstyle{
    \left\{\bar{{\bf S}}_1({\bf C})+\bar{{\bf S}}_2({\bf C})A^*\right\}\left\{b(Z)-h_1({\bf C};\beta_1)-h_2({\bf C};\beta_2)A^*  \right\}+\bar{{\bf S}}_2({\bf C})h_2({\bf C};\beta_2)\sigma_{\varepsilon}^{2\dag}.%}
\end{align*}
The following theorem shows that the GMM estimator solving $\sum_{i=1}^n\bar{\bf U}({\bf O}_i;\beta)={\bf 0}$ is consistent and asymptotically normal under certain regularity conditions. 
%\begin{comment}
\begin{theorem}\label{thm:mediator_can}
    Suppose the following all hold: (i) measurement error model (\ref{eqn:me6}) and mediator regression models (\ref{eqn:om7}) are both correctly specified for all $b_k$, (ii) $\beta^{\dag}$ is in the interior of a compact set $B$, (iii) $\bar{\bf U}({\bf O};\beta)$ is continuous at each $\beta\in B$ with probability one, (iv) $\bar{\bf U}({\bf O};\beta)$ is continuously differentiable in a neighborhood $\mathcal{N}$ of $\beta^{\dag}$ with probability approaching one, (v) $E\{\sup_{\beta\in B}\lVert\bar{\bf U}({\bf O};\beta^{\dag})\rVert\}<\infty$,  (v) $E\{\lVert \bar{\bf U}({\bf O};\beta^{\dag})\rVert^2\}<\infty$, (vi) $E\{\sup_{\beta\in\mathcal{N}}\lVert\nabla_{\beta}\bar{\bf U}({\bf O};\beta^{\dag})\rVert\}<\infty$, and (vi) $\bar{{\bf G}}\equiv E\{\nabla_{\beta}\bar{\bf U}({\bf O};\beta^{\dag})\}$ is nonsingular. Let $\hat{\beta}$ be the estimator solving the equations $\sum_{i=1}^n\allowbreak \bar{\bf U}(\allowbreak {\bf O}_i;\allowbreak \beta)\allowbreak =\allowbreak {\bf 0}$. Then $n^{1/2}(\hat{\beta}-\beta^{\dag})\rightsquigarrow \mathcal{N}\{0,\bar{{\bf G}}^{-1}\bar{{\bf \Omega}}(\bar{{\bf G}}^{-1})^T\}$, 
    %$(\bar{{\bf G}}^T\bar{{\bf G}})^{-1}\bar{{\bf G}}^T\bar{{\bf \Omega}}\bar{{\bf G}}(\bar{{\bf G}}^T\bar{{\bf G}})^{-1}$, 
    where $\bar{{\bf \Omega}}\equiv E\{\bar{\bf U}({\bf O}_i;\beta)^{\otimes 2}\}$.
\end{theorem}
%\end{comment}
If the models for $b_k$ are all variationally independent, then their corresponding estimating equations can be stacked and their parameters can be estimated jointly by solving the stacked estimating equations. The sandwich estimator can be used to estimate their asymptotic variance. If the models are not variationally independent, then one must take care to ensure that they are compatible and that shared parameters are appropriately encoded across the models in order for the estimating equation stacking approach to be valid. In the linear model case, this class of estimators contains the classical method-of-moments estimator \citep{fuller2009measurement} (which is equivalent to regression calibration in this setting \citep{carroll2006measurement}) as a special case when $\bar{\bf S}_1({\bf C})+\bar{\bf S}_2({\bf C})A^*=[1,{\bf C}^T,A^*]^T$. %Theorem S4 in Section S3 of the supporting web materials gives the optimal choice of $\bar{{\bf S}}\equiv\{\bar{{\bf S}}_1,\bar{{\bf S}}_2\}$ with respect to asymptotic variance.

The next theorem gives the optimal choice of $\bar{{\bf S}}\equiv\{\bar{{\bf S}}_1,\bar{{\bf S}}_2\}$ with respect to asymptotic variance. 
Let $\hat{\beta}(\bar{{\bf S}})$ be the estimator solving $\sum_{i=1}^n\bar{\bf U}_{\bar{{\bf S}}}({\bf O}_i;\beta)={\bf 0}$ corresponding to the moment functions in Theorem \ref{thm:mediator_can}, and define 
\begin{align*}
\Delta^*(\beta) \equiv &\; b(Z) - h_1({\bf C};\beta_1) - h_2({\bf C};\beta_2)A^*,
\end{align*}
\begin{align*}
{ d({\bf C})\equiv E\left\{\Delta^*(\beta)^2\mid {\bf C}\right\}\left[E\left\{\Delta^*(\beta)^2A^{*2}\mid {\bf C}\right\} - \sigma_{\varepsilon}^{\dag 4}h_2({\bf C};\beta_2)^2\right] - E\left\{\Delta^*(\beta)^2A^*\mid {\bf C}\right\}^2},
\end{align*}
\begin{align*}
\bar{{\bf S}}_1^*({\bf C})\equiv d({\bf C})^{-1}\left[\begin{array}{c}
{\scriptscriptstyle\left[\sigma_{\varepsilon}^{\dag 4}h_2({\bf C};\beta_2)^2 - \mathrm{Cov}\left\{\Delta^*(\beta)^2A^*, A^*\mid {\bf C}\right\}\right]\nabla_{\beta_1}h_1({\bf C};\beta_1)}\\
{\scriptscriptstyle\left(E\left\{\Delta^*(\beta)^2A^*\mid {\bf C}\right\}\left\{E(A^{*2}\mid {\bf C}) + \sigma_{\varepsilon}^{\dag 2}\right\} - \left[E\left\{\Delta^*(\beta)^2A^{*2}\mid {\bf C}\right\} - \sigma_{\varepsilon}^{\dag 4}h_2({\bf C};\beta_2)^2\right]E(A^*\mid {\bf C})\right)\nabla_{\beta_2}h_2({\bf C};\beta_2)^T}
\end{array}\right],
\end{align*}
\begin{align*}
\bar{{\bf S}}^*_2({\bf C})\equiv d({\bf C})^{-1}\left[\begin{array}{c}{\scriptscriptstyle
\mathrm{Cov}\left\{\Delta^*(\beta)^2, A^*\mid {\bf C}\right\}\nabla_{\beta_1}h_1({\bf C};\beta_1)}\\
{\scriptscriptstyle\left[E\left\{\Delta^*(\beta)^2A^*\mid {\bf C}\right\}E(A^*\mid {\bf C}) - E\left\{\Delta^*(\beta)^2\mid {\bf C}\right\}\left\{E(A^{*2}\mid {\bf C}) + \sigma_{\varepsilon}^{\dag 2}\right\}\right]\nabla_{\beta_2}h_2({\bf C};\beta_2)^T}
\end{array}\right].
\end{align*}
\begin{theorem}\label{thm:mediator_efficiency}
    Under the conditions in Theorem \ref{thm:mediator_can}, $\hat{\beta}(\bar{{\bf S}}^*)$ attains the minimum asymptotic variance, $E\left\{\bar{\bf U}_{\bar{{\bf S}}^*}({\bf O};\beta)^{\otimes 2}\right\}$,
    of all estimators in the class of GMM estimators indexed by $\bar{{\bf S}}$ solving $\sum_{i=1}^n\bar{\bf U}_{\bar{{\bf S}}}({\bf O}_i,\beta)={\bf 0}$.
\end{theorem}
The terms $E(A^*\mid {\bf C})$ and $E(A^{*2}\mid {\bf C})$ can be easily modeled. The terms $\Delta^*(\beta)$ and $h_2({\bf C};\beta_2)$ clearly depend on $\beta$. These can be estimated iteratively with these terms in $\bar{{\bf S}}$ updated with each iteration as in weighted least squares. The na\"ive estimate ignoring measurement error can be used as the initial value. Conditional expectations and covariances involving $\Delta^*(\beta)$ can then be modeled. The terms $\nabla_{\beta_1}h_1({\bf C};\beta_1)$ and $\nabla_{\beta_2}h_2({\bf C};\beta_2)$ do not depend on $\beta$ when $h_1$ or $h_2$ are linear in $\beta$, respectively. Otherwise, they can be updated iteratively as with $\Delta^*(\beta)$ and $h_2({\bf C};\beta_2)$. Clearly, these take a complicated form and may not be practical to implement. Homoscedasticity of $b(Z)$ with respect to $A$ and $\bf C$ alone does not help much to simplify these expressions. If we additionally assume no measurement error, $\bar{{\bf S}}^*({\bf C})$ reduces to \[[\bar{{\bf S}}_1^*({\bf C}),\bar{{\bf S}}_2^*({\bf C})] \propto \left[\begin{array}{cc}
    \mathrm{Var}(A^*\mid {\bf C})\nabla_{\beta_1}h_1({\bf C;\beta_1}) & {\bf 0} \\
    {\bf 0} & \nabla_{\beta_2}h_2({\bf C};\beta_2)]
\end{array}\right] ,\]
%\begin{align*}
%    \bar{{\bf S}}_1^*({\bf C})+\bar{{\bf S}}_2^*({\bf C})A^* \propto [\mathrm{Var}(A^*\mid {\bf C})\nabla_{\beta_1}h_1({\bf C;\beta_1})^T,A^*\nabla_{\beta_2}h_2({\bf C};\beta_2)^T]^T,
%\end{align*}
\begin{comment}
Under homoscedasticity in the model for $b(Z)$, we have $\sigma^{*2}_Z \equiv E\{\Delta^*(\beta)^2\mid A^*,{\bf C}\} = E\{\Delta(\beta)^2\} + \sigma_{\varepsilon}^{\dag 2}$, where $\Delta(\beta)\equiv b(Z) - h_1({\bf C};\beta_1) - h_2({\bf C};\beta_2)A$ is the residual of the error-free mediator model (whose variance we are assuming to be constant). In this case, $\bar{{\bf S}}^*$ reduces to
\[\bar{{\bf S}}_1^*({\bf C}) = d^*({\bf C})^{-1}\left[\begin{array}{c}
    \left\{\mathrm{Var}(A^*\mid {\bf C}) - \sigma_{\varepsilon}^{\dag 4}\right\}\nabla_{\beta_1}h_1({\bf C};\beta_1)   \\
      -\left\{\sigma_{\varepsilon}^{\dag 2}h_2({\bf C};\beta_2) + 1\right\}\sigma_{\varepsilon}^{\dag 2}E(A^*\mid {\bf C})\nabla_{\beta_2}h_2({\bf C};\beta_2)
\end{array}\right],\]
\[\bar{{\bf S}}_2^*(c) = d^*({\bf C})^{-1}\left[\begin{array}{c}
    0   \\
      \left\{\mathrm{Var}(A^*\mid {\bf C}) + \sigma_{\varepsilon}^{\dag 2}\right\}\nabla_{\beta_2}h_2({\bf C};\beta_2)
\end{array}\right],\]
where $d^*({\bf C})\equiv E\{\Delta^*(\beta)^2\}\mathrm{Var}(A^*\mid {\bf C}) - \sigma_{\varepsilon}^{\dag 4}h_2({\bf C};\beta_2)^2$, 
\end{comment}
which is more straightforward to implement and closely resembles the method-of-moments estimator in the linear model case. %Substituting zero for $\sigma_{\varepsilon}^{\dag 2}$ yields \[\bar{{\bf S}}_1({\bf C}) + \bar{{\bf S}}_2({\bf C})A^* \propto [\nabla_{\beta_1}h_1({\bf C};\beta_1)^T, \nabla_{\beta_2}h_2({\bf C};\beta_2)^TA^*]^T.\] 
Therefore, if the measurement error variance is not too large, this should serve as a reasonable choice for $\bar{{\bf S}}$ in terms of efficiency. The resulting estimator will still be consistent and asymptotically normal regardless of the size of the measurement error.

When the measurement error variance $\sigma_{\varepsilon}^{\dag 2}$ is unknown (as is typically the case), one can use the estimated variance obtained by the method described in Section 3. In fact, the outcome regression and mediator regression parameters as well as the measurement error variance can all be estimated jointly by stacking all of their corresponding estimating equations. These can be further augmented by the following equation for the NIE:
\begin{align*}
    \sum_{i=1}^n \sum_{k=1}^Ka'^{I(k>k^*)}g_k({\bf C}_i;\theta_k^{\dag})h_{k,2}({\bf C}_i;\beta_{k,2})(a'-a'') - n\psi_{\mathrm{NIE}} = 0,
\end{align*}
where we now make the dependence of the models for functions of the mediator on $k$ explicit. Otherwise, the parameter estimates can be substituted into $n^{-1}\allowbreak \sum_{i=1}^n \allowbreak \sum_{k=1}^K\allowbreak a'^{I(k>k^*)}\allowbreak g_k({\bf C}_i;\theta_k^{\dag})\allowbreak h_{k,2}({\bf C}_i;\beta_{k,2})\allowbreak (a'-a'')$ to estimate the NIE, and its asymptotic variance can be derived using the delta method. Alternatively, inference can be based on the bootstrap. For nonlinear models not satisfying (\ref{eqn:om7}), one can use regression calibration or simulation extrapolation (SIMEX) to estimate the models for $b_k(Z)$.

\section{Proofs}
\label{sec:proofs}
\begin{proof}[Proof of Theorem \ref{thm:can}]
Letting $W_k=\hat{W}_k\equiv I_{\dim(U_k(O;\phi_k))}$ for all $n$ and $k=1,2$, the estimator solving the estimating equation $n^{-1}\sum_{i=1}^nU_k(O_i;\phi)=0$ %based on (3) 
is identical to the GMM estimator
\begin{align*}
    \hat{\phi}=\arg\min_{\phi}\left\{\frac{1}{n}\sum_{i=1}^nU_k(O_i;\phi)\right\}\hat{W}_k\left\{\frac{1}{n}\sum_{i=1}^nU_k(O_i;\phi)\right\}.
\end{align*}
Clearly $U_k(O;\phi)$ is linear in $\phi$ and hence an everywhere continuously differentiable function. Letting $\theta_0^*\equiv \theta_0^{\dag}-\theta_{C_1}^{\dag T}\Gamma_1^{-1}\gamma_0$, $\theta_{C_1}^{* T}\equiv \theta_{C_1}^{\dag T}\Gamma_1^{-1}$, and $\theta_{C_2}^{* T}\equiv \theta_{C_2}^{\dag T}-\theta_{C_1}^{\dag T}\Gamma_1^{-1}\Gamma_2$, under the measurement error model in Assumption 2, we have unbiasedness of $U_1(O;\theta)$ for $\theta^*\equiv (\theta_0^*, \theta_{C_1}^{*T}, \theta_{C_2}^{*T}, \theta_Z^{\dag})^T$:
\begin{align*}
    &   E\left\{S\left(Z,C_2\right)\left(Y-\theta_0^*-\theta_{C_1}^{* T}C_1^*-\theta_{C_2}^{* T}C_2-\theta_Z^{\dag}Z\right)\right\}\\
    =&   E\left(S\left(Z,C_2\right)\left[E\left(Y\mid Z,C\right)-\theta_0^*-\theta_{C_1}^{* T}\left\{\gamma_0+\Gamma_1 E\left(C_1\mid Z,C_2\right)+\Gamma_2 C_2\right\}-\theta_{C_2}^{* T}C_2-\theta_Z^{\dag}Z\right]\right)\\
    =&  E\left(S\left(Z,C_2\right)\left[\theta_0^{\dag}+\theta_{C_1}^{\dag T}C_1+\theta_{C_2}^{\dag T}C_2+\theta_Z^{\dag}Z\right.\right.\\
    &\left.\left.-\theta_0^*-\theta_{C_1}^{* T}\left\{\gamma_0+\Gamma_1 E\left(C_1\mid Z,C_2\right)+\Gamma_2 C_2\right\}-\theta_{C_2}^{* T}C_2-\theta_Z^{\dag}Z\right]\right)\\
    =&  E\left[S\left(Z,C_2\right)\left\{\theta_{C_1}^{\dag T}E\left(C_1\mid Z,C_2\right)-\theta_{C_1}^{* T}\Gamma_1E\left(C_1\mid Z,C_2\right)\right\}\right]\\
    =&  0.
\end{align*}
Under the measurement error model in Assumption 1, $\gamma_0=0_{\dim(C_1)}$, $\Gamma_1=I_{\dim(C_1)}$, and $\Gamma_2=0_{\dim(C_1)\times\dim(C_2)}$ such that $\theta^*=\theta^{\dag}$. In this case, we have unbiasedness of the last term in $U_2(O;\theta,\mu)$ for $\phi_2^*$:% $\theta^*\equiv (\theta_0^*, \theta_{C_1}^{*T}, \theta_{C_2}^{*T}, \theta_Z^{\dag})^T$:
\begin{align*}
    E\left(\theta_0^{\dag} + \theta_{C_1}^{\dag T}C^*_{1,i} + \theta_{C_2}^{\dag T}C_{2,i} - \mu^{\dag}\right)
    = \theta_0^{\dag} + \theta_{C_1}^{\dag T}E\left(C^*_{1}\right) + \theta_{C_2}^{\dag T}E\left(C_{2}\right) - E\left\{Y(0)\right\} = 0
\end{align*}
in addition to the unbiasedness of rest of the terms shown above.

By (b) and (c), we also have
\begin{align*}
    & \; E\left\{\left\lVert U_1\left(O;\theta^*\right)\right\rVert^2\right\}\\
    =& \; E\left\{\lVert S(Z,C_2)\rVert^2\left(Y-\theta_0^*-\theta_{C_1}^{*T}C_1^*-\theta_{C_2}^{*T}C_2-\theta_Z^{\dag}Z\right)^2\right\}\\
    \leq & \; E\left\{\lVert S(Z,C_2)\rVert^4\right\}^{1/2}E\left\{\left(Y-\theta_0^*-\theta_{C_1}^{*T}C_1^*-\theta_{C_2}^{*T}C_2-\theta_Z^{\dag}Z\right)^4\right\}^{1/2}\\
    < & \; \infty
\end{align*}
and
\begin{align*}
    &E\left\{\left\lVert U_2\left(O;\theta^*\right)\right\rVert^2\right\}\\
    =& \; E\left\{\left\lVert U_1\left(O;\theta^*\right)\right\rVert^2\right\}+ E\left\{\left(\theta_0^*+\theta_{C_1}^{*T}C_1^*+\theta_{C_2}^{*T}C_2-\mu\right)^2\right\},
\end{align*}
where we have just shown the first term to be finite. For the second term, we have 
\begin{align*}
    & E\left\{\left(\theta_0^*+\theta_{C_1}^{*T}C_1^*+\theta_{C_2}^{*T}C_2-\mu\right)^2\right\}\\
    =& \; E\left[\left\{E(Y\mid Z=0,C)+\theta_{C_1}^{*T}(C_1^*-C_1)-\mu\right\}^2\right]\\
    =& \;E\left[\left\{E(Y\mid Z=0, C)-\mu\right\}^2\right] + E\left[\left\{\theta_{C_1}^{*T}\left(C_1^*-C_1\right)\right\}^2\right]\\
    & \; +2E\left[\left\{E(Y\mid Z=0,C)-\mu\right\}\theta_{C_1}^{*T}(C_1^*-C_1)\right]\\
    \leq & \; E\left[\left\{E(Y\mid Z=0, C) - \mu\right\}^2\right] + \lVert\theta_{C_1}\rVert^2E\left(\lVert C_1^*-C_1\rVert^2\right)\\
    \leq & \; \mathrm{Var}\left\{E(Y\mid Z=0,C)\right\} + \lVert\theta_{C_1}\rVert^2E\left\{\left(\lVert C_1^*\rVert + \lVert C_1\rVert\right)^2\right\}\\
    < & \; \infty,
\end{align*}
hence $E\left\{\left\lVert U_2\left(O;\theta^*\right)\right\rVert^2\right\}<\infty$. Additionally, for Euclidean norm of matrix $A$ with $j,k$ entry $a_{jk}$ denoted $\lVert A\rVert=(\sum_{j,k}a_{jk}^2)^{1/2}$, we have
\begin{align*}
    &E\left\{\sup_{\phi_1\in \mathcal{N}(\phi_1^*)}\left\lVert \nabla_{\phi_1}U_1\left(O;\phi_1\right)\right\rVert\right\}\\
%    =& E\left(\left\lVert G\right\rVert\right)\\
    =& \; E\left\{\lVert S(Z,C_2)\rVert\left(1+\lVert C_1^{*}\rVert^2+\lVert C_2\rVert^2+Z^2\right)^{1/2}\right\}\\
    \leq & \; E\left\{\lVert S(Z,C_2)\rVert^2\right\}^{1/2}E\left(1+\lVert C_1^{*}\rVert^2+\lVert C_2\rVert^2+Z^2\right)^{1/2}\\
    < & \; \infty,
\end{align*}
and 
\begin{align*}
    &E\left\{\sup_{\phi_2\in \mathcal{N}(\phi_2^*)}\left\lVert \nabla_{\phi_2}U_2\left(O;\phi_2\right)\right\rVert\right\}\\
%    =& E\left(\left\lVert G\right\rVert\right)\\
    =& \; E\left[\left\{\lVert S(Z,C_2)\rVert^2\left(1+\lVert C_1^{*}\rVert^2+\lVert C_2\rVert^2+Z^2\right) + 1 + \lVert C_1^{*}\rVert^2+\lVert C_2\rVert^2 + 1\right\}^{1/2}\right]\\
    \leq& \; E\left(\left[\left\{\lVert S(Z,C_2)\rVert^2+1\right\}\left\{1+\lVert C_1^{*}\rVert^2+\lVert C_2\rVert^2+\max(Z^2, 1)\right\}\right]^{1/2}\right)\\
    \leq & \; E\left[\left\{\lVert S(Z,C_2)\rVert^2+1\right\}^2\right]^{1/2}E\left\{1+\lVert C_1^{*}\rVert^2+\lVert C_2\rVert^2+\max(Z^2,1)\right\}^{1/2}\\
    < & \; \infty.
\end{align*}
$G_k^TW_kG_k=G_k^TG_k$ is nonsingular for $k=1,2$ since $G_1$ is nonsingular. Thus, by Theorem 3.4 in \cite{newey1994large}, for $k=1,2$,
\begin{align*}
\sqrt{n}\left(\hat{\phi}_k-\phi_k^*\right)&\rightsquigarrow N\left\{0,(G_k^TG_k)^{-1}G_k^T\Omega G_k(G_k^TG_k)^{-1}\right\}\\
&= N\left\{0,G_k^{-1}\Omega_k (G_k^{-1})^T\right\}.
\end{align*}
\end{proof}

\begin{proof}[Proof of Theorem \ref{thm:efficiency-linear}]
Let %$\tau\equiv (W,S)$, 
\[m(O;W,S)\equiv E\{S(Z,C_2)(1,C_1^{*T},C_2^T,Z)\}^TWS(Z,C_2)(Y-\theta_0^*-\theta_{C_1}^{*T}C_1^*-\theta_2^{*T}C_2-\theta_Z^TZ),\]
\[D(W,S)\equiv E\{S(Z,C_2)(1,C_1^{*T},C_2^T,Z)\}^TWE\{S(Z,C_2)(1,C_1^{*T},C_2^T,Z)\}\]
such that $D(W,S)^{-1}E\{m(O;W,S)m(O;W,S)^T\}\{D(W,S)^{-1}\}^T$ equals the asymptotic variance of the GMM estimator
\begin{align*}
    \hat{\theta}=\arg\min_{\theta}\left\{\frac{1}{n}\sum_{i=1}^nU_S(O_i;\theta)\right\}\hat{W}\left\{\frac{1}{n}\sum_{i=1}^nU_S(O_i;\theta)\right\}
\end{align*}
with $\hat{W}\xrightarrow{p} W$. For all $S$ and $W$, \[S^*(Z,C_2)\equiv \sigma^{-2}(Z,C_2)[1, E(C_1^*\mid Z,C_2)^T, C_2^T, Z]^T\] and \[W^*\equiv E\{\sigma^{-2}(Z,C_2)[1, E(C_1^*\mid Z,C_2)^T, C_2^T, Z]^T[1, E(C_1^*\mid Z,C_2)^T, C_2^T, Z]\}^{-1}\] satisfy
\begin{align*}
&    E\{m(O;W,S)m(O;W^*,S^*)^T\}\\
=&  E\{S(Z,C_2)(1,C_1^{*T},C_2^T,Z)\}^TW\\
&\times E\{S(Z,C_2)(Y-\theta_0^*-\theta_{C_1}^{*T}C_1^*-\theta_2^{*T}C_2-\theta_Z^{\dag}Z)^2\sigma^{-2}(Z,C_2)[1,E(C_1^*\mid Z,C_2)^T,C_2^T,Z]\}\\
&\times E\{\sigma^{-2}(Z,C_2)[1,E(C_1^*\mid Z,C_2)^T,C_2^T,Z]^T[1,E(C_1^*\mid Z,C_2)^T,C_2^T,Z]\}^{-1}\\
&\times E\{\sigma^{-2}(Z,C_2)[1,E(C_1^*\mid Z,C_2)^T,C_2^T,Z]^T[1,C_1^{*T},C_2^T,Z]^T\}\\
=&  E\{S(Z,C_2)(1,C_1^{*T},C_2^T,Z)\}^TWE\{S(Z,C_2)(1,C_1^{*T},C_2^T,Z)\}\\
=&  D(W,S).
\end{align*}
Since $n^{-1}\sum_{i=1}^nU_{S^*}(O_i;\theta)$ is a just-identified estimating equation, the above GMM estimator is equivalent for all sequences of positive definite matrices $\hat{W}$. Thus, the estimator solving $n^{-1}\sum_{i=1}^nU_{S^*}(O_i;\theta)$, which is equivalent to the above GMM estimator with $S=S^*$ and $\hat{W}=I$ for all $n$, has asymptotic variance equal to \[D(W^*,S^*)^{-1}E\{m(O;W^*,S^*)m(O;W^*,S^*)^T\}\{D(W^*,S^*)^{-1}\}^T\]
(which is also straightforward to confirm). Thus, by Theorem 5.3 in \cite{newey1994large}, the estimator solving
\[
\frac{1}{n}\sum_{i=1}^nS^*(Z,C_2)(Y_i - \theta_0 - \theta_{C_1}^T{\bf C}^*_{1,i} - \theta_{C_2}^TC_{2,i} - \theta_Z Z_i) = 0
\]
is efficient with respect to asymptotic variance in the class of estimators indexed by $S$.
\end{proof}

\begin{proof}[Proof of Theorem \ref{thm:sigma_can}]
Letting $W=\hat{W}\equiv 1$ for all $n$, the estimator solving the estimating equation $\sum_{i=1}^n\dot{U}(O_i;\sigma_{\varepsilon}^2)=0$ is identical to the GMM estimator \[\hat{\sigma}^2_{\varepsilon}\equiv\argmin_{\sigma^2_{\varepsilon}\in [\delta_{\ell},\delta_u]}\left\{\frac1n\sum_{i=1}^n\dot{U}\left(O_i;\sigma^2_{\varepsilon}\right)\right\}^2.\]
Under outcome model (\ref{eqn:om4}) and the measurement error model in Assumption 4, we have unbiasedness of $\dot{U}(O;\theta)$ for $\sigma^{2\dag}_{\varepsilon}$:
\begin{align*}
&\; E\left(T(Z,C_{2})\left[C_{1}^{*}\left\{Y-g_1(Z,C_{2};\theta_1^{\dag})-g_2(Z,C_{2};\theta_2^{\dag})C_{1}^*\right\}+g_2(Z,C_{2};\theta_2^{\dag})\sigma^{2\dag}_{\varepsilon}\right]\right)\\
=&\; E\left[T(Z,C_{2})C_{1}\left\{Y-g_1(Z,C_{2};\theta_1^{\dag})-g_2(Z,C_{2};\theta_2^{\dag})C_{1}\right\}\right]\\
&\;- E\left\{T(Z,C_{2})C_{1}g_2(Z,C_{2};\theta_2^{\dag})\varepsilon\right\}\\
&\; +E\left[T(Z,C_{2})\varepsilon\left\{Y-g_1(Z,C_{2};\theta_1^{\dag})-g_2(Z,C_{2};\theta_2^{\dag})C_{1}\right\}\right]\\
&\; -E\left\{T(Z,C_{2})g_2(Z,C_{2};\theta_2^{\dag})\varepsilon^2\right\}\\
&\; +E\left\{T(Z,C_{2})g_2(Z,C_{2};\theta_2^{\dag})\right\}\sigma^{2\dag}_{\varepsilon}\\
=&\; E\left[T(Z,C_{2})C_{1}\left\{E(Y\mid Z,C)-g_1(Z,C_{2};\theta_1^{\dag})-g_2(Z,C_{2};\theta_2^{\dag})C_{1}\right\}\right]\\
&\;- E\left\{T(Z,C_{2})C_{1}g_2(Z,C_{2};\theta_2^{\dag})E(\varepsilon\mid Z,C)\right\}\\
&\; +E\left[T(Z,C_{2})E(\varepsilon\mid Z,C,Y)\left\{Y-g_1(Z,C_{2};\theta_1^{\dag})-g_2(Z,C_{2};\theta_2^{\dag})C_{1}\right\}\right]\\
&\; -E\left\{T(Z,C_{2})g_2(Z,C_{2};\theta_2^{\dag})E\left(\varepsilon^2\mid Z,C_2\right)\right\}\\
&\; +E\left\{T(Z,C_{2})g_2(Z,C_{2};\theta_2^{\dag})\right\}\sigma^{2\dag}_{\varepsilon}\\
=&\; \dot{G}\left\{\sigma^{2\dag}_{\varepsilon}-E\left(\varepsilon^2\right)\right\}\\
%-E\left\{T(Z,C_{2})g_2(Z,C_{2};\theta_2^{\dag})\right\}E\left(\varepsilon^2\right)+E\left\{T(Z,C_{2})g_2(Z,C_{2};\theta_2^{\dag})\right\}\sigma^{2\dag}_{\varepsilon}\\
=&\; 0.
%=&\; \mathrm{vec}\Big\{E\left\{T(Z,C_{2})C_{1}^{*T}E(Y\mid Z,C_1^*,C_2)\right\}\\
%&\; \left.-E\left(T(Z,C_{2})\left[C_{1}^{*T}\left\{g_1(Z,C_{2};\theta_1^{\dag})+g_2(Z,C_{2};\theta_2^{\dag})^TC_{1}^*\right\}+g_2(Z,C_{2};\theta_2^{\dag})^TLL^T\right]\right)\right\}\\
\end{align*}
Since $\dot{U}$ is linear in $\sigma_{\varepsilon}^2$, $\sigma_{\varepsilon}^{2\dag}$ is the unique solution provided $E\{T(Z,C_{2})g_2(Z,C_{2};\theta_2^{\dag})\}\neq 0$. By (b), $\sigma_{\varepsilon}^{2\dag}$ is in the interior of a compact subset of the parameter space. $\dot{U}(O;\sigma^2_{\varepsilon})$ is clearly everywhere continuously differentiable in $\sigma^2_{\varepsilon}$. By the above derivation, we have
\begin{align*}
    E\left\{\sup_{\sigma_{\varepsilon}^{2}\in(\delta_{\ell},\delta_u)}\lVert\dot{U}(O;\sigma_{\varepsilon}^{2})\rVert\right\} =& \; \sup_{\sigma_{\varepsilon}^{2}\in(\delta_{\ell},\delta_u)}\lvert\dot{G}(\sigma_{\varepsilon}^{2}-\sigma_{\varepsilon}^{2\dag})\rvert\leq\lvert\dot{G}\rvert(\delta_u-\delta_{\ell})<\infty.
\end{align*}
$\nabla_{\sigma_{\varepsilon}^2}\dot{U}(O;\sigma_{\varepsilon}^2)=T(Z,C_{2})g_2(Z,C_{2};\theta_2^{\dag})$, which does not depend on $\sigma_{\varepsilon}^2$, so \[E\left\{\sup_{\sigma_{\varepsilon}^2\in(\delta_{\ell},\delta_u)}\lVert\nabla_{\sigma_{\varepsilon}^2}\dot{U}(O;\sigma_{\varepsilon}^2)\rVert\right\}=E\lvert T(Z,C_{2})g_2(Z,C_{2};\theta_2^{\dag})\rvert<\infty.\]
Since $\dot{G}\neq 0$ and $W=1$, $\dot{G}^TW\dot{G}$ is nonsingular. Thus, by Theorem 3.4 of \cite{newey1994large}, 
\[\sqrt{n}(\hat{\sigma}_{\varepsilon}^2-\sigma_{\varepsilon}^{2\dag})\rightsquigarrow N(0,\dot{\Omega}/\dot{G}^2)\]
since $(\dot{G}^TW\dot{G})^{-1}\dot{G}^TW\dot{\Omega}W\dot{G}(\dot{G}^TW\dot{G})^{-1}=\dot{\Omega}/\dot{G}^2$.

\end{proof}

\begin{proof}[Proof of Theorem \ref{thm:can_c1}]
Letting $W_k=\hat{W}_k\equiv I_{\dim(U_k(O;\phi_k))}$ for all $n$ and $k=1,2$, the estimator solving the estimating equation $n^{-1}\sum_{i=1}^n\check{\bf U}_k(O_i;\phi)=0$ %based on (3) 
is identical to the GMM estimator
\begin{align*}
    \hat{\phi}=\arg\min_{\phi}\left\{\frac{1}{n}\sum_{i=1}^n\check{\bf U}_k(O_i;\phi)\right\}\hat{W}_k\left\{\frac{1}{n}\sum_{i=1}^n\check{\bf U}_k(O_i;\phi)\right\}.
\end{align*}
Clearly $\check{\bf U}_k(O;\phi)$ is linear in $\phi$ and hence an everywhere continuously differentiable function. Under the measurement error model in Assumption 4, we have unbiasedness of $\check{\bf U}_1(O;\theta)$ for $\theta^*\equiv (\theta_0^{\dag}, \theta_{C_1}^{\dag}, \theta_{C_2}^{\dag T}, \theta_Z^{\dag})^T$:
\begin{align*}
    &\;   E\left[\left\{S_1(Z,C_2)C^*_1+S_2(Z,C_2)\right\}\left(Y-\theta_0^{\dag}-\theta_{C_1}^{\dag}C_1^*-\theta_{C_2}^{\dag T}C_2-\theta_Z^{\dag}Z\right)\right]\\
    =&\;   E\left[\left\{S_1(Z,C_2)C_1+S_2(Z,C_2)\right\}\left(Y-\theta_0^{\dag}-\theta_{C_1}^{\dag}C_1-\theta_{C_2}^{\dag T}C_2-\theta_Z^{\dag}Z\right)\right]\\
    &-   \theta_{C_1}^{\dag}E\left[\left\{S_1(Z,C_2)C_1+S_2(Z,C_2)\right\}\varepsilon\right]\\
    &+   E\left\{S_1(Z,C_2)\varepsilon(Y-\theta_0^{\dag}-\theta_{C_1}^{\dag}C_1-\theta_{C_2}^{\dag T}C_2-\theta_Z^{\dag}Z)\right\}\\
    &-   \theta_{C_1}^{\dag}E\left\{S_1(Z,C_2)\varepsilon^2\right\}\\
    =&\;   E\left[\left\{S_1(Z,C_2)C_1+S_2(Z,C_2)\right\}\left\{E\left(Y\mid C,Z\right)-\theta_0^{\dag}-\theta_{C_1}^{\dag}C_1-\theta_{C_2}^{\dag T}C_2-\theta_Z^{\dag}Z\right\}\right]\\
    &-   \theta_{C_1}^{\dag}E\left[\left\{S_1(Z,C_2)C_1+S_2(Z,C_2)\right\}E\left(\varepsilon\mid C,Z\right)\right]\\
    &+   E\left[S_1(Z,C_2)\varepsilon\left\{E\left(Y\mid C,Z,\varepsilon\right)-\theta_0^{\dag}-\theta_{C_1}^{\dag}C_1-\theta_{C_2}^{\dag T}C_2-\theta_Z^{\dag}Z\right\}\right]\\
    &-   \theta_{C_1}^{\dag}E\left\{S_1(Z,C_2)E\left(\varepsilon^2\mid C,Z\right)\right\}\\
    =&\;   E\left[S_1(Z,C_2)\varepsilon\left\{E\left(Y\mid C,Z\right)-\theta_0^{\dag}-\theta_{C_1}^{\dag}C_1-\theta_{C_2}^{\dag T}C_2-\theta_Z^{\dag}Z\right\}\right]\\
    &-   \theta_{C_1}^{\dag}E\left(\varepsilon^2\right)E\left\{S_1(Z,C_2)\right\}\\
    =&\; 0.
\end{align*}
We additionally have unbiasedness of the last term in $\check{\bf U}_2(O;\theta,\mu)$ for $\phi_2^*$:% $\theta^*\equiv (\theta_0^*, \theta_{C_1}^{*T}, \theta_{C_2}^{*T}, \theta_Z^{\dag})^T$:
\begin{align*}
    E\left(\theta_0^{\dag} + \theta_{C_1}^{\dag}{\bf C}^*_{1,i} + \theta_{C_2}^{\dag T}C_{2,i} - \mu^{\dag}\right)
    = \theta_0^{\dag} + \theta_{C_1}^{\dag}E\left(C^*_{1}\right) + \theta_{C_2}^{\dag T}E\left(C_{2}\right) - E\left\{Y(0)\right\} = 0
\end{align*}
in addition to the unbiasedness of rest of the terms shown above.

By (b) and (c), we also have
\begin{align*}
    &\; E\left\{\left\lVert \check{\bf U}_1\left(O;\theta^{\dag}\right)\right\rVert^2\right\}\\
    =& \; E\left\{\lVert S_1(Z,C_2)C^*_1+S_2(Z,C_2)\rVert^2\left(Y-\theta_0^{\dag}-\theta_{C_1}^{\dag}C_1^*-\theta_{C_2}^{\dag T}C_2-\theta_Z^{\dag}Z\right)^2\right\}\\
    \leq & \; E\left\{\lVert S_1(Z,C_2)C^*_1+S_2(Z,C_2)\rVert^4\right\}^{1/2}E\left\{\left(Y-\theta_0^{\dag}-\theta_{C_1}^{\dag}C_1^*-\theta_{C_2}^{\dag T}C_2-\theta_Z^{\dag}Z\right)^4\right\}^{1/2}\\
    < & \; \infty
\end{align*}
and
\begin{align*}
    &\; E\left\{\left\lVert \check{\bf U}_2\left(O;\theta^{\dag}\right)\right\rVert^2\right\}\\
    =& \; E\left\{\left\lVert \check{\bf U}_1\left(O;\theta^{\dag}\right)\right\rVert^2\right\}+ E\left\{\left(\theta_0^{\dag}+\theta_{C_1}^{\dag}C_1^*+\theta_{C_2}^{\dag T}C_2-\mu^{\dag}\right)^2\right\},
\end{align*}
where we have just shown the first term to be finite. Finiteness of the second term follows exactly as in the proof of Theorem 1, hence $E\left\{\left\lVert U_2\left(O;\theta^*\right)\right\rVert^2\right\}<\infty$. Additionally, for Euclidean norm of matrix $A$ with $j,k$ entry $a_{jk}$ denoted $\lVert A\rVert=(\sum_{j,k}a_{jk}^2)^{1/2}$, we have
\begin{align*}
    &\; E\left\{\sup_{\phi_1\in \mathcal{N}(\phi_1^*)}\left\lVert \nabla_{\phi_1}\check{\bf U}_1\left(O;\phi_1\right)\right\rVert\right\}\\
%    =& E\left(\left\lVert G\right\rVert\right)\\
    =& \; E\left\{\lVert S_1(Z,C_2)C^*_1+S_2(Z,C_2)\rVert\left(1+ C_1^{*2}+\lVert C_2\rVert^2+Z^2\right)^{1/2}\right\}\\
    \leq & \; E\left\{\lVert S_1(Z,C_2)C^*_1+S_2(Z,C_2)\rVert^2\right\}^{1/2}E\left(1+ C_1^{*2}+\lVert C_2\rVert^2+Z^2\right)^{1/2}\\
    < & \; \infty,
\end{align*}
and 
\begin{align*}
    &\; E\left\{\sup_{\phi_2\in \mathcal{N}(\phi_2^*)}\left\lVert \nabla_{\phi_2}\check{\bf U}_2\left(O;\phi_2\right)\right\rVert\right\}\\
%    =& E\left(\left\lVert G\right\rVert\right)\\
    =& \; E\left[\left\{\lVert S_1(Z,C_2)C^*_1+S_2(Z,C_2)\rVert^2\left(1+ C_1^{*2}+\lVert C_2\rVert^2+Z^2\right) + 1 + C_1^{*2}+\lVert C_2\rVert^2 + 1\right\}^{1/2}\right]\\
    \leq& \; E\left(\left[\left\{\lVert S_1(Z,C_2)C^*_1+S_2(Z,C_2)\rVert^2+1\right\}\left\{1+C_1^{*2}+\lVert C_2\rVert^2+\max(Z^2, 1)\right\}\right]^{1/2}\right)\\
    \leq & \; E\left[\left\{\lVert S_1(Z,C_2)C^*_1+S_2(Z,C_2)\rVert^2+1\right\}^2\right]^{1/2}E\left\{1+C_1^{*2}+\lVert C_2\rVert^2+\max(Z^2,1)\right\}^{1/2}\\
    < & \; \infty.
\end{align*}
$\check{G}_k^TW_k\check{G}_k=\check{G}_k^T\check{G}_k$ is nonsingular for $k=1,2$ since $G_1$ is nonsingular. Thus, by Theorem 3.4 in \cite{newey1994large}, for $k=1,2$,
\begin{align*}
\sqrt{n}\left(\check{\phi}_k-\phi_k^*\right)&\rightsquigarrow N\left\{0,(\check{G}_k^T\check{G}_k)^{-1}\check{G}_k^T\check{\Omega} \check{G}_k(\check{G}_k^T\check{G}_k)^{-1}\right\}\\
&= N\left\{0,\check{G}_k^{-1}\check{\Omega}_k (\check{G}_k^{-1})^T\right\}.
\end{align*}
\end{proof}

\begin{proof}[Proof of Theorem \ref{thm:efficiency-nonlinear}]
Let $S\equiv \{S_1,S_2\}$, %$\tau\equiv (W,S)$, 
\begin{align*}
m(O;W,S)\equiv& \; %E\left[\left\{S_1(Z,C_2)C^*_1+S_2(Z,C_2)\right\}(1,C_1^{*},C_2^T,Z)\right]^TW\\
%& \; \times
\left\{S_1(Z,C_2)C^*_1+S_2(Z,C_2)\right\}(Y-\theta_0^{\dag}-\theta_{C_1}^{\dag}C_1^*-\theta_2^{\dag T}C_2-\theta_Z^{\dag}Z),\\
D(W,S)\equiv& \; %E\left[\left\{S_1(Z,C_2)C^*_1+S_2(Z,C_2)\right\}(1,C_1^{*},C_2^T,Z)\right]^TW\\
%& \; \times 
E\left[\left\{S_1(Z,C_2)C^*_1+S_2(Z,C_2)\right\}(1,C_1^{*},C_2^T,Z)\right]
\end{align*}
such that $D(W,S)^{-1}E\{m(O;W,S)m(O;W,S)^T\}\{D(W,S)^{-1}\}^T$ equals the asymptotic variance of the GMM estimator
\begin{align*}
    \hat{\theta}=\arg\min_{\theta}\left\{\frac{1}{n}\sum_{i=1}^n\check{\bf U}_S(O_i;\theta)\right\}^T\left\{\frac{1}{n}\sum_{i=1}^n\check{\bf U}_S(O_i;\theta)\right\}.
\end{align*}
The optimal $S^*$ in the class of $S$ functions satisfies the following equation for all $S$:
\begin{comment}
\begin{align*}
    & \; -E\left\{\frac{\partial}{\partial\theta}\check{\bf U}_S(O;\theta^{\dag})\right\} \\
    =& \; E\left\{h_1(Z,C_2)E\left(C_1^*[1,C_1^*,C_2^T,Z]\mid Z,C_2\right) + h_2(Z,C_2)[1,E(C_1^*\mid Z,C_2),C_2^T,Z]\right\}
\end{align*}
\end{comment}
%and
\begin{align*}
    & \; E\left\{\check{\bf U}_S(O;\theta^{\dag})\check{\bf U}_{S^*}(O;\theta^{\dag})^T\right\}\\
    =& \; E\left(E\left[\left\{S_1(Z,C_2)C_1^*+S_2(Z,C_2)\right\}C_1^*\sigma^{*2}(Z,C_1^*,C_2)\mid Z,C_2\right]S_1^*(Z,C_2)\right.\\
    &\; + \left.E\left[\left\{S_1(Z,C_2)C_1^*+S_2(Z,C_2)\right\}\sigma^{*2}(Z,C_1^*,C_2)\mid Z,C_2\right]S_2^*(Z,C_2)\right)\\
    =& \; E\left(E\left[\left\{S_1(Z,C_2)C_1^*+S_2(Z,C_2)\right\}[1,C_1^*,C_2^T,Z]\mid Z,C_2\right]\right)\\
    =& \; -E\left\{\frac{\partial}{\partial\theta}\check{\bf U}_S(O;\theta^{\dag})\right\}.
\end{align*}
Thus, for all $S$ such that $E\{S_1(Z,C_2)\}=0$, $S^*$ satisfies
\begin{align*}
    0 =& \; E\left\{S_1(Z,C_2)E\left(C_1^*\left[[1,C_1^*,C_2^T,Z]\right.\right.\right.\\
    & \; - \left.\left.\left.\left\{S_1^*(Z,C_2)C_1^*+S_2(Z,C_2)\right\}^T\sigma^{*2}(Z,C_1^*,C_2)\right]\mid Z,C_2\right)\right\}\\
    & \; + E\left(S_2(Z,C_2)E\left[[1,C_1^*,C_2^T,Z]\right.\right.\\
    & \; - \left.\left.\left\{S_1^*(Z,C_2)C_1^*+S_2^*(Z,C_2)\right\}^T\sigma^{*2}(Z,C_1^*,C_2)\mid Z,C_2\right]\right)
\end{align*}
    $\iff$
\begin{align*}
    0 =& \; E\left\{S_1(Z,C_2)E\left(C_1^*\left[[1,C_1^*,C_2^T,Z]\right.\right.\right.\\
    & \; - \left.\left.\left.\left\{S_1^*(Z,C_2)C_1^*+S_2(Z,C_2)\right\}^T\sigma^{*2}(Z,C_1^*,C_2)\right]\mid Z,C_2\right)\right\}\\
    0 =& \; E\left(S_2(Z,C_2)E\left[[1,C_1^*,C_2^T,Z]\right.\right.\\
    & \; - \left.\left.\left\{S_1^*(Z,C_2)C_1^*+S_2^*(Z,C_2)\right\}^T\sigma^{*2}(Z,C_1^*,C_2)\mid Z,C_2\right]\right)
\end{align*}
    $\iff$
\begin{align*}
    K^T =& \; E\left(C_1^*\left[[1,C_1^*,C_2^T,Z] - \left\{S_1^*(Z,C_2)C_1^*+S_2(Z,C_2)\right\}^T\sigma^{*2}(Z,C_1^*,C_2)\right]\mid Z,C_2\right)\\
    0 =& \; E\left[[1,C_1^*,C_2^T,Z] - \left\{S_1^*(Z,C_2)C_1^*+S_2^*(Z,C_2)\right\}^T\sigma^{*2}(Z,C_1^*,C_2)\mid Z,C_2\right]\\
    &\text{for some constant vector $K$ (since $S_1$ is constrained to have mean zero)}
\end{align*}
    $\iff$
\begin{align*}
    & \; E\left(C_1^*[1,C_1^*,C_2^T,Z]\mid Z,C_2\right) - K^T\\ 
    =& \; E\left\{C_1^{*2}\sigma^{*2}(Z,C_1^*,C_2)\mid Z,C_2\right\}S_1^*(Z,C_2)^T + E\left\{C_1^{*}\sigma^{*2}(Z,C_1^*,C_2)\mid Z,C_2\right\}S_2^*(Z,C_2)^T,\\
    & \; [1,E(C_1^*\mid Z,C_2),C_2^T,Z] \\
    =& \; E\left\{C_1^*\sigma^{*2}(Z,C_1^*,C_2)\mid Z,C_2\right\}S_1^*(Z,C_2)^T+E\left\{\sigma^{*2}(Z,C_1^*,C_2)\mid Z,C_2\right\}S_2^*(Z,C_2)^T\\
\end{align*}
    $\iff$
\begin{align*}
    & \left[\begin{array}{cc}
        E\left\{C_1^{*2}\sigma^{*2}(Z,C_1^*,C_2)\mid Z,C_2\right\} & E\left\{C_1^{*}\sigma^{*2}(Z,C_1^*,C_2)\mid Z,C_2\right\} \\
        E\left\{C_1^*\sigma^{*2}(Z,C_1^*,C_2)\mid Z,C_2\right\} & E\left\{\sigma^{*2}(Z,C_1^*,C_2)\mid Z,C_2\right\}
    \end{array}\right]
    \left[\begin{array}{c}
        S_1^*(Z,C_2)^T \\
        S_2^*(Z,C_2)^T
    \end{array}\right]\\
    =& 
    \left[\begin{array}{c}
        E\left(C_1^*[1,C_1^*,C_2^T,Z]\mid Z,C_2\right) - K^T \\
        \left[1,E(C_1^*\mid Z,C_2),C_2^T,Z\right]
    \end{array}\right]
\end{align*}
    $\iff$
\begin{align*}
    & \left[\begin{array}{c}
        S_1^*(Z,C_2)^T \\
        S_2^*(Z,C_2)^T
    \end{array}\right]\\
    =& 
    \left[\begin{array}{cc}
        E\left\{C_1^{*2}\sigma^{*2}(Z,C_1^*,C_2)\mid Z,C_2\right\} & E\left\{C_1^{*}\sigma^{*2}(Z,C_1^*,C_2)\mid Z,C_2\right\} \\
        E\left\{C_1^*\sigma^{*2}(Z,C_1^*,C_2)\mid Z,C_2\right\} & E\left\{\sigma^{*2}(Z,C_1^*,C_2)\mid Z,C_2\right\}
    \end{array}\right]^{-1}\\
    & \; \times
    \left\{\left[\begin{array}{cccc}
        E\left(C_1^*\mid Z,C_2\right) & E\left(C_1^{*2}\mid Z,C_2\right) & E\left(C_1^*\mid Z,C_2\right)C_2^T & E\left(C_1^*\mid Z,C_2\right)Z \\
        %E\left(C_1^*[1,C_1^*,C_2^T,Z]\mid Z,C_2\right) - K \\
        %\left[1,E(C_1^*\mid Z,C_2),C_2^T,Z\right] \\
        1 & E(C_1^*\mid Z,C_2) & C_2^T & Z
    \end{array}\right]
    - \left[\begin{array}{c}
        K^T \\
        0^T
    \end{array}\right]\right\}
    \end{align*}
    $\iff$
\begin{align*}
    & \; S_1^*(Z,C_2)\\
    =& \; \left[E\left\{C_1^{*2}\sigma^{*2}(Z,C_1^*,C_2)\mid Z,C_2\right\}\sigma^{2}(Z,C_2)\right.\\ 
    & \; \left.- E\left\{C_1^{*}\sigma^{*2}(Z,C_1^*,C_2)\mid Z,C_2\right\}E\left\{C_1^*\sigma^{*2}(Z,C_1^*,C_2)\mid Z,C_2\right\}\right]^{-1}\\
    & \; \times \left[\left[\begin{array}{c}
        E(C_1^*\mid Z,C_2)\sigma^{2}(Z,C_2) - E\left\{C_1^*\sigma^{*2}(Z,C_1^*,C_2)\mid Z,C_2\right\}\\
        E(C_1^{*2}\mid Z,C_2)\sigma^{2}(Z,C_2) - E(C_1^*\mid Z,C_2)E\left\{C_1^*\sigma^{*2}(Z,C_1^*,C_2)\mid Z,C_2\right\}\\
        C_2 E(C_1^{*}\mid Z,C_2)\sigma^{2}(Z,C_2) - C_2 E\left\{C_1^*\sigma^{*2}(Z,C_1^*,C_2)\mid Z,C_2\right\}\\
        Z E(C_1^{*}\mid Z,C_2)\sigma^{2}(Z,C_2) - Z E\left\{C_1^*\sigma^{*2}(Z,C_1^*,C_2)\mid Z,C_2\right\}\\
    \end{array}\right]\right.\\
    & \; \left.\begin{array}{c}
        \mathrm{ } \\
        \mathrm{ } \\
        \mathrm{ } \\
        \mathrm{ }
    \end{array} - K\sigma^2(Z,C_2)\right]\\
    =& \; \left[\mathrm{Var}\left\{C_1^{*}\sigma^{*2}(Z,C_1^*,C_2)\mid Z,C_2\right\} - \mathrm{Cov}\left\{C_1^{*2}\sigma^{*2}(Z,C_1^*,C_2),\sigma^{*2}(Z,C_1^*,C_2)\mid Z,C_2\right\}\right]^{-1}\\
    & \; \times \left[\left[\begin{array}{c}
        -\mathrm{Cov}\left\{C_1^*,\sigma^{*2}(Z,C_1^*,C_2)\mid Z,C_2\right\}\\
        \mathrm{Cov}\left\{C_1^{*},C_1^*\sigma^{*2}(Z,C_1^*,C_2)\mid Z,C_2\right\} - \mathrm{Cov}\left\{C_1^{*2},\sigma^{*2}(Z,C_1^*,C_2)\mid Z,C_2\right\}\\
        -C_2 \mathrm{Cov}\left\{C_1^{*},\sigma^{*2}(Z,C_1^*,C_2)\mid Z,C_2\right\}\\
        -Z \mathrm{Cov}\left\{C_1^{*},\sigma^{*2}(Z,C_1^*,C_2)\mid Z,C_2\right\}
    \end{array}\right]\right.\\
    & \; \left.\begin{array}{c}
        \mathrm{ } \\
        \mathrm{ } \\
        \mathrm{ } \\
        \mathrm{ }
    \end{array} - K\sigma^2(Z,C_2)\right],
\end{align*}
\begin{align*}
    & \; S_2^*(Z,C_2)\\
    =& \; \left[E\left\{C_1^{*2}\sigma^{*2}(Z,C_1^*,C_2)\mid Z,C_2\right\}\sigma^{2}(Z,C_2)\right.\\ 
    & \; \left.- E\left\{C_1^{*}\sigma^{*2}(Z,C_1^*,C_2)\mid Z,C_2\right\}E\left\{C_1^*\sigma^{*2}(Z,C_1^*,C_2)\mid Z,C_2\right\}\right]^{-1}\\
    & \; \times \left[\begin{array}{c}
        -E(C_1^*\mid Z,C_2)\sigma^{2}(Z,C_2) + E\left\{C_1^{*2}\sigma^{*2}(Z,C_1^*,C_2)\mid Z,C_2\right\}\\
        -E(C_1^{*2}\mid Z,C_2)E\left\{C_1^*\sigma^{*2}(Z,C_1^*,C_2)\mid Z,C_2\right\}\\
        \quad\quad + E(C_1^{*}\mid Z,C_2)E\left\{C_1^{*2}\sigma^{*2}(Z,C_1^*,C_2)\mid Z,C_2\right\}\\
        -C_2 E(C_1^{*}\mid Z,C_2)E\left\{C_1^*\sigma^{*2}(Z,C_1^*,C_2)\mid Z,C_2\right\} + C_2 E\left\{C_1^*\sigma^{*2}(Z,C_1^*,C_2)\mid Z,C_2\right\}\\
        -Z E(C_1^{*}\mid Z,C_2)E\left\{C_1^*\sigma^{*2}(Z,C_1^*,C_2)\mid Z,C_2\right\} + Z E\left\{C_1^*\sigma^{*2}(Z,C_1^*,C_2)\mid Z,C_2\right\}\\
    \end{array}\right]\\
    =& \; \left[\mathrm{Var}\left\{C_1^{*}\sigma^{*2}(Z,C_1^*,C_2)\mid Z,C_2\right\} - \mathrm{Cov}\left\{C_1^{*2}\sigma^{*2}(Z,C_1^*,C_2),\sigma^{*2}(Z,C_1^*,C_2)\mid Z,C_2\right\}\right]^{-1}\\
    & \; \times \left[\begin{array}{c}
        \mathrm{Cov}\left\{C_1^*,C_1^*\sigma^{*2}(Z,C_1^*,C_2)\mid Z,C_2\right\}\\
        -\mathrm{Cov}\left\{C_1^{*},C_1^*\sigma^{*2}(Z,C_1^*,C_2)\mid Z,C_2\right\} + \mathrm{Cov}\left\{C_1^{*2},C_1^*\sigma^{*2}(Z,C_1^*,C_2)\mid Z,C_2\right\}\\
        C_2 \mathrm{Cov}\left\{C_1^{*},C_1^*\sigma^{*2}(Z,C_1^*,C_2)\mid Z,C_2\right\}\\
        Z \mathrm{Cov}\left\{C_1^{*},C_1^*\sigma^{*2}(Z,C_1^*,C_2)\mid Z,C_2\right\}
    \end{array}\right],
\end{align*}
where the constant vector $K$ is chosen to satisfy the constraint $E\{S_1^*(Z,C_2)\}=0$. If $\sigma^{*2}(Z,C_1^*,C_2)$ is mean independent of $C_1^*$ given $Z$ and $C_2$, then \[E\left\{\sigma^{*2}(Z,C_1^*,C_2)\mid Z,C_2\right\}=\sigma^2(Z,C_2),\] and $S^*$ reduces to:
\begin{align*}
    S_1^*(Z,C_2) =& \; \left\{\sigma^2(Z,C_2)\mathrm{Var}(C_1^*\mid Z,C_2)\right\}^{-1}\left[\begin{array}{c}
        K_1 \\
        \mathrm{Var}(C_1^*\mid Z,C_2) + K_2 \\
        K_3 \\
        K_4 \\
    \end{array}\right]\\
    =& \; \left[\begin{array}{c}
        0 \\
        \sigma^{-2}(Z,C_2)\left\{1 + K_2\mathrm{Var}(C_1^*\mid Z,C_2)^{-1}\right\}\\
        0 \\
        0 \\
    \end{array}\right]\\
    =& \; \left[\begin{array}{c}
        0 \\
        \sigma^{-2}(Z,C_2)\left[1 - \frac{E\left\{\sigma^{-2}(Z,C_2)\right\}\mathrm{Var}(C_1^*\mid Z,C_2)^{-1}}{E\left\{\sigma^{-2}(Z,C_2)\mathrm{Var}(C_1^*\mid Z,C_2)^{-1}\right\}}\right]\\
        0 \\
        0 \\
    \end{array}\right],\\
    S_2^*(Z,C_2) =& \; \sigma^{-2}(Z,C_2)\left[\begin{array}{c}
        1 \\
        0 \\
        C_2 \\
        Z \\
    \end{array}\right].
\end{align*}
Furthermore, if $\sigma^2(Z,C_2)=\sigma^2$ for some constant $\sigma^2$, i.e., if $Y$ is homoscedastic with respect to $\{Z,C_1^*,C_2\}$, then $S^*$ reduces to
\begin{align*}
    S_1^*(Z,C_2) =& \; \left[\begin{array}{c}
        0 \\
        \sigma^{-2}\left[1 - \frac{\mathrm{Var}(C_1^*\mid Z,C_2)^{-1}}{E\left\{\mathrm{Var}(C_1^*\mid Z,C_2)^{-1}\right\}}\right]\\
        0 \\
        0 \\
    \end{array}\right],\\
    S_2^*(Z,C_2) =& \; \sigma^{-2}\left[\begin{array}{c}
        1 \\
        0 \\
        C_2 \\
        Z \\
    \end{array}\right].
\end{align*}

Thus, under any of the above sets of conditions (heteroscedasticity, homoscedasticity with respect to $C_1^*$, homoscedasticity with respect to $Z$, $C_1^*$, and $C_2$) the estimator solving $n^{-1}\sum_{i=1}^n\check{\bf U}_{S^*}(O_i;\theta)$ has asymptotic variance equal to \[D(W^*,S^*)^{-1}E\{m(O;W^*,S^*)m(O;W^*,S^*)^T\}\{D(W^*,S^*)^{-1}\}^T\]
(which is also straightforward to confirm). By Theorem 5.3 in \cite{newey1994large}, the GMM estimator solving
\[
\frac{1}{n}\sum_{i=1}^n\left\{S_1^*(Z,C_2)C_1^* + S_2^*(Z,C_2)\right\}(Y_i - \theta_0 - \theta_{C_1}^T{\bf C}^*_{1,i} - \theta_{C_2}^TC_{2,i} - \theta_Z Z_i) = 0,
\]
for any of the above $S^*$ under their respective heteroscedasticity/homoscedasticity conditions, is efficient with respect to asymptotic variance in the class of estimators indexed by $S$.
\end{proof}

\begin{proof}[Proof of Theorem \ref{thm:partially-linear-can}]
Letting $W_k=\hat{W}_k\equiv I_{\dim(U_k(O;\phi_k))}$ for all $n$ and $k=1,2$, the estimator solving the estimating equation $n^{-1}\sum_{i=1}^n\tilde{\bf U}_k(O_i;\phi)=0$ %based on (3) 
is identical to the GMM estimator
\begin{align*}
    \hat{\phi}=\arg\min_{\phi}\left\{\frac{1}{n}\sum_{i=1}^n\tilde{\bf U}_k(O_i;\phi)\right\}\hat{W}_k\left\{\frac{1}{n}\sum_{i=1}^n\tilde{\bf U}_k(O_i;\phi)\right\}.
\end{align*}
Under the measurement error model in Assumption 1, we have unbiasedness of $\tilde{\bf U}_1(O;\theta)$ for $\theta^{\dag}$:
\begin{align*}
    & \;  E\left[S\left(Z,C_2\right)\left\{Y-g_1(Z,C_2;\theta_1^{\dag})-g_2(Z,C_2;\theta_2^{\dag})^TC_1^*\right\}\right]\\
    =& \;  E\left[S\left(Z,C_2\right)\left\{E\left(Y\mid Z,C\right)-g_1(Z,C_2;\theta_1^{\dag})-g_2(Z,C_2;\theta_2^{\dag})^TE(C_1^*\mid Z,C_2)\right\}\right]\\
    =& \;  E\left[S\left(Z,C_2\right)\left\{g_2(Z,C_2;\theta_2^{\dag})^TC_1-g_2(Z,C_2;\theta_2^{\dag})^TE(C_1^*\mid Z,C_2)\right\}\right]\\
    =& \;  E\left[S\left(Z,C_2\right)g_2(Z,C_2;\theta_2^{\dag})^T\left\{E(C_1\mid Z,C_2)-E(C_1^*\mid Z,C_2)\right\}\right]\\
    =& \; 0.
\end{align*}
We also have unbiasedness of the last terms in $\tilde{\bf U}_2(O;\theta,\mu)$ for $\phi_2^*$:% $\theta^*\equiv (\theta_0^*, \theta_{C_1}^{*T}, \theta_{C_2}^{*T}, \theta_Z^{\dag})^T$:
\begin{align*}
    &\; E\left\{g_1(z_j,C_2;\theta_1^{\dag})+g_2(z_j,C_2;\theta_2^{\dag})^TC_1^* - \mu_j^{\dag}\right\}\\
    =& \; E\left\{g_1(z_j,C_2;\theta_1^{\dag})+g_2(z_j,C_2;\theta_2^{\dag})^TE(C_1^*\mid Z,C_2) - \mu_j^{\dag}\right\}\\
    =& \; E\left\{g_1(z_j,C_2;\theta_1^{\dag})+g_2(z_j,C_2;\theta_2^{\dag})^TE(C_1\mid Z,C_2) - \mu_j^{\dag}\right\}\\
    =& \; E\left\{g_1(z_j,C_2;\theta_1^{\dag})+g_2(z_j,C_2;\theta_2^{\dag})^TC_1 - \mu_j^{\dag}\right\}\\
    =& \; E\left\{E(Y\mid z_j,C)\right\} - E\left\{Y(z_j)\right\}\\
    =& \; 0
\end{align*}
in addition to the unbiasedness of rest of the terms shown above.

$\tilde{G}_k^T\tilde{W}_k\tilde{G}_k=\tilde{G}_k^T\tilde{G}_k$ is nonsingular for $k=1,2$ since $\tilde{G}_1$ is nonsingular. Thus, by Theorem 3.4 in \cite{newey1994large}, for $k=1,2$,
\begin{align*}
\sqrt{n}\left(\tilde{\phi}_k-\phi_k^*\right)&\rightsquigarrow N\left\{0,(\tilde{G}_k^T\tilde{G}_k)^{-1}\tilde{G}_k^T\tilde{\Omega} \tilde{G}_k(\tilde{G}_k^T\tilde{G}_k)^{-1}\right\}\\
&= N\left\{0,\tilde{G}_k^{-1}\tilde{\Omega}_k (\tilde{G}_k^{-1})^T\right\}.
\end{align*}
\end{proof}

\begin{proof}[Proof of Theorem \ref{thm:partially-linear-efficiency}]
Let %$\tau\equiv (W,S)$, 
\begin{align*}
m(O;W,S)
\equiv& \; E\left\{S(Z,C_2)\left[\nabla_{\theta_1}g_1(Z,C_2)^T\vert_{\theta_1^{\dag}},C_1^{*T}\nabla_{\theta_2}^T g_2(Z,C_2)\vert_{\theta_2^{\dag}}\right]\right\}^TW\\
& \; \times S(Z,C_2)\left\{Y-g_1(Z,C_2;\theta_1^{\dag})-g_2(Z,C_2;\theta_2^{\dag})^TC_1^*\right\},\\
D(W,S)
\equiv& \; E\left\{S(Z,C_2)\left[\nabla_{\theta_1}g_1(Z,C_2)^T\vert_{\theta_1^{\dag}},C_1^{*T}\nabla_{\theta_2}^T g_2(Z,C_2)\vert_{\theta_2^{\dag}}\right]\right\}^TW\\
&\; \times E\left\{S(Z,C_2)\left[\nabla_{\theta_1}g_1(Z,C_2)^T\vert_{\theta_1^{\dag}},C_1^{*T}\nabla_{\theta_2}^T g_2(Z,C_2)\vert_{\theta_2^{\dag}}\right]\right\}
\end{align*}
such that $D(W,S)^{-1}E\{m(O;W,S)m(O;W,S)^T\}\{D(W,S)^{-1}\}^T$ equals the asymptotic variance of the GMM estimator
\begin{align*}
    \tilde{\theta}=\arg\min_{\theta}\left\{\frac{1}{n}\sum_{i=1}^n\tilde{\bf U}_S(O_i;\theta)\right\}\tilde{W}\left\{\frac{1}{n}\sum_{i=1}^n\tilde{\bf U}_S(O_i;\theta)\right\}
\end{align*}
with $\tilde{W}\xrightarrow{p} W$. For all $S$ and $W$, \[S^*(Z,C_2)\equiv \sigma^{-2}(Z,C_2)[\nabla_{\theta_1}g_1(Z,C_2)^T,E(C_1^*\mid Z,C_2)^T\nabla_{\theta_2}^Tg_2(Z,C_2)]^T\] and \[W^*\equiv E\left\{\sigma^{-2}(Z,C_2)[\nabla_{\theta_1}g_1(Z,C_2)^T,E(C_1^*\mid Z,C_2)^T\nabla_{\theta_2}^Tg_2(Z,C_2)]^{\otimes 2}\right\}^{-1}\] satisfy
\begin{align*}
& \;   E\{m(O;W,S)m(O;W^*,S^*)^T\}\\
=& \; E\left\{S(Z,C_2)\left[\nabla^T_{\theta_1}g_1(Z,C_2)\vert_{\theta_1^{\dag}},C_1^{*T}\nabla_{\theta_2}^T g_2(Z,C_2)\vert_{\theta_2^{\dag}}\right]\right\}^TW\\
&\;\times E\left[S(Z,C_2)\left\{Y-g_1(Z,C_2;\theta_1^{\dag})-g_2(Z,C_2;\theta_2^{\dag})^TC_1^*\right\}^2\sigma^{-2}(Z,C_2)\right.\\
&\quad\times\left.\left[\nabla^T_{\theta_1}g_1(Z,C_2)\vert_{\theta_1^{\dag}},E(C_1^*\mid Z,C_2)^{T}\nabla_{\theta_2}^T g_2(Z,C_2)\vert_{\theta_2^{\dag}}\right]\right]\\
&\;\times E\left\{\sigma^{-2}(Z,C_2)\left[\nabla^T_{\theta_1}g_1(Z,C_2)\vert_{\theta_1^{\dag}},E(C_1^*\mid Z,C_2)^T\nabla_{\theta_2}^Tg_2(Z,C_2)\vert_{\theta_2^{\dag}}\right]^{\otimes 2}\right\}^{-1}\\
&\;\times E\left\{\sigma^{-2}(Z,C_2)\left[\nabla^T_{\theta_1}g_1(Z,C_2)\vert_{\theta_1^{\dag}},E(C_1^*\mid Z,C_2)^T\nabla_{\theta_2}^Tg_2(Z,C_2)\vert_{\theta_2^{\dag}}\right]^T\right.\\
&\quad \left.\times\left[\nabla_{\theta_1}^Tg_1(Z,C_2)\vert_{\theta_1^{\dag}},C_1^{*T}\nabla_{\theta_2}^T g_2(Z,C_2)\vert_{\theta_2^{\dag}}\right]\right\}\\
=& \; E\left\{S(Z,C_2)\left[\nabla^T_{\theta_1}g_1(Z,C_2)\vert_{\theta_1^{\dag}},C_1^{*T}\nabla_{\theta_2}^T g_2(Z,C_2)\vert_{\theta_2^{\dag}}\right]\right\}^TW\\
&\; \times E\left\{S(Z,C_2)\left[\nabla^T_{\theta_1}g_1(Z,C_2)\vert_{\theta_1^{\dag}},E(C_1^*\mid Z,C_2)^{T}\nabla_{\theta_2}^T g_2(Z,C_2)\vert_{\theta_2^{\dag}}\right]\right\}\\
=& \; D(W,S).
\end{align*}
Since $n^{-1}\sum_{i=1}^n\tilde{\bf U}_{S^*}(O_i;\theta)$ is a just-identified estimating equation, the above GMM estimator is equivalent for all sequences of positive definite matrices $\tilde{W}$. Thus, the estimator solving $n^{-1}\sum_{i=1}^n\tilde{\bf U}_{S^*}(O_i;\theta)$, which is equivalent to the above GMM estimator with $S=S^*$ and $\tilde{W}=I$ for all $n$, has asymptotic variance equal to \[D(W^*,S^*)^{-1}E\{m(O;W^*,S^*)m(O;W^*,S^*)^T\}\{D(W^*,S^*)^{-1}\}^T\]
(which is also straightforward to confirm). Thus, by Theorem 5.3 in \cite{newey1994large}, the GMM estimator solving
\[
\frac{1}{n}\sum_{i=1}^nS^*(Z_i,C_{2,i})\left\{Y_i-g_1(Z_i,C_{2,i};\theta_1)-g_2(Z_i,C_{2,i};\theta_2)^TC_{1,i}^*\right\} = 0
\]
is efficient with respect to asymptotic variance in the class of estimators indexed by $S$.
\end{proof}

\begin{proof}[Proof of Theorem \ref{thm:mediator_can}]
Letting $W_k=\hat{W}_k\equiv I_{\dim(\bar{U}(O;\beta))}$ for all $n$, the estimator solving the estimating equation $\sum_{i=1}^n\bar{U}_k(O_i;\beta)=0$ %based on (3) 
is identical to the GMM estimator
\begin{align*}
    \hat{\beta}=\arg\min_{\beta}\left\{\frac{1}{n}\sum_{i=1}^n\bar{U}_k(O_i;\beta)\right\}\hat{W}_k\left\{\frac{1}{n}\sum_{i=1}^n\bar{U}_k(O_i;\beta)\right\}.
\end{align*}
Under measurement error model (\ref{eqn:me6}), we have unbiasedness of $\bar{U}(O;\beta)$ for $\beta^{\dag}$:
\begin{align*}
    & \;  E\left[\left\{\bar{S}_1(C)+\bar{S}_2(C)A^*\right\}\left\{b_k(Z)-h_1(C;\beta_1^{\dag})-h_2(C;\beta_2^{\dag})A^*  \right\}+\bar{S}_2(C)h_2(C;\beta_2)\sigma_{\varepsilon}^{2\dag}\right]\\
    =& \;  E\left(\bar{S}_1\left(C\right)\left[E\left\{b_k(Z)\mid A,C\right\}-h_1(C;\beta_1^{\dag})-h_2(C;\beta_2^{\dag})E(A^*\mid C)\right]\right)\\
    &+ \;  E\left[\bar{S}_2(C)A\left\{b_k(Z)-h_1(C;\beta_1^{\dag})-h_2(C;\beta_2^{\dag})A\right\}\right]\\
    &- \;  E\left\{\bar{S}_2(C)Ah_2(C;\beta_2^{\dag})\varepsilon\right\}\\
    &+ \;  E\left[\bar{S}_2(C)\varepsilon\left\{b_k(Z)-h_1(C;\beta_1^{\dag})-h_2(C;\beta_2^{\dag})A\right\}\right]\\
    &- \;  E\left\{\bar{S}_2(C) h_2(C;\beta_2^{\dag})\varepsilon^2\right\}\\
    &+ \;  E\left\{\bar{S}_2(C)h_2(C;\beta_2)\sigma_{\varepsilon}^{2\dag}\right\}\\
    =& \;  E\left[\bar{S}_1\left(C\right)h_2(A,C;\beta_2^{\dag})\left\{A-E(A\mid C)\right\}\right]\\
    &+ \;  E\left(\bar{S}_2(C)A\left[E\left\{b_k(Z)\mid A,C\right\}-h_1(C;\beta_1^{\dag})-h_2(C;\beta_2^{\dag})A\right]\right)\\
    &- \;  E\left\{\bar{S}_2(C)Ah_2(C;\beta_2^{\dag})E(\varepsilon\mid A,C)\right\}\\
    &+ \;  E\left[\bar{S}_2(C)E(\varepsilon\mid A,C,Z)\left\{b_k(Z)-h_1(C;\beta_1^{\dag})-h_2(C;\beta_2^{\dag})A\right\}\right]\\
    &- \;  E\left\{\bar{S}_2(C) h_2(C;\beta_2^{\dag})E(\varepsilon^2\mid C)\right\}\\
    &+ \;  E\left\{\bar{S}_2(C)h_2(C;\beta_2)\right\}\sigma_{\varepsilon}^{2\dag}\\
    =& \;  E\left\{\bar{S}_2(C) h_2(C;\beta_2^{\dag})\right\}\left\{\sigma_{\varepsilon}^{2\dag}-E(\varepsilon^2)\right\}\\
    =& \; 0.
\end{align*}

$\bar{G}^T\bar{W}\bar{G}=\bar{G}^T\bar{G}$ is nonsingular since $\bar{G}$ is nonsingular. Thus, by Theorem 3.4 in \cite{newey1994large}, 
\begin{align*}
\sqrt{n}\left(\bar{\beta}-\beta^*\right)&\rightsquigarrow N\left\{0,(\bar{G}^T\bar{G})^{-1}\bar{G}^T\bar{\Omega} \bar{G}(\bar{G}^T\bar{G})^{-1}\right\}\\
&= N\left\{0,\bar{G}^{-1}\bar{\Omega} (\bar{G}^{-1})^T\right\}.
\end{align*}
\end{proof}

\begin{proof}[Proof of Theorem \ref{thm:mediator_efficiency}]
Let %$\tau\equiv (W,S)$, 
\begin{align*}
m(O;\bar{S})
\equiv& \; \bar{U}_{\bar{S}}(O;\beta)\\ 
=& \; \left\{\bar{{ S}}_1({ C})+\bar{{ S}}_2({ C})A^*\right\}\left\{b_k(Z)-h_1({ C};\beta_1)-h_2({ C};\beta_2)A^*  \right\}+\bar{{ S}}_2({ C})h_2({ C};\beta_2)\sigma_{\varepsilon}^{2\dag},\\
D(\bar{S})
\equiv& \; -E\left\{\nabla_{\beta}\bar{U}(O;\beta)\right\}\\
=& \; E\left[\left\{\bar{S}_1(C)+\bar{S}_2(C)A^*\right\}\left[\nabla_{\beta_1}h_1(C;\beta_1)^T, A^*\nabla_{\beta_2}h_2(C;\beta_2)^T\right]\right]\\
    &+ \sigma^{2\dag}E\left\{\bar{S}_2(C)\left[0^T, \nabla_{\beta_2}h_2(C;\beta_2)^T\right]\right\}
\end{align*}
such that $D(\bar{S})^{-1}E\{m(O;\bar{S})m(O;\bar{S})^T\}\{D(\bar{S})^{-1}\}^T$ equals the asymptotic variance of the GMM estimator
\begin{align*}
    \bar{\beta}=\arg\min_{\beta}\left\{\frac{1}{n}\sum_{i=1}^n\bar{ U}_{\bar{S}}(O_i;\beta)\right\}^{\otimes 2}.
\end{align*}
Let $\Delta^*(\beta) \equiv b(Z) - h_1(C;\beta_1) - h_2(C;\beta_2)A^*$, and replace $\bar{S}_1$ and $\bar{S}_2$ with $\ell$ and $m$, respectively, for notational convenience. 
We have
\begin{align*}
    &-E\left\{\nabla_{\beta}\bar{U}(O;\beta)\right\}\\ 
    =& \; E\left[\left\{\ell(C)+m(C)A^*\right\}\left[\nabla_{\beta_1}h_1(C;\beta_1)^T, A^*\nabla_{\beta_2}h_2(C;\beta_2)^T\right]\right]\\
    &+ \sigma^{2\dag}E\left\{m(C)\left[0^T, \nabla_{\beta_2}h_2(C;\beta_2)^T\right]\right\}\\
    =& \; E\left[\left[\left\{\ell(C)+m(C)A^*\right\}\nabla_{\beta_1}h_1(C;\beta_1)^T,\right.\right.\\ 
    &\left.\left.\left\{\ell(C)+m(C)A^*\right\}A^*\nabla_{\beta_2}h_2(C;\beta_2)^T+ \sigma^{2\dag}m(C)\nabla_{\beta_2}h_2(C;\beta_2)^T\right]\right]\\
    &= \; E\left\{\left[\ell(C), m(C)\right]\left[\begin{array}{cc}
        \nabla_{\beta_1}h_1(C;\beta_1)^T & A^*\nabla_{\beta_2}h_2(C;\beta_2)^T \\
        A^*\nabla_{\beta_1}h_1(C;\beta_1)^T & \left(A^{*2} + \sigma^{2\dag}\right)\nabla_{\beta_2}h_2(C;\beta_2)^T
    \end{array}\right]\right\}\\
    &\equiv \; E\left\{\left[\ell(C), m(C)\right]\left[\begin{array}{cc}
        V_1 & V_2 \\
        W_1 & W_2
\end{array}\right]\right\}\\
\end{align*}
and
\begin{align*}
    & E\left\{\bar{U}_{\bar{S}}(O;\beta)\bar{U}_{{\bar{S}}^*}(O;\beta)^T\right\}\\
    =& \; E\left[\Delta^*(\beta)^2\left\{\ell(C)+m(C)A^*\right\}\left\{\ell^*(C)+m^*(C)A^*\right\}^T\right]\\
    &+ E\left[\left\{\ell(C)+m(C)A^*\right\}\Delta^*(\beta)\sigma^{2\dag}h_2(C;\beta_2)m^*(C)^T\right]\tag{1}\\
    &+ E\left[m(C)h_2(C;\beta_2)\sigma^{2\dag}\Delta^*(\beta)\left\{\ell^*(C)+m^*(C)A^*\right\}^T\right]\tag{2}\\
    &+ \sigma^{\dag 4}E\left\{h_2(C;\beta_2)^2m(C)m^*(C)^T\right\}.
\end{align*}
We also have
\begin{align*}
    & E\left[\left\{\ell(C)+m(C)A^*\right\}\left\{b(Z)-h_1(C;\beta_1)-h_2(C;\beta_2)A^*\right\}\mid C\right]\\
    =& \; E\left[\left\{\ell(C)+m(C)A\right\}\left\{b(Z)-h_1(C;\beta_1)-h_2(C;\beta_2)A\right\}\mid C\right]\\
    &- E\left[\left\{\ell(C)+m(C)A\right\}h_2(C;\beta_2)\varepsilon\mid C\right]\\
    &+ E\left[m(C)\varepsilon\left\{b(Z)-h_1(C;\beta_1)-h_2(C;\beta_2)A\right\}\mid C\right]\\
    &- E\left\{m(C)\varepsilon^2h_2(C;\beta_2)\mid C\right\}\\
    =& \; -m(C)h_2(C;\beta_2)\sigma^{\dag 2},
\end{align*}
and
\begin{align*}
    & E\left[\left\{\ell^*(C)+m^*(C)A^*\right\}\left\{b(Z)-h_1(C;\beta_1)-h_2(C;\beta_2)A^*\right\}\mid C\right]\\
    =& \; E\left[\left\{\ell^*(C)+m^*(C)A\right\}\left\{b(Z)-h_1(C;\beta_1)-h_2(C;\beta_2)A\right\}\mid C\right]\\
    &- E\left[\left\{\ell^*(C)+m^*(C)A\right\}h_2(C;\beta_2)\varepsilon\mid C\right]\\
    &+ E\left[m^*(C)\varepsilon\left\{b(Z)-h_1(C;\beta_1)-h_2(C;\beta_2)A\right\}\mid C\right]\\
    &- E\left\{m^*(C)\varepsilon^2h_2(C;\beta_2)\mid C\right\}\\
    =& \; -m^*(C)h_2(C;\beta_2)\sigma^{\dag 2},
\end{align*}
so 
\begin{align*}
    (1) =& \; E\left(E\left[\left\{\ell(C)+m(C)A^*\right\}\Delta^*(\beta)\mid C\right]\sigma^{\dag 2}h_2(C;\beta_2)m^*(C)^T\right)\\
    %=& \; -E\left\{m(C)h_2(C;\beta_2)\sigma^{\dag 4}h_2(C;\beta_2)m^*(C)^T\right\}\\
    =& \; -\sigma^{\dag 4}E\left\{h_2(C;\beta_2)^2m(C)m^*(C)^T\right\}
\end{align*}
and
\begin{align*}
    (2) &= E\left(m(C)h_2(C;\beta_2)\sigma^{2\dag}E\left[\Delta^*(\beta)\left\{\ell^*(C)+m^*(C)A^*\right\}^T\mid C\right]\right)\\
    &= -\sigma^{\dag 4}E\left\{h_2(C;\beta_2)^2m(C)m^*(C)^T\right\}.
\end{align*}
Thus,
\begin{align*}
    E\left\{\bar{U}_{\bar{S}}(O;\beta)\bar{U}_{{\bar{S}}^*}(O;\beta)^T\right\}
    =& \; E\left[\Delta^*(\beta)^2\left\{\ell(C)+m(C)A^*\right\}\left\{\ell^*(C)+m^*(C)A^*\right\}^T\right]\\
    &-\sigma^{\dag 4}E\left\{h_2(C;\beta_2)^2m(C)m^*(C)^T\right\}.
\end{align*}
Decomposing $\ell(C)=[\ell_1(C)^T, \ell_2(C)^T]^T$ and $m(C)=[m_1(C)^T, m_2(C)^T]^T$,
the equation $-E\left\{\nabla_{\beta}\bar{U}(O;\beta)\right\} = E\left\{\bar{U}_{\bar{S}}(O;\beta)\bar{U}_{{\bar{S}}^*}(O;\beta)^T\right\}$ implies, for $k=1,2$:
\begin{align*}
    &-E\left\{\ell(C)V_k+m(C)W_k\right\}\\
    =& \; E\left[\Delta^*(\beta)^2\left\{\ell(C)+m(C)A^*\right\}\left\{\ell^*_k(C)+m^*_k(C)A^*\right\}^T - \sigma^{\dag 4}h_2(C;\beta_2)^2m(C)m_k^*(C)^T\right]
\end{align*}
$\Longleftrightarrow$
\begin{align*}
    0 =&\; E\left(\ell(C)E\left[\Delta^*(\beta)^2\left\{\ell_k^*(C)+m_k^*(C)A^*\right\}^T + V_k \mid C\right]\right.\\
    & \; \left. + \; m(C)E\left[\Delta^*(\beta)^2A^*\left\{\ell_k^*(C)+m_k^*(C)A^*\right\}^T + W_k - \sigma^{\dag 4}h_2(C;\beta_2)^2m_k^*(C)^T\mid C\right]\right)
\end{align*}
$\Longleftrightarrow$
\begin{align*}
    0 =&\; E\left[\Delta^*(\beta)^2\left\{\ell_k^*(C)+m_k^*(C)A^*\right\}^T + V_k \mid C\right]\\
    0 =& \; E\left[\Delta^*(\beta)^2A^*\left\{\ell_k^*(C)+m_k^*(C)A^*\right\}^T + W_k - \sigma^{\dag 4}h_2(C;\beta_2)^2m_k^*(C)^T\mid C\right]
\end{align*}
$\Longleftrightarrow$
\begin{align*}
    0 =&\; E\left\{\Delta^*(\beta)^2\mid C\right\}\ell_k^*(C)^T+E\left\{\Delta^*(\beta)^2A^*\mid C\right\}m_k^*(C)^T + E(V_k\mid C)\\
    0 =& \; E\left\{\Delta^*(\beta)^2A^*\mid C\right\}\ell_k^*(C)^T+\left[E\left\{\Delta^*(\beta)^2A^{*2}\mid C\right\} - \sigma^{\dag 4}h_2(C;\beta_2)^2\right]m_k^*(C)^T + E(W_k\mid C)
\end{align*}
$\Longleftrightarrow$
\[\left[\begin{array}{cc}
    \scriptstyle E\left\{\Delta^*(\beta)^2\mid C\right\} & \scriptstyle E\left\{\Delta^*(\beta)^2A^*\mid C\right\} \\
    \scriptstyle E\left\{\Delta^*(\beta)^2A^*\mid C\right\} & \scriptstyle E\left\{\Delta^*(\beta)^2A^{*2}\mid C\right\} - \sigma^{\dag 4}h_2(C;\beta_2)^2
\end{array}\right]
\left[\begin{array}{c}
    \ell_k^*(C)^T \\
    m_k^*(C)^T
\end{array}\right] = -\left[\begin{array}{c}
    E(V_k\mid C) \\
    E(W_k\mid C)
\end{array}\right]\]
$\Longleftrightarrow$
\begin{align*}
\left[\begin{array}{c}
    \ell_k^*(C)^T \\
    m_k^*(C)^T
\end{array}\right] &= -\left[\begin{array}{cc}\scriptstyle
    E\left\{\Delta^*(\beta)^2\mid C\right\} & \scriptstyle E\left\{\Delta^*(\beta)^2A^*\mid C\right\} \\
    \scriptstyle E\left\{\Delta^*(\beta)^2A^*\mid C\right\} & \scriptstyle E\left\{\Delta^*(\beta)^2A^{*2}\mid C\right\} - \sigma^{\dag 4}h_2(C;\beta_2)^2
\end{array}\right]^{-1}\left[\begin{array}{c}
    E(V_k\mid C) \\
    E(W_k\mid C)
\end{array}\right]\\
&= -d(C)\left[\begin{array}{cc}
    \scriptstyle E\left\{\Delta^*(\beta)^2A^{*2}\mid C\right\} - \sigma^{\dag 4}h_2(C;\beta_2)^2 & -\scriptstyle E\left\{\Delta^*(\beta)^2A^*\mid C\right\} \\
    -\scriptstyle E\left\{\Delta^*(\beta)^2A^*\mid C\right\} & \scriptstyle E\left\{\Delta^*(\beta)^2\mid C\right\}
\end{array}\right]\left[\begin{array}{c}
    E(V_k\mid C) \\
    E(W_k\mid C)
\end{array}\right]\\
\end{align*}
where 
\begin{align*}
    E(V_1\mid C) &= \nabla_{\beta_1}h_1(C;\beta_1)^T\\ 
    E(V_2\mid C) &= \nabla_{\beta_2}h_2(C;\beta_2)^TE(A^*\mid C)\\ 
    E(W_1\mid C) &= \nabla_{\beta_1}h_1(C;\beta_1)^TE(A^*\mid C)\\ 
    E(W_2\mid C) &= \nabla_{\beta_2}h_2(C;\beta_2)^T\{E(A^{*2}\mid C)+\sigma^{\dag 2}\}
\end{align*}
provided 
\begin{align*}
0 \neq d(C) \equiv & \; E\left\{\Delta^*(\beta)^2\mid C\right\}E\left\{\Delta^*(\beta)^2A^{*2}\mid C\right\} - E\left\{\Delta^*(\beta)^2\mid C\right\}\sigma^{\dag 4}h_2(C;\beta_2)^2\\
& \; - \; E\left\{\Delta^*(\beta)^2A^*\mid C\right\}^2.
\end{align*}
Thus,
\begin{align*}
\left[\begin{array}{c}
    {\scriptstyle\ell_1^*(C)^T} \\
    {\scriptstyle m_1^*(C)^T}
\end{array}\right] = d(C)^{-1}\left[\begin{array}{c}
{\scriptscriptstyle\left[\sigma_{\varepsilon}^{\dag 4}h_2(C;\beta_2)^2 - \mathrm{Cov}\left\{\Delta^*(\beta)^2A^*, A^*\mid C\right\}\right]\nabla_{\beta_1}h_1(C;\beta_1)^T}\\
{\scriptscriptstyle
\mathrm{Cov}\left\{\Delta^*(\beta)^2, A^*\mid C\right\}\nabla_{\beta_1}h_1(C;\beta_1)^T}
\end{array}\right],
\end{align*}
\begin{align*}
\left[\begin{array}{c}
    {\scriptstyle\ell_2^*(C)^T} \\
    {\scriptstyle m_2^*(C)^T}
\end{array}\right] = {\scriptstyle d(C)^{-1}}\left[\begin{array}{c}
{\scriptscriptstyle\left(E\left\{\Delta^*(\beta)^2A^*\mid C\right\}\left\{E(A^{*2}\mid C) + \sigma_{\varepsilon}^{\dag 2}\right\} - \left[E\left\{\Delta^*(\beta)^2A^{*2}\mid C\right\} - \sigma_{\varepsilon}^{\dag 4}h_2(C;\beta_2)^2\right]E(A^*\mid C)\right)\nabla_{\beta_2}h_2(C;\beta_2)^T}\\
{\scriptscriptstyle\left[E\left\{\Delta^*(\beta)^2A^*\mid C\right\}E(A^*\mid C) - E\left\{\Delta^*(\beta)^2\mid C\right\}\left\{E(A^{*2}\mid C) + \sigma_{\varepsilon}^{\dag 2}\right\}\right]\nabla_{\beta_2}h_2(C;\beta_2)^T}
\end{array}\right],
\end{align*}
and
\begin{align*}
\bar{S}_1^*(C) &= \left[\begin{array}{c}
    \ell_1^*(C) \\
    \ell_2^*(C)
\end{array}\right]\\ 
&= {\scriptstyle d(C)^{-1}}\left[\begin{array}{c}
{\scriptscriptstyle\left[\sigma_{\varepsilon}^{\dag 4}h_2(C;\beta_2)^2 - \mathrm{Cov}\left\{\Delta^*(\beta)^2A^*, A^*\mid C\right\}\right]\nabla_{\beta_1}h_1(C;\beta_1)^T}\\
{\scriptscriptstyle\left(E\left\{\Delta^*(\beta)^2A^*\mid C\right\}\left\{E(A^{*2}\mid C) + \sigma_{\varepsilon}^{\dag 2}\right\} - \left[E\left\{\Delta^*(\beta)^2A^{*2}\mid C\right\} - \sigma_{\varepsilon}^{\dag 4}h_2(C;\beta_2)^2\right]E(A^*\mid C)\right)\nabla_{\beta_2}h_2(C;\beta_2)^T}
\end{array}\right],
\end{align*}
\begin{align*}
\bar{S}^*_2(C) &= \left[\begin{array}{c}
    m_1^*(C) \\
    m_2^*(C)
\end{array}\right]\\ 
&= {\scriptstyle d(C)^{-1}}\left[\begin{array}{c}{\scriptscriptstyle
\mathrm{Cov}\left\{\Delta^*(\beta)^2, A^*\mid C\right\}\nabla_{\beta_1}h_1(C;\beta_1)^T}\\
{\scriptscriptstyle\left[E\left\{\Delta^*(\beta)^2A^*\mid C\right\}E(A^*\mid C) - E\left\{\Delta^*(\beta)^2\mid C\right\}\left\{E(A^{*2}\mid C) + \sigma_{\varepsilon}^{\dag 2}\right\}\right]\nabla_{\beta_2}h_2(C;\beta_2)^T}
\end{array}\right].
\end{align*}

%Since $n^{-1}\sum_{i=1}^n\bar{ U}_{\bar{S}^*}(O_i;\beta)$ is a just-identified estimating equation, the above GMM estimator is equivalent for all sequences of positive definite matrices $\bar{W}$. 
Thus, the estimator solving $n^{-1}\sum_{i=1}^n\bar{ U}_{\bar{S}^*}(O_i;\beta)$ %, which is equivalent to the above GMM estimator with $\bar{S}=\bar{S}^*$ and $\bar{W}=I$ for all $n$, 
has asymptotic variance equal to \[D(\bar{S}^*)^{-1}E\{m(O;\bar{S}^*)m(O;\bar{S}^*)^T\}\{D(\bar{S}^*)^{-1}\}^T\]
(which is also straightforward to confirm). Thus, by Theorem 5.3 in \cite{newey1994large}, the GMM estimator solving
\[
\frac{1}{n}\sum_{i=1}^n\left[\left\{\bar{{ S}}_1({ C_i})+\bar{{ S}}_2({ C_i})A^*_i\right\}\left\{b_k(Z_i)-h_1({ C_i};\beta_1)-h_2({ C_i};\beta_2)A^*_i  \right\}+\bar{{ S}}_2({ C_i})h_2({ C_i};\beta_2)\sigma_{\varepsilon}^{2\dag}\right] = 0
\]
is efficient with respect to asymptotic variance in the class of estimators indexed by $\bar{S}$.

\end{proof}

\section{Simulation study data generating process}
\label{sec:sims-supp}
Simulated data were generated according to the following data generating process.

\subsection*{Mediation settings}

Covariates:\\
\indent Age mother (years) $\sim N(24,9)$\\
\indent Low education $\sim \mathrm{Bern}(0.7)$\\
\indent Age at testing (months) $\sim N(24, 12.25)$\\

\noindent Exposure: $A \sim N(\mu_e, \sigma^2_a = 0.01)$\\
\indent Under the rank condition with strong constructed IV:
\[\mu_e=\gamma_0 + \gamma_1 \times \mathrm{agemother} + \gamma_{1,2} \times \mathrm{agemother}^2 + \gamma_2 \times \mathrm{lowedu} + \gamma_3 \times \mathrm{age} + \gamma_4 \times \mathrm{agemother} \times \mathrm{lowedu},\]
\indent\indent where $\gamma_0= 0,  \gamma_1   =0.1,\gamma_{1,2} =0.3,          \gamma_2  = -0.5, \gamma_3   =0.21, %0.25, 
\gamma_4= -0.7$.\\
\indent Under the rank condition with weak constructed IV: $\mu_e$ is the same as above but with $\gamma_{1,2} =0.001$ and $\gamma_4=-.001$. This leads to weak IV $F$ test statistics only exceeding 10 about 29--40\% of the time across measurement error scenarios (compared with the approximately 14\% we see when there is no IV relevance). \\
\indent Under violation of the rank condition:
\[\mu_e = \gamma_0 + \gamma_1 \times \mathrm{agemother} + \gamma_2 \times \mathrm{lowedu} + \gamma_3 \times \mathrm{age}.\]
\\
\noindent Measurement error: $U \sim N(\mu_u, \sigma^2_u)$,\\
\indent where $\mu_u= 0$ and $\sigma^2_u \in (0.0043, 0.0025, 0.0011)$ such that the conditional reliability ratio $\sigma^2_a/(\sigma^2_a+\sigma^2_u)\in(0.7, 0.8, 0.9)$.\\
%(0.70, 0.80, 0.90)$.\\
%\indent Under classical measurement error: $\mu_u= 0$\\
%\indent Under violation of classical measurement error: $\mu_u= \mathrm{lowedu}-0.5\times\mathrm{agemother}$\\
%\indent $\sigma^2_u \in (0.70, 0.80, 0.90)$\\

\noindent Mediator: $M \sim N(\mu_m,0.01)$,\\
\indent where $\mu_m= \beta_0 + \beta_1 \times A + \beta_2 \times \mathrm{agemother} + \beta_3 \times \mathrm{lowedu} + \beta_4 \times \mathrm{age}$,\\
\indent
$\beta_0=0, 
\beta_1 \in (0,-0.3,-0.01),
\beta_2 = 0.1,
\beta_3 = 0.1,
\beta_4   = 0.1$.\\

\noindent Outcome: $Y\sim N(\mu_y,1)$,\\
\indent where
\[\mu_y= \theta_0 + \theta_1 \times A + \theta_2 \times M +\theta_3\times A\times M+ \theta_4 \times A^2+ \theta_5 \times \mathrm{agemother} + \theta_6 \times \mathrm{lowedu} + \theta_7 \times\mathrm{age},\]
\indent $\theta_0 =0,
\theta_1  \in (0,1.5),
\theta_2  \in (0,0.5) ,
\theta_3  \in (0,0.05),
\theta_4  \in (0,0.1),
\theta_5  = 0.1,
\theta_6  = 0.2,
\theta_7  = 6.$

\noindent Constructed IV GAM models setting 1: spline term for agemother allowing for agemother$\times$lowedu interaction.\\
\indent For outcome model:
\[A^* \sim s(M)+s(\mathrm{agemother},\mathrm{lowedu})+\mathrm{age}\]
\indent For mediator model:
\[A^* \sim s(\mathrm{agemother},\mathrm{lowedu})+\mathrm{age}\]

\noindent Constructed IV GAM models setting 2: spline term for agemother not including agemother$\times$lowedu interaction.\\
\indent For outcome model:
\[A^* \sim s(M)+s(\mathrm{agemother})+\mathrm{lowedu}+\mathrm{age}\]
\indent For mediator model:
\[A^* \sim s(\mathrm{agemother})+\mathrm{lowedu}+\mathrm{age}\]

\noindent Constructed IV GAM models setting 3: only including spline term for age, which is linear in true data generating process and not including agemother$\times$lowedu interaction.\\
\indent For outcome model:
\[A^* \sim s(M)+\mathrm{agemother}+\mathrm{lowedu}+s(\mathrm{age})\]
\indent For mediator model:
\[A^* \sim \mathrm{agemother}+\mathrm{lowedu}+s(\mathrm{age})\]

Weak constructed IV results are shown in Table S7. Across conditional reliability ratios, the biases of the various constructed IV approaches are similar to that of the naive approach and the standard deviations are inflated, as expected. The approach that uses the constructed IV for the mediator model and the method of moments using the estimated measurement error variance for the outcome model (IVZ-MoMY) performs a bit better than the others across all settings. The coverage probabilities of the constructed IV approaches are much closer to the nominal 95\% than the naive approach and the sensitivity analysis approaches, except for when the true reliability ratio is known. Thus, while constructed IV methods may yield confidence intervals with fairly reasonable coverage when the constructed IV is weak, bias and variance may still be inflated, which can ultimately make confidence intervals fairly uninformative.

Results from varying the functional form of the model used to fit the constructed IV are shown in Tables S7 and S8, where Table S7 shows results using functional form 1 and Table S8 shows results using functional forms 2 and 3. Under functional form 2, the models we use for the conditional mean of $A^*$ given $\bf C$ as well as given $\bf C$ and $M$ are mis-specified in that they do not include an interaction between ``agemother'' and ``lowedu''. However, we still see minimal bias and reasonable coverage probabilities for all effects across all reliability ratios. Under functional form 3, the models we use for the conditional mean of $A^*$ given $\bf C$ as well as given $\bf C$ and $M$ are mis-specified in that they also do not include an interaction between ``agemother'' and ``lowedu'', and among the covariates in $\bf C$, they only include a spline term for ``age'', which is linear in the true model for $A^*$ given $\bf C$. In this case, we see enormous bias and standard deviation as well as over-coverage, since the rank condition is practically violated by virtue of the model failing to capture the nonlinearity in the conditional mean of $A^*$ given $\bf C$ and/or $M$. This highlights the importance of correctly capturing nonlinearities in the model used for the constructed IV.

\subsection*{Average treatment effect setting}
\noindent For ATE estimation, we use nearly the same data generating mechanism, but target the ATE of $M$ on $Y$, i.e., viewing $M$ as the exposure of interest and $A^*$ as an error-prone confounder. The only difference is that we use the non-classical measurement error model, which is the same as above but with $\mu_u= \mathrm{lowedu}-0.5\times\mathrm{agemother}$.

Results from the average treatment effect setting are shown in Table S9. In this case, we see reduced bias and similar standard error for the constructed IV approach compared to the naive approach. In the method of moments sensitivity analysis approach, the magnitude of the bias corresponds to how closely the reliability ratio is to being correctly specified, though the bias is never especially large. There is reasonable coverage probability across all methods, with the constructed IV method having slightly reduced coverage probability, which may be attributable to Monte Carlo error. Overall, under this particular data generating mechanism, it does not seem that the measurement error induces a large amount of bias.

\section{Additional tables and figures}
\label{sec:additional-tabs}

%\newpage

\begin{figure}[p!]
 \begin{center}
 %\caption*{\label{fig:sims1}}
 \includegraphics[width=\textwidth]{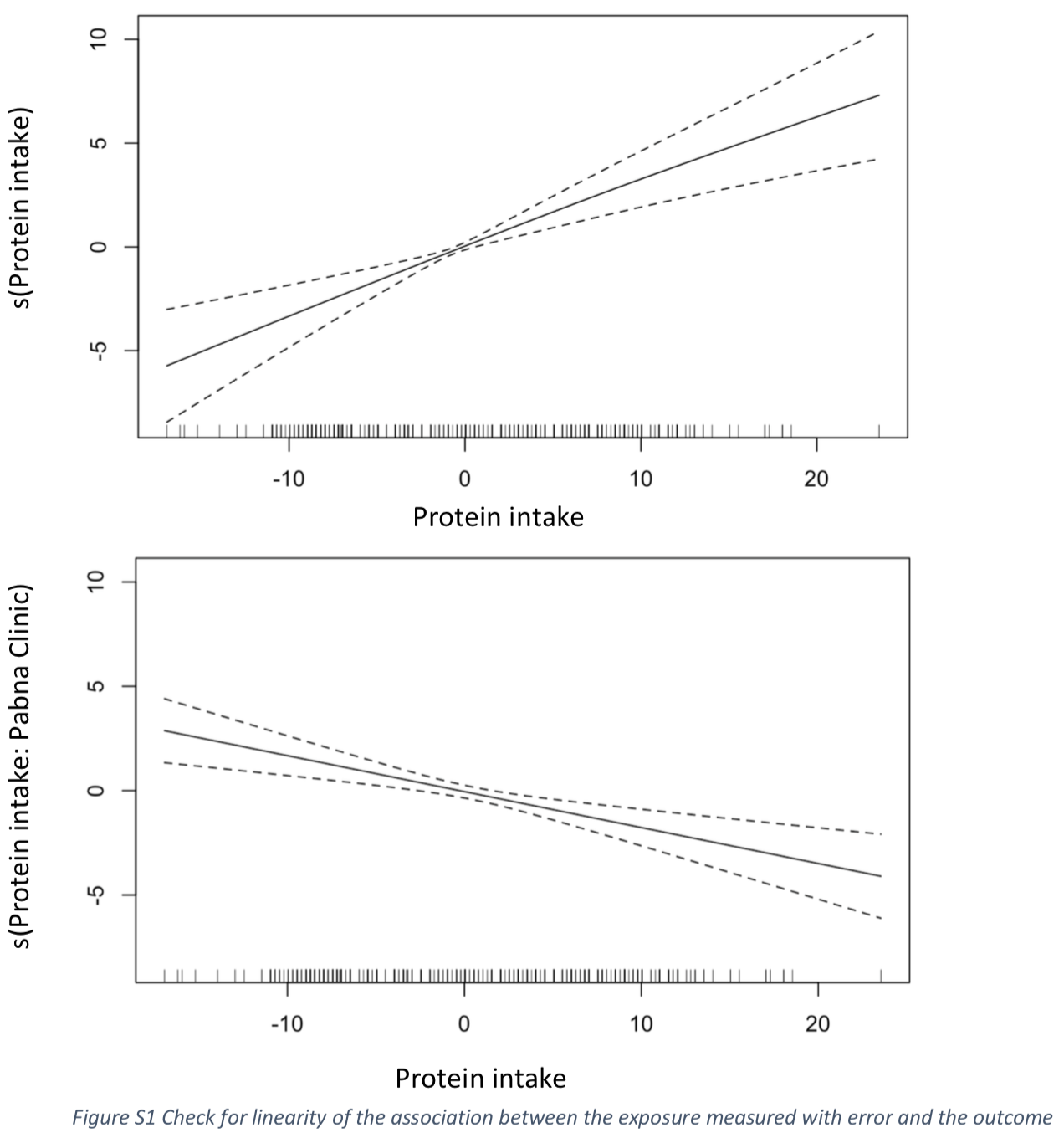}
 \end{center}
\end{figure}

\newpage

\begin{figure}[p!]
 \begin{center}
 %\caption*{}
 \label{fig:sims2}
 \includegraphics[width=\textwidth]{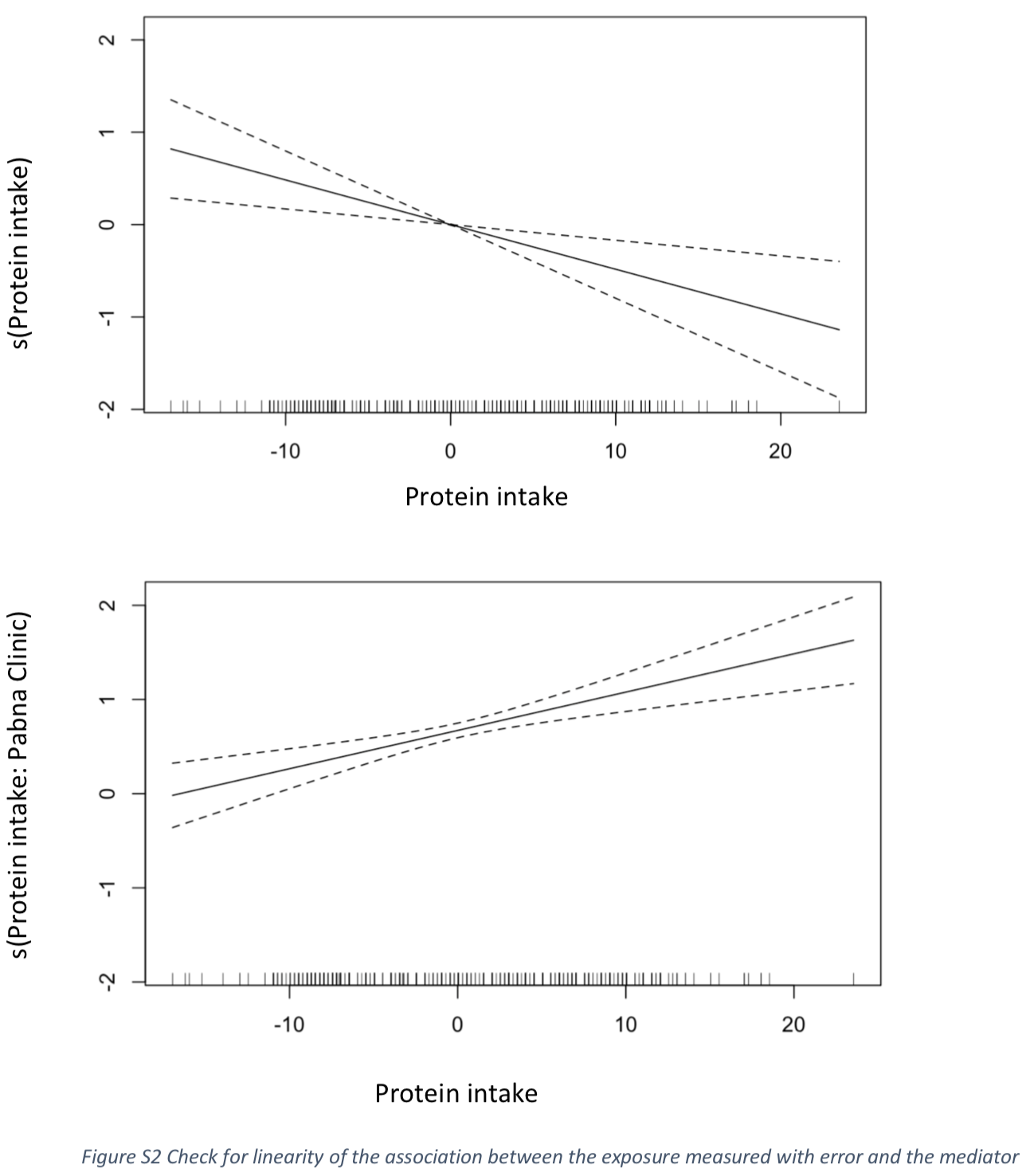}
 \end{center}
\end{figure}

\newpage

\begin{figure}[p!]
 %\caption*{}
 \begin{center}
 \label{tab:S1}
 \includegraphics[width=\textwidth]{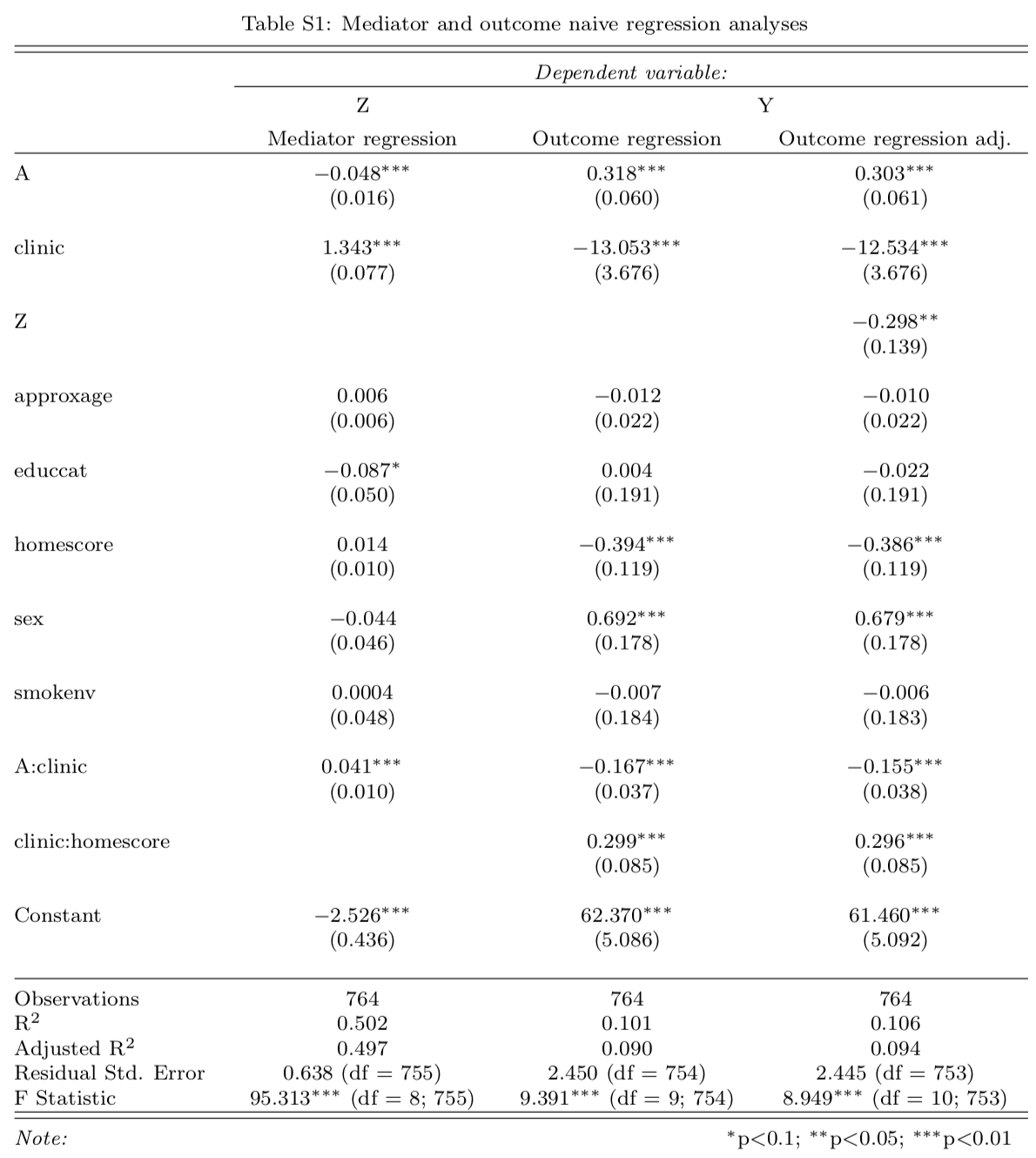}
 \end{center}
\end{figure}

\newpage

\begin{figure}[p!]
 \begin{center}
 \label{tab:S2}
 %\caption*{}
 \includegraphics[width=\textwidth]{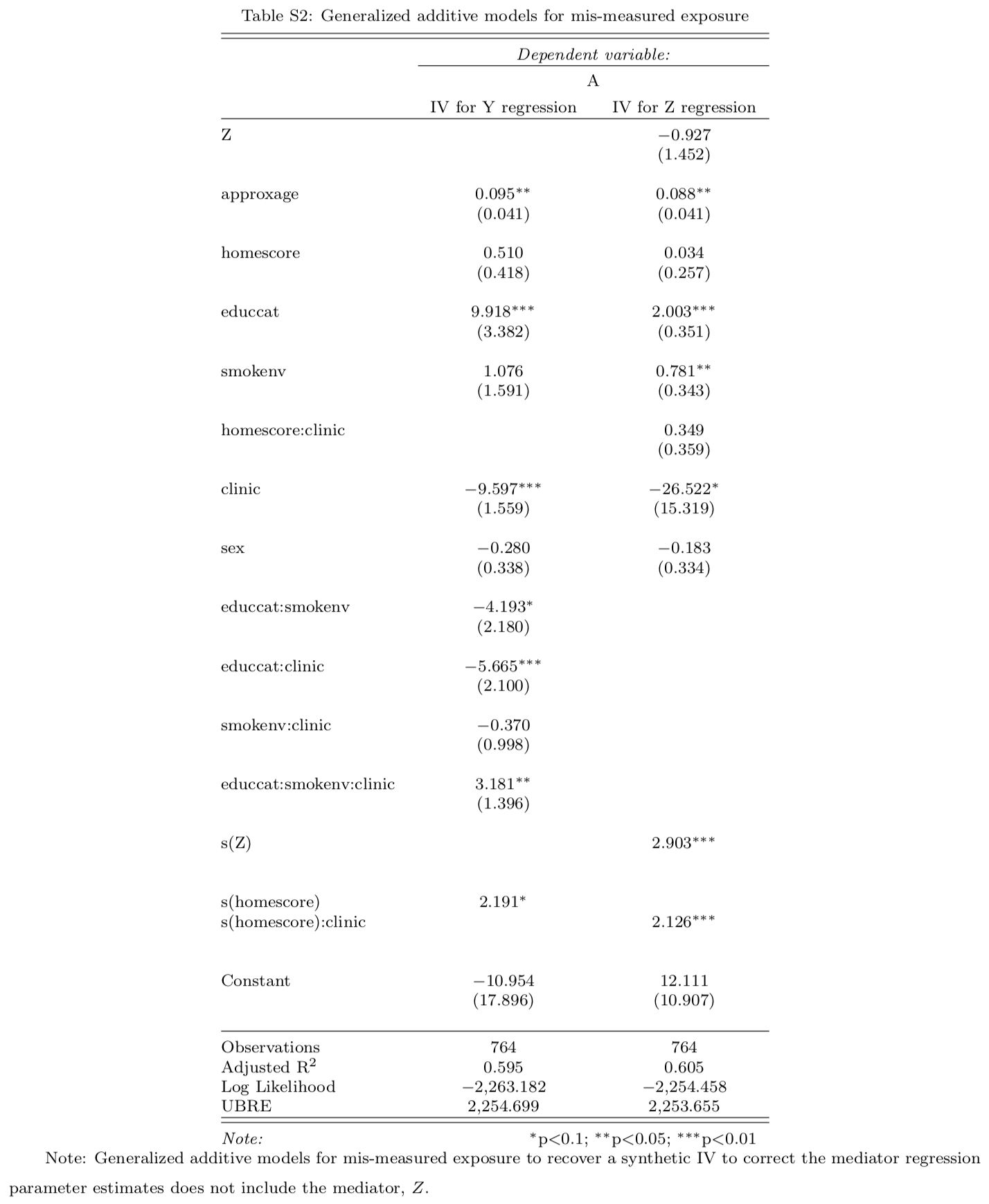}
 \end{center}
\end{figure}

\newpage

\begin{figure}[p!]
 \begin{center}
 \label{fig:S3}
 %\caption*{}
 \includegraphics[width=.85\textwidth]{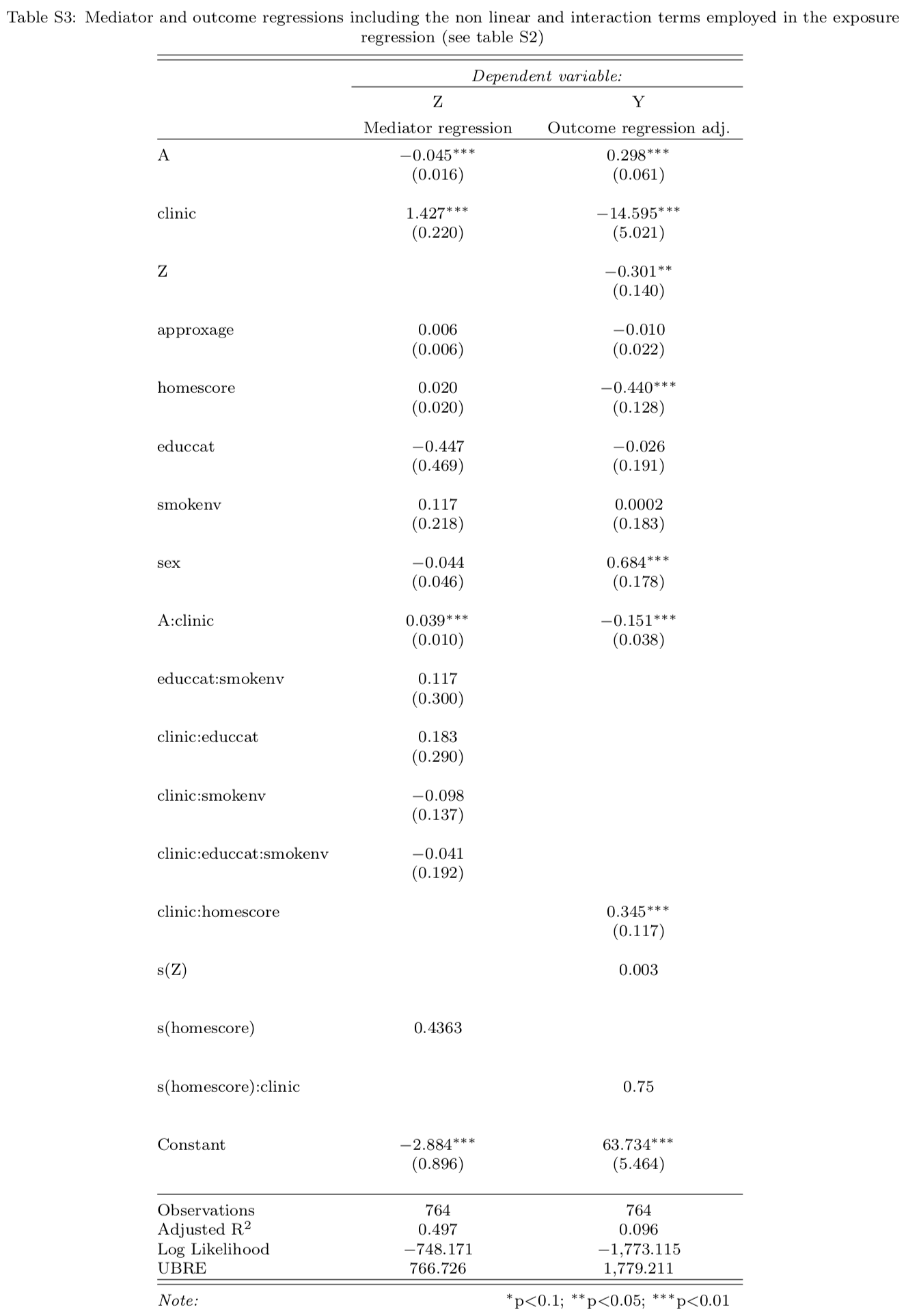}
 \end{center}
\end{figure}

\newpage

\begin{figure}[p!]
 \begin{center}
 \label{fig:S4}
 %\caption*{}
 \includegraphics[width=.85\textwidth]{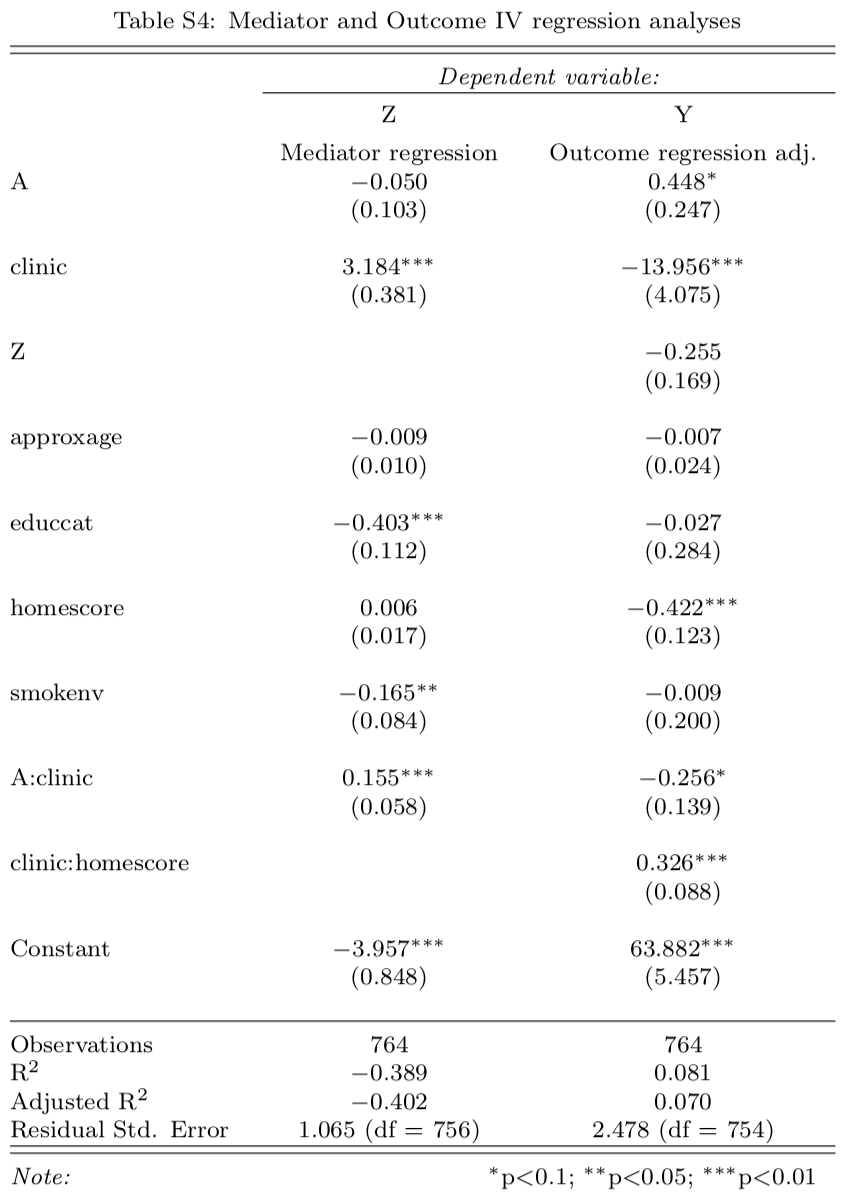}
 \end{center}
\end{figure}

\newpage

\begin{figure}[p!]
 \begin{center}
 \label{fig:sims}
 %\caption*{}
 \includegraphics[width=\textwidth]{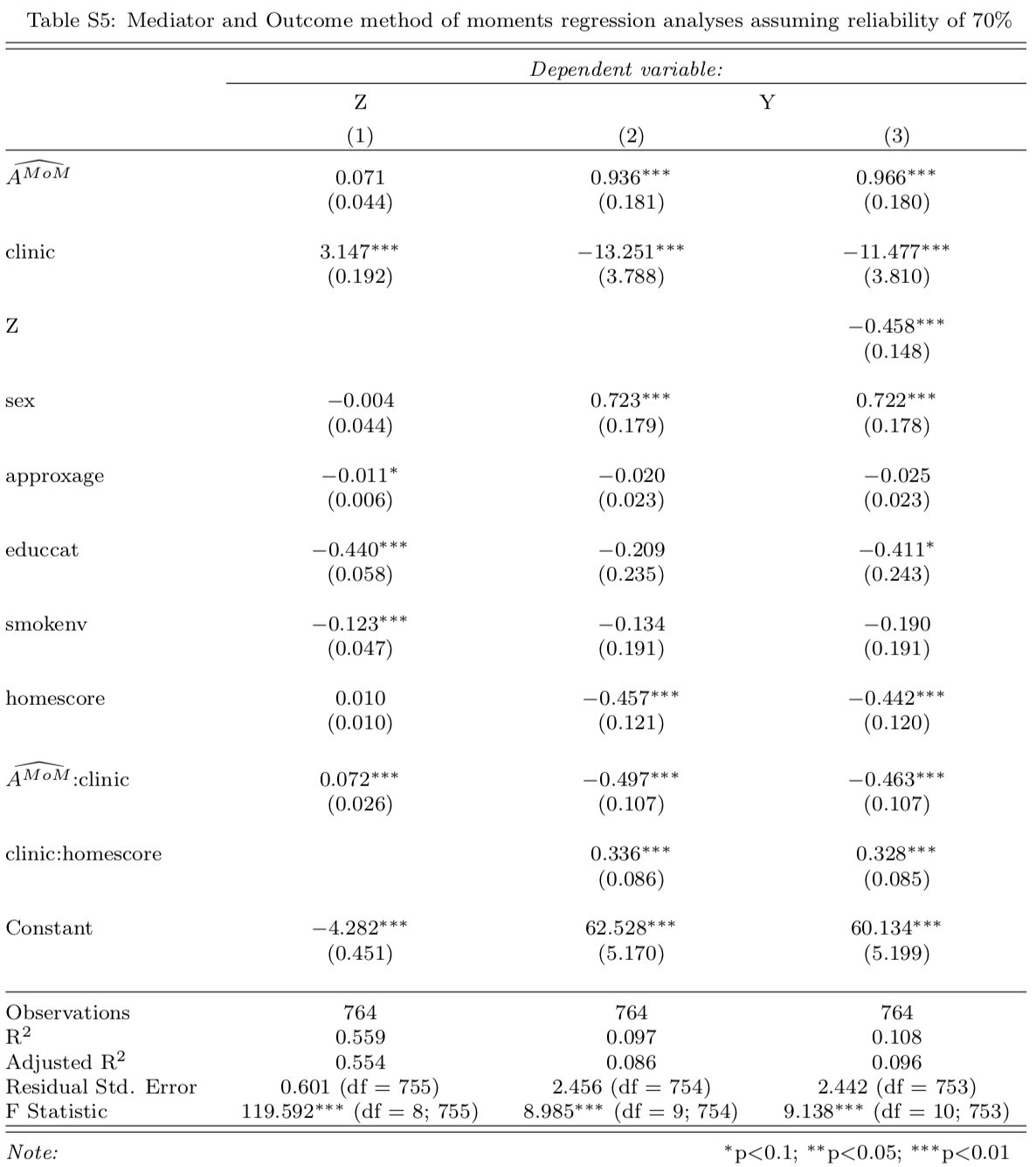}
 \end{center}
\end{figure}

\newpage

\begin{comment}
\begin{figure}[p!]
 \begin{center}
 \label{fig:sims}
 \caption*{}
 \includegraphics[page=4,width=\textwidth]{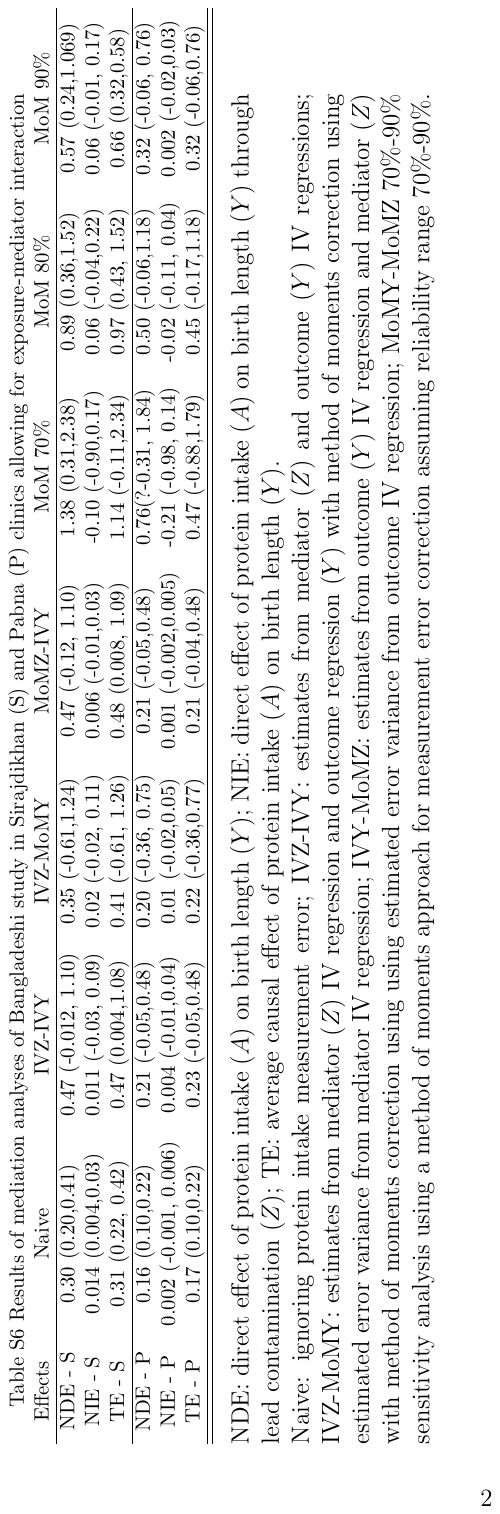}
 \end{center}
\end{figure}
\end{comment}

\setcounter{table}{5}

\begin{table}[p!]
\label{tab:site-results}
\caption{Results of measurement error-naive and measurement error-adjusted mediation analyses of Bangladeshi study in Sirajdikhan and Pabna clinics (including exposure-mediator interaction).}
\centering
\begin{tabular}{ll r@{.}l@{ } r@{.}l@{ } r@{.}l r@{.}l@{ } r@{.}l@{ } r@{.}l r@{.}l@{ } r@{.}l@{ } r@{.}l}
\hline \hline
Site    &   Estimator & \multicolumn{6}{l}{NDE} & \multicolumn{6}{l}{NIE}  & \multicolumn{6}{l}{TE} \\
\hline
Sirajdikhan &   Naive     & 0&30  & (0&20,   & 0&41)  & 0&014 & (0&004,  & 0&03)  & 0&31  & (0&22,  & 0&42) \\
&   IVZ-IVY   & 0&47 & (-0&012, & 1&10)  & 0&011 & (-0&03,  & 0&09)  & 0&47  & (0&004, & 1&08) \\
&   IVZ-MoMY  & 0&35 & (-0&61,  & 1&24)  & 0&02  & (-0&02,  & 0&11)  & 0&41  & (-0&61, & 1&26) \\
&   MoMZ-IVY  & 0&47 & (-0&12,  & 1&10)  & 0&006 & (-0&01,  & 0&03)  & 0&48  & (0&008, & 1&09) \\
&   MoM 70\%  & 1&38 & (0&31,   & 2&38)  & -0&10  & (-0&90,  & 0&17)  & 1&14  & (-0&11, & 2&34) \\
&   MoM 80\%  & 0&89 & (0&36,   & 1&52)  & 0&06  & (-0&04,  & 0&22)  & 0&97  & (0&43,  & 1&52) \\
&   MoM 90\%  & 0&57 & (0&24,   & 1&069) & 0&06  & (-0&01,  & 0&17)  & 0&66  & (0&32,  & 0&58) \\
\hline
Pabna   &   Naive     & 0&16 & (0&10,   & 0&22)  & 0&002 & (-0&001, & 0&006) & 0&17  & (0&10,  & 0&22) \\
&   IVZ-IVY   & 0&21 & (-0&05,  & 0&48)  & 0&004 & (-0&01,  & 0&04)  & 0&23  & (-0&05, & 0&48) \\
&   IVZ-MoMY  & 0&20  & (-0&36,  & 0&75)  & 0&01  & (-0&02,  & 0&05)  & 0&22  & (-0&36, & 0&77) \\
&   MoMZ-IVY  & 0&21 & (-0&05,  & 0&48)  & 0&001 & (-0&002, & 0&005) & 0&21  & (-0&04, & 0&48) \\
&   MoM 70\%  & 0&76 & (-0&31,  & 1&84)  & -0&21 & (-0&98,  & 0&14)  & 0&47  & (-0&88, & 1&79) \\
&   MoM 80\%  & 0&50  & (-0&06,  & 1&18)  & -0&02 & (-0&11,  & 0&04)  & 0&45  & (-0&17, & 1&18) \\
&   MoM 90\%  & 0&32 & (-0&06,  & 0&76)  & 0&002 & (-0&02,  & 0&03)  & 0&32  & (-0&06, & 0&76) \\
\hline
\end{tabular}
\end{table}

\newpage

%\section*{Web Appendix I: Additional simulation results tables}

\newpage

\begin{table}[p!]
\caption{\label{tab:sims-weak}Simulation results in the no interaction case with weak constructed IV (NDE = 1.5, NIE = -0.15, TE = 1.35).}
\centering
\resizebox{\textwidth}{!}{
\begin{tabular}{ll r@{.}l r@{.}l r@{.}l r@{.}l r@{.}l r@{.}l r@{.}l r@{.}l r@{.}l r@{.}l r@{.}l r@{.}l }
\\
\hline\hline
\multicolumn{2}{l}{Reliability ratio:}          & \multicolumn{6}{c}{50\%}    & \multicolumn{6}{c}{70\%}    & \multicolumn{6}{c}{80\%}    & \multicolumn{6}{c}{90\%}    \\
 &  & \multicolumn{2}{c}{Bias}  & \multicolumn{2}{c}{S.D.} & \multicolumn{2}{c}{C.P.} & \multicolumn{2}{c}{Bias}  & \multicolumn{2}{c}{S.D.} & \multicolumn{2}{c}{C.P.} & \multicolumn{2}{c}{Bias}  & \multicolumn{2}{c}{S.D.} & \multicolumn{2}{c}{C.P.} & \multicolumn{2}{c}{Bias}  & \multicolumn{2}{c}{S.D.} & \multicolumn{2}{c}{C.P.} \\
 \hline
Naive     & NDE      & -0&76 &	0&20 &	0&06 &	-0&46 &	0&24 &	0&45 &	-0&31 &	0&26 &	0&76 &	-0&15 &	0&28 &	0&87  \\
%-0&47 & 0&23     & 0&56     & -0&32 & 0&25     & 0&84     & -0&17 & 0&27     & 0&89     \\
          & NIE      & 0&12 &	0&05 &	0&28 &	0&08 &	0&06 &	0&75 &	0&06 &	0&07 &	0&88 &	0&03 &	0&08 &	0&92  \\
          %0&08 & 0&06     & 0&77     & 0&05 & 0&08     & 0&87     & 0&03 & 0&09     & 0&91      \\
          & TE       & -0&65 &	0&20 &	0&13 &	-0&38 &	0&24 &	0&60 &	-0&25 &	0&26 &	0&79 &	-0&12 &	0&28 &	0&89  \\
          %-0&39  & 0&24     & 0&64     & -0&26 & 0&25     & 0&85     & -0&13 & 0&27     & 0&93     \\
IVZ-IVY & NDE      & -0&69 &	1&02 &	1&00 &	-0&49 &	1&31 &	1&00 &	-0&38 &	1&35 &	1&00 &	0&08 &	3&56 &	1&00 \\
%-0&02 & 0&35      & 0&95     & -0&01 & 0&35      & 0&95     & -0&01 & 0&35      & 0&95     \\
          & NIE      & 0&10 &	0&10 &	0&92 &	0&08 &	0&13 &	0&98 &	0&08 &	0&18 &	0&99 &	-0&01 &	0&54 &	1&00 \\
          %0&01  & 0&10      & 0&89     & 0&01 & 0&10      & 0&89     & 0&00 & 0&10      & 0&89     \\
          & TE       & -0&59 &	0&97 &	1&00 &	-0&41 &	1&22 &	1&00 &	-0&30 &	1&23 &	1&00 &	0&08 &	3&05 &	1&00   \\
          %-0&01  & 0&32     & 0&97     & -0&01  & 0&32     & 0&97     & 0&00  & 0&32     & 0&96     \\
IVZ-MoMY  & NDE      & -0&55 &	0&59 &	0&85 &	-0&28 &	0&67 &	0&94 &	-0&15 &	0&71 &	0&95 &	-0&04 &	0&75 &	0&98  \\
%0&01 & 0&34      & 0&97     & 0&00 & 0&32      & 0&96     & 0&00 & 0&31      & 0&95     \\
          & NIE      & 0&05 &	0&09 &	1&00 &	0&01 &	0&12 &	1&00 &	-0&02 &	0&13 &	1&00 &	-0&04 &	0&15 &	1&00    \\
          %0&01  & 0&09      & 0&91     & 0&00 & 0&09      & 0&91     & 0&00 & 0&10      & 0&91      \\
          & TE       & -0&50 &	0&54 &	1&00 &	-0&27 &	0&59 &	1&00 &	-0&17 &	0&62 &	1&00 &	-0&08 &	0&64 &	1&00  \\
          %0&01  & 0&34     & 0&97     & 0&01  & 0&32     & 0&96     & 0&00  & 0&30     & 0&96     \\
MoMZ-IVY  & NDE      & -0&69 &	1&02 &	1&00 &	-0&49 &	1&31 &	1&00 &	-0&38 &	1&35 &	1&00 &	0&08 &	3&56 &	1&00   \\
%-0&02 & 0&35     & 0&95     & -0&01 & 0&35     & 0&95     & -0&01 & 0&35      & 0&95     \\
          & NIE      & -1&95 &	1&87 &	0&93 &	0&20 &	2&54 &	0&97 &	-0&29 &	0&92 &	0&98 &	-0&52 &	3&87 &	0&98   \\
          %0&00 & 0&16     & 0&94     & -0&01 & 0&12     & 0&93     & 0&00 & 0&12     & 0&93      \\
          & TE       & 0&51 &	1&97 &	0&86 &	-0&29 &	2&58 &	0&92 &	-0&66 &	1&41 &	0&95 &	-0&44 &	2&44 &	0&97   \\
          %-0&02 & 0&32     & 0&96     & -0&02 & 0&31     & 0&95     & -0&01     & 0&31      & 0&94     \\
MoM-70   & NDE      & 0&08 &	0&44 &	0&94 &	3&05 &	1&22 &	0&26 &	11&29 &	12&15 &	0&14 &	-50&90 &	457&11 &	1&00 \\
%0&03  & 0&35     & 0&96     & 0&40  & 0&41     & 0&87     & 0&84   & 0&49     & 0&67     \\
          & NIE      & -0&01 &	0&10 &	0&96 &	-1&01 &	0&46 &	0&22 &	-6&50 &	11&61 &	0&13 &	69&22 &	461&34 &	1&00   \\
          %-0&00     & 0&10     & 0&94     & -0&07 & 0&13     & 0&92     & -0&18  & 0&16     & 0&79     \\
          & TE       & 0&08 &	0&42 &	0&95 &	2&04 &	0&92 &	0&36 &	4&79 &	1&74 &	0&10 &	18&32 &	66&04 &	0&41    \\
          %0&03  & 0&34     & 0&95     & 0&32  & 0&39     & 0&88     & 0&66   & 0&44     & 0&78     \\
MoM-80   & NDE      & -0&55 &	0&26 &	0&45 &	0&04 &	0&36 &	0&94 &	0&42 &	0&43 &	0&82 &	0&87 &	0&50 &	0&58  \\
%-0&23 & 0&29      & 0&93     & 0&01  & 0&32     & 0&96     & 0&28  & 0&37     & 0&86     \\
          & NIE      & 0&09 &	0&06 &	0&62 &	0&00 &	0&10 &	0&94 &	-0&08 &	0&12 &	0&90 &	-0&20 &	0&15 &	0&75    \\
          %0&04 & 0&08     & 0&91     & 0&00     & 0&10     & 0&93     & -0&05 & 0&12     & 0&93     \\
          & TE       & -0&45 &	0&26 &	0&53 &	0&04 &	0&35 &	0&92 &	0&34 &	0&41 &	0&81 &	0&67 &	0&46 &	0&66 \\
          %-0&19 & 0&29      & 0&93     & 0&02  & 0&32     & 0&95     & 0&23  & 0&35     & 0&90     \\
MoM-90   & NDE      & -0&65 &	0&23 &	0&19 &	-0&22 &	0&30 &	0&87 &	0&03 &	0&34 &	0&94 &	0&30 &	0&38 &	0&83 \\
%-0&37 & 0&25     & 0&75     & -0&19 & 0&28     & 0&92     & 0&00     & 0&31     & 0&96     \\
          & NIE      & 0&11 &	0&05 &	0&45 &	0&05 &	0&08 &	0&92 &	0&00 &	0&10 &	0&93 &	-0&05 &	0&11 &	0&94 \\
          %0&06  & 0&07     & 0&84     & 0&03  & 0&08     & 0&91     & 0&00     & 0&10     & 0&92     \\
          & TE       & -0&55 &	0&23 &	0&30 &	-0&17 &	0&30 &	0&88 &	0&03 &	0&33 &	0&91 &	0&25 &	0&36 &	0&85    \\
          %-0&31 & 0&26     & 0&80     & -0&16 & 0&28     & 0&93     & 0&00     & 0&30     & 0&95    \\
\hline
\end{tabular}}
\end{table}

\begin{table}
\caption{\label{tab:sims-varying-func-form}Simulation results in the no interaction case with rank condition holding and varying constructed IV functional form (NDE = 1.5, NIE = -0.15, TE = 1.35).}
\centering
\resizebox{\textwidth}{!}{
\begin{tabular}{lll r@{.}l r@{.}l r@{.}l r@{.}l r@{.}l r@{.}l r@{.}l r@{.}l r@{.}l }
\\
\hline\hline
Constr.~IV    &   \multicolumn{2}{l}{Reliability ratio:}          & \multicolumn{6}{c}{70\%}    & \multicolumn{6}{c}{80\%}    & \multicolumn{6}{c}{90\%}    \\
func.~form &&  & \multicolumn{2}{c}{Bias}  & \multicolumn{2}{c}{S.D.} & \multicolumn{2}{c}{C.P.} & \multicolumn{2}{c}{Bias}  & \multicolumn{2}{c}{S.D.} & \multicolumn{2}{c}{C.P.} & \multicolumn{2}{c}{Bias}  & \multicolumn{2}{c}{S.D.} & \multicolumn{2}{c}{C.P.} \\
 \hline
2   &   IVZ-IVY & NDE      & -0&05 &	0&42 &	0&92 &	-0&04 &	0&42 &	0&92 &	-0&04 &	0&42 &	0&92    \\
          &   & NIE      & 0&01 &	0&10 &	0&92 &	0&01 &	0&10 &	0&92 &	0&01 &	0&10 &	0&92 \\
          &   & TE       & -0&04 &	0&41 &	0&92 &	-0&04 &	0&41 &	0&92 &	-0&03 &	0&41 &	0&92  \\
&   IVZ-MoMY  & NDE      & 0&02 &	0&41 &	0&96 &	0&01 &	0&37 &	0&95 &	-0&01 &	0&35 &	0&94    \\
          &   & NIE      & 0&00 &	0&10 &	0&94 &	0&00 &	0&10 &	0&92 &	0&01 &	0&10 &	0&91 \\
          &   & TE       & 0&03 &	0&41 &	0&91 &	0&01 &	0&37 &	0&92 &	-0&01 &	0&35 &	0&93    \\
&   MoMZ-IVY  & NDE      & -0&05 &	0&42 &	0&92 &	-0&04 &	0&42 &	0&92 &	-0&04 &	0&42 &	0&92  \\
          &   & NIE      & 0&02 &	0&27 &	0&94 &	0&00 &	0&13 &	0&94 &	0&00 &	0&11 &	0&93 \\
          &   & TE       & -0&03 &	0&46 &	0&94 &	-0&05 &	0&41 &	0&94 &	-0&04 &	0&40 &	0&93  \\
3   &   IVZ-IVY & NDE      & 18&93 &	17&57 &	1&00 &	18&81 &	18&08 &	1&00 &	18&93 &	17&90 &	1&00    \\
%-0&05 &	0&42 &	0&92 &	-0&04 &	0&42 &	0&92 &	-0&04 &	0&42 &	0&92    \\
          &   & NIE      & 75&96 &	69&75 &	1&00 &	74&24 &	70&30 &	1&00 &	71&35 &	66&72 &	1&00   \\
          %0&01 &	0&10 &	0&92 &	0&01 &	0&10 &	0&92 &	0&01 &	0&10 &	0&92 \\
          &   & TE       & 94&89 &	86&00 &	1&00 &	93&05 &	87&29 &	1&00 &	90&28 &	83&77 &	1&00   \\
          %-0&04 &	0&41 &	0&92 &	-0&04 &	0&41 &	0&92 &	-0&03 &	0&41 &	0&92  \\
&   IVZ-MoMY  & NDE      & -6&17 &	4&67 &	0&99 &	-5&78 &	4&16 &	0&99 &	-5&44 &	3&83 &	0&99  \\
%0&02 &	0&41 &	0&96 &	0&01 &	0&37 &	0&95 &	-0&01 &	0&35 &	0&94    \\
          &   & NIE      & -24&79 &	22&61 &	1&00 &	-24&45 &	23&97 &	1&00 &	-23&28 &	20&66 &	1&00    \\
          %0&00 &	0&10 &	0&94 &	0&00 &	0&10 &	0&92 &	0&01 &	0&10 &	0&91 \\
          &   & TE       & -30&96 &	26&74 &	1&00 &	-30&23 &	27&57 &	1&00 &	-28&72 &	23&92 &	1&00    \\
          %0&03 &	0&41 &	0&91 &	0&01 &	0&37 &	0&92 &	-0&01 &	0&35 &	0&93    \\
&   MoMZ-IVY  & NDE      & -25&98 &	17&57 &	1&00 &	-25&78 &	18&08 &	1&00 &	-24&76 &	17&90 &	1&00    \\
%-0&05 &	0&42 &	0&92 &	-0&04 &	0&42 &	0&92 &	-0&04 &	0&42 &	0&92  \\
          &   & NIE      & -17&25 &	15&50 &	1&00 &	-18&09 &	19&93 &	1&00 &	-19&34 &	25&49 &	1&00    \\
          %0&02 &	0&27 &	0&94 &	0&00 &	0&13 &	0&94 &	0&00 &	0&11 &	0&93 \\
          &   & TE       & 1&68 &	2&93 &	0&99 &	0&71 &	9&34 &	0&99 &	-0&41 &	19&38 &	0&99   \\
          %-0&03 &	0&46 &	0&94 &	-0&05 &	0&41 &	0&94 &	-0&04 &	0&40 &	0&93  \\
\hline
\end{tabular}}
\end{table}

\begin{table}
\caption{\label{tab:sims-ate}Simulation results for the ATE (=0.5) in the no interaction case with rank condition holding.}
\centering
\begin{tabular}{l r@{.}l r@{.}l r@{.}l r@{.}l r@{.}l r@{.}l r@{.}l r@{.}l r@{.}l }
\\
\hline\hline
Reliability ratio:          & \multicolumn{6}{c}{70\%}    & \multicolumn{6}{c}{80\%}    & \multicolumn{6}{c}{90\%}    \\
 & \multicolumn{2}{c}{Bias}  & \multicolumn{2}{c}{S.D.} & \multicolumn{2}{c}{C.P.} & \multicolumn{2}{c}{Bias}  & \multicolumn{2}{c}{S.D.} & \multicolumn{2}{c}{C.P.} & \multicolumn{2}{c}{Bias}  & \multicolumn{2}{c}{S.D.} & \multicolumn{2}{c}{C.P.} \\
 \hline
Naive      & -0&12 &	0&30 &	0&93 &	-0&08 &	0&31 &	0&92 &	-0&05 &	0&31 &	0&91    \\
%-0&47 & 0&23     & 0&56     & -0&32 & 0&25     & 0&84     & -0&17 & 0&27     & 0&89     \\
Constr.~IV      & -0&02 &	0&33 &	0&90 &	-0&01 &	0&33 &	0&90 &	-0&01 &	0&33 &	0&90 \\
%-0&02 & 0&35      & 0&95     & -0&01 & 0&35      & 0&95     & -0&01 & 0&35      & 0&95     \\
MoM-70      & 0&00 &	0&31 &	0&93 &	0&08 &	0&32 &	0&94 &	0&17 &	0&33 &	0&92    \\
%0&01 & 0&34      & 0&97     & 0&00 & 0&32      & 0&96     & 0&00 & 0&31      & 0&95     \\
MoM-80      & -0&06 &	0&31 &	0&92 &	-0&01 &	0&31 &	0&91 &	0&05 &	0&32 &	0&94  \\
%-0&02 & 0&35     & 0&95     & -0&01 & 0&35     & 0&95     & -0&01 & 0&35      & 0&95     \\
MoM-90      & -0&09 &	0&31 &	0&93 &	-0&05 &	0&31 &	0&91 &	-0&01 &	0&31 &	0&91   \\
%0&03  & 0&35     & 0&96     & 0&40  & 0&41     & 0&87     & 0&84   & 0&49     & 0&67     \\
\hline
\end{tabular}
\end{table}

%Insert Tables 3 \& 4 from simulation study here\\
\begin{figure}[p!]
 %\caption*{\label{fig:simstables}}
 \begin{center}
\includegraphics[angle=90,origin=c,page=2,width=\textwidth]{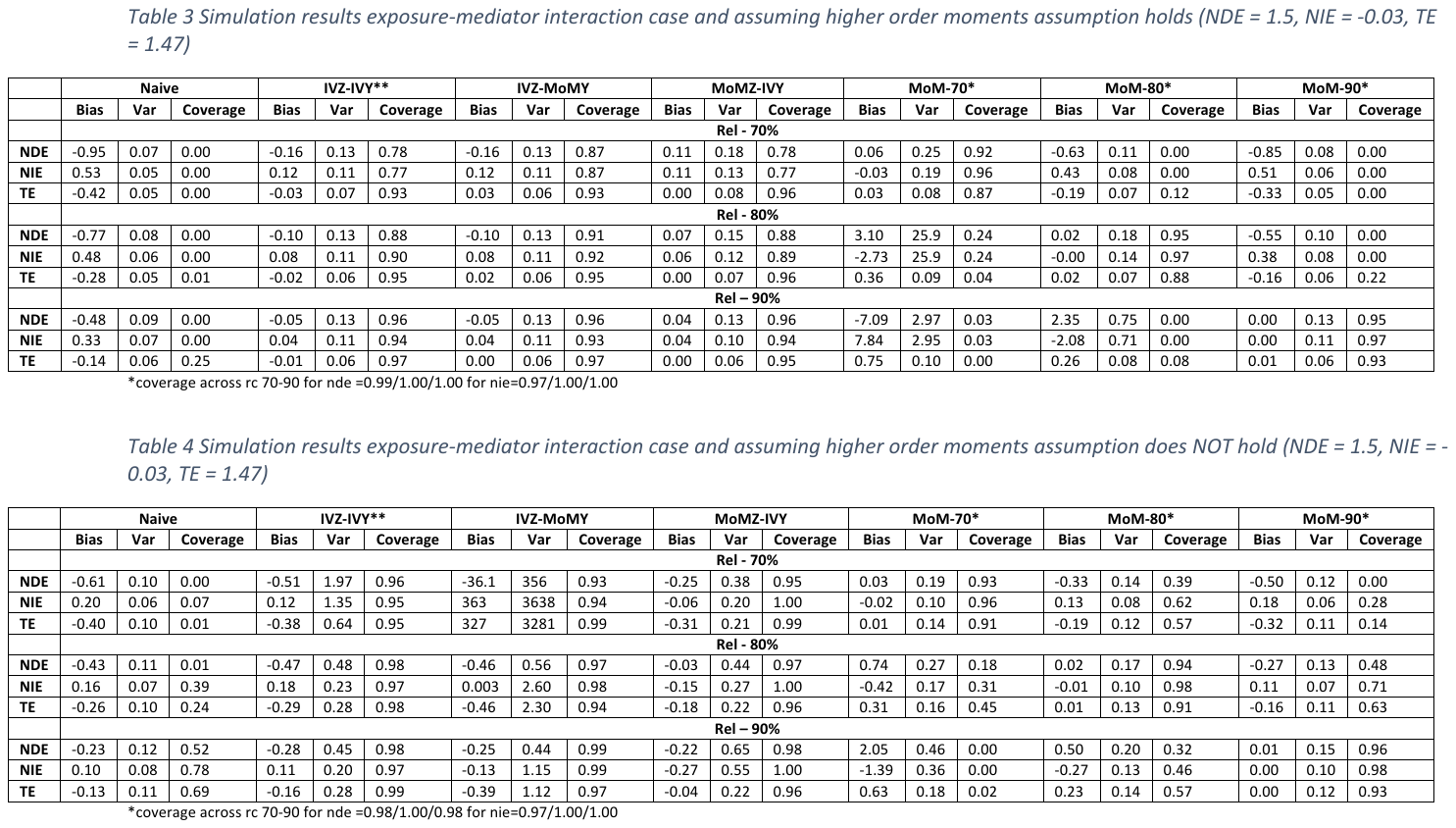}
 \end{center}
\end{figure}

\clearpage

\pagenumbering{gobble}

\end{document}